\documentclass[
  aps,prb,            
  reprint,            
  superscriptaddress, 
  longbibliography,   
]{revtex4-2}

\usepackage[utf8]{inputenc}      
\usepackage[T1]{fontenc}         
\usepackage{graphicx}            
\usepackage{amsmath,amssymb}     
\usepackage{dcolumn}             
\usepackage{bm}                  
\usepackage{hyperref}   
\usepackage{subcaption}
\usepackage{dsfont} 
\usepackage{physics}           
\usepackage{blkarray}

\newcommand*{\Comb}[2]{{}^{#1}C_{#2}}%

\usepackage{xcolor}
\usepackage[singlelinecheck=false]{caption}
\usepackage{comment}

\begin{document}

\title{Hidden $Z_{2}\times Z_{2}$ subspace symmetry protection for quantum scars}

\author{Ayush Sharma}
\email{ayush.sharma@tifr.res.in}
\affiliation{Department of Theoretical Physics, Tata Institute of Fundamental Research, Homi Bhabha Road, Navy Nagar, Mumbai 400005, India}

\author{Vikram Tripathi}
\email{vtripathi@theory.tifr.res.in}

\affiliation{Department of Theoretical Physics, Tata Institute of Fundamental Research, Homi Bhabha Road, Navy Nagar, Mumbai 400005, India}


\date{\today}

\begin{abstract}
We study the paradigmatic spin-1 XY chain under open boundary conditions, which hosts exact quantum many-body scars generated by an emergent Spectrum Generating Algebra (SGA). We show that the scar subspace possesses a symmetry-protected trivial (SPt) character that we attribute to a hidden $Z_{2}\times Z_{2}$ symmetry of another model, namely the commutant Hamiltonian, for which the scars are the ground states. We construct a Lieb-Schultz-Mattis (LSM) type twist operator, which, for scar states, takes the value $-1,$ and, for ergodic states, approaches zero in the thermodynamic limit.  A complementary understanding of the stability of the scars under different perturbations is obtained by analyzing the Loschmidt echo and Quantum Fisher Information (QFI) of the scars. Finite-size scaling analysis of the QFI reveals that the scars are much more sensitive to perturbations as compared to the nearby thermal states. Based on the analysis of QFI and different LSM twist operators, we obtain a classification of different SGA-preserving and SGA-breaking perturbations.
\end{abstract}

\maketitle


\section{Introduction}

A variety of mechanisms are known to violate the Eigenstate Thermalization Hypothesis (ETH)~\cite{PhysRevE.50.888, MarkSrednicki_1999, MarkSrednicki_1996}, including many-body localization (MBL)~\cite{RevModPhys.91.021001}, integrability~\cite{Calabrese_2016}, Hilbert-space fragmentation~\cite{Moudgalya_2022}, dipole-conserving dynamics~\cite{PhysRevX.10.011047, PhysRevB.101.174204}, and quantum many-body scars (QMBS)~\cite{Moudgalya_2022, PhysRevB.98.155134}.

There is intense interest in QMBS because they constitute a vanishing fraction of eigenstates that deviate from ETH — yet leave sharp dynamical signatures for certain initial states, evidenced, for example, in Rydberg-atom simulators~\cite{Bernien2017}.  Numerous scar-hosting models have been discovered, and explanatory frameworks have been developed, including the Shiraishi–Mori embedding~\cite{PhysRevLett.119.030601}, Generalized Projector embedding~\cite{Omiya_embedding_PXP, Fractionalization_pavesway}, quasi-symmetry approaches~\cite{PhysRevLett.126.120604}, group-invariant sectors~\cite{PhysRevLett.125.230602}, and “tunnels-to-tower” constructions~\cite{PhysRevResearch.2.043305}, commutant algebra~\cite{PhysRevX.12.011050}, and spectrum generating algebra (SGA)~\cite{PhysRevB.102.085140}. 
In particular, SGA provides a natural construction of scar towers with some symmetries—such towers are physically important as they leave a clear dynamical imprint on the time evolution of the expectation of local observables: equally spaced scar levels produce pronounced revivals in fidelity. 
SGA-based constructions often lead to analytically tractable scars and align with the widespread $SU(2)$ structure observed in a variety of QMBS models~\cite{PhysRevB.102.085140,PhysRevB.102.075132, PhysRevB.98.235155, PhysRevB.98.235156, PhysRevB.101.195131, PhysRevB.101.024306}. Here we study the spin-1 XY chain with open boundary conditions~\cite{PhysRevB.101.024306} where the scars are constructed using the SGA.

We argue that the existence of SGA guarantees the preservation of the scar subspace but not of individual scars themselves. In contrast, the individual scars are protected against perturbations under the more restricted set of commutant symmetries~\cite{PhysRevX.12.011050, mohapatra2025unravelingadditionalquantummanybody}—the set of operators that commute with each local term of the Hamiltonian (Bond algebra) \footnote{It should be clarified here that the definition of the Bond algebra is important, and we will permit it to contain extensive local operators~\cite{PhysRevX.14.041069} that should be regarded as a single bond operator.}.

We find that the scar states exhibit a symmetry-protected trivial (SPt) character. To identify the symmetry group responsible for protecting this SPt structure, we construct a commutant Hamiltonian, defined as a Hamiltonian built entirely from operators in the commutant algebra, whose ground state manifold consists of the scars. It turns out that the commutant Hamiltonian is integrable, as it possesses an extensive set of exponentially many conserved local charges given by elements of the bond algebra~\cite{PRXQuantum.5.040330, PhysRevX.12.011050}. We argue that it is precisely the symmetries of this commutant Hamiltonian that endow the scar subspace with its SPt character, which turns out to be a $Z_{2} \times Z_{2}$ symmetry corresponding to a sublattice symmetry and a flipping symmetry, which essentially flips the tower of the scars.

To detect the SPt character of the scars, we construct a Lieb–Schultz–Mattis (LSM) type twist operator that measures the polarization of the conserved charge of the commutant Hamiltonian. We observe that all scar states lie at $-1$ on the unit circle, whereas the ergodic eigenstates cluster around zero. While SPt phases are known to occur in various models, they are typically associated with ground states. In contrast, in the present setting the SPt character is not tied to the ground state of the physical Hamiltonian, making this realization qualitatively distinct from conventional examples.

Our analysis further reveals that the individual scar states are protected both by the spectrum-generating algebra (SGA) and by the $U(1)$ symmetry of the original Hamiltonian. Interestingly, even upon explicitly breaking the $U(1)$ symmetry, the scar states persist, as does the SGA structure, although in this case the generators of the SGA must be appropriately redefined. 

In a recent parallel work~\cite{matsui2025symmetryprotectedtopologicalscarsubspaces}, it has been shown that the scars in the AKLT model also possess a $Z_{2}\times Z_{2}$ symmetry protection (inherited from the SPt character of the ground state), which can be detected using the string order parameter~\cite{Kennedy1992}. However, in our case, the string order parameter and LSM twist operator are different from those associated with the SPt ground state in the Haldane gap regime. Secondly, in our case, the symmetry protection is not inherited from the ground state, for the scar towers can survive even in parameter regimes where the ground state is non-SPt. This difference between the works lies in the fact that the root state in our case may or may not be the ground state for certain parameter regimes, but one still gets the SPt character independent of the ground state properties.

To understand the stability of scar states, we use the Loschmidt echo along with the fact that the short-time expansion of the Loschmidt echo is inversely proportional to the Quantum Fisher Information (QFI). We show that the QFI for scarred states scales super extensively, $\sim N^{2}$ which has also been reported in a recent work~\cite{PhysRevLett.129.020601} in the context of the PXP model. We obtain QFI analytically for the scar towers from the expressions of the $k$-body reduced density matrix and decomposition of operators onto the SGA generators. We find three regimes for the scaling of QFI depending on the type of perturbation operators we consider: (a) a super-linear scaling $\sim L^{2}$ (for extensive local operators that preserve the scar subspace), (b) a linear scaling $\sim L$ (for extensive local operators that do not preserve the scar subspace), and finally (c) a constant QFI $\sim O(1)$ (for intensive local operators that do not preserve the scar subspace). We thus classify the perturbations as scar-preserving or not, probed by the QFI of the perturbation. We would also like to point out that this classification does not hold for any operator, but rather for an operator in a certain basis.
 
Differences in the scaling behavior of the QFI for distinct operators provide a quantitative measure of the sensitivity of scar states to perturbations in the regimes discussed above. We find that scar states with the largest QFI are the most sensitive under generic perturbations. Moreover, we observe that the QFI scaling of asymptotic quantum many-body scars (AQMBS)~\cite{PhysRevLett.131.190401} closely matches that of the exact scars, indicating a similar multipartite entanglement structure—an observation that, to our knowledge, has not been previously reported.

The outline of the rest of the paper is as follows: In Sec.~\eqref{sec:2_model}, We study a spin-1 XY model and its extensions incorporating disorder and specific classes of long-range interactions. We find that the scar ansatz originally formulated for the short-range model remains applicable in these more general settings. Numerical results are presented to substantiate the existence of scars. We also introduce the commutant Hamiltonian, which is central for understanding the SPt character of scars. In Sec.~\eqref{sec: spt}, the construction of an appropriate Lieb–Schultz–Mattis twist operator that diagnoses the symmetry protection of the scarred states is explored. In Sec.~\eqref{sec:qfi}, we report the finite-size scaling behavior of the quantum Fisher information and an investigation of its dynamical stability, which we relate to the quantum Fisher information through the Loschmidt echo. In Sec~\eqref{sec:conclusion} we summarize our work and discuss the possible directions for future research.

\section{Model}\label{sec:2_model}


We begin our analysis with the following spin-1 XY chain, which is a modification of the chain introduced by Schecter et al.~\cite{PhysRevLett.123.147201} in the context of QMBS, with open boundary conditions (OBC),

\begin{align}\label{eq:Hamiltonian-Spin1-XY}
    H & = \sum_{\alpha \in odd} \sum_{i} J_{\alpha, i} \left(S^{x}_{i}S^{x}_{i+\alpha}+S^{y}_{i}S^{y}_{i+\alpha}\right) \nonumber \\
       & \qquad\qquad + \sum_{i} D_{i} \left( S^{z}_{i}\right)^{2} +h\sum_{i}S^{z}_{i}.
\end{align}

where \(S_{i}^{\alpha}\) (\(\alpha=x,y,z\)) are spin-1 operators at site \(i\).
We work in a fixed \(U(1)\) sector of total magnetization \(M_z=\sum_i S_i^z=m\).
The longer-range XY couplings \(J_3\) (here third-neighbor, \(i\to i+3\)), $J_{5}$ etc. explicitly break a hidden nonlocal symmetry~\cite{AtsuhiroKitazawa_2003} of the nearest-neighbor OBC chain but does not affect the scar states of the model, and hence with these terms present, Wigner–Dyson statistics is obeyed within each magnetization symmetry sector \footnote{With only nearest-neighbor XY interactions and special boundaries, certain nonlocal algebraic structures can survive and obscure bulk level-statistics. Adding longer-range XY terms removes these accidental structures while preserving the global \(U(1)\) symmetry as well as the scar structure.}. The long-range model considered here is non-integrable, as shown in the supplemental material.

\begin{figure*}[t!]
    \centering

    \begin{subfigure}[t]{0.49\linewidth}
        \centering
        \includegraphics[width=\linewidth]{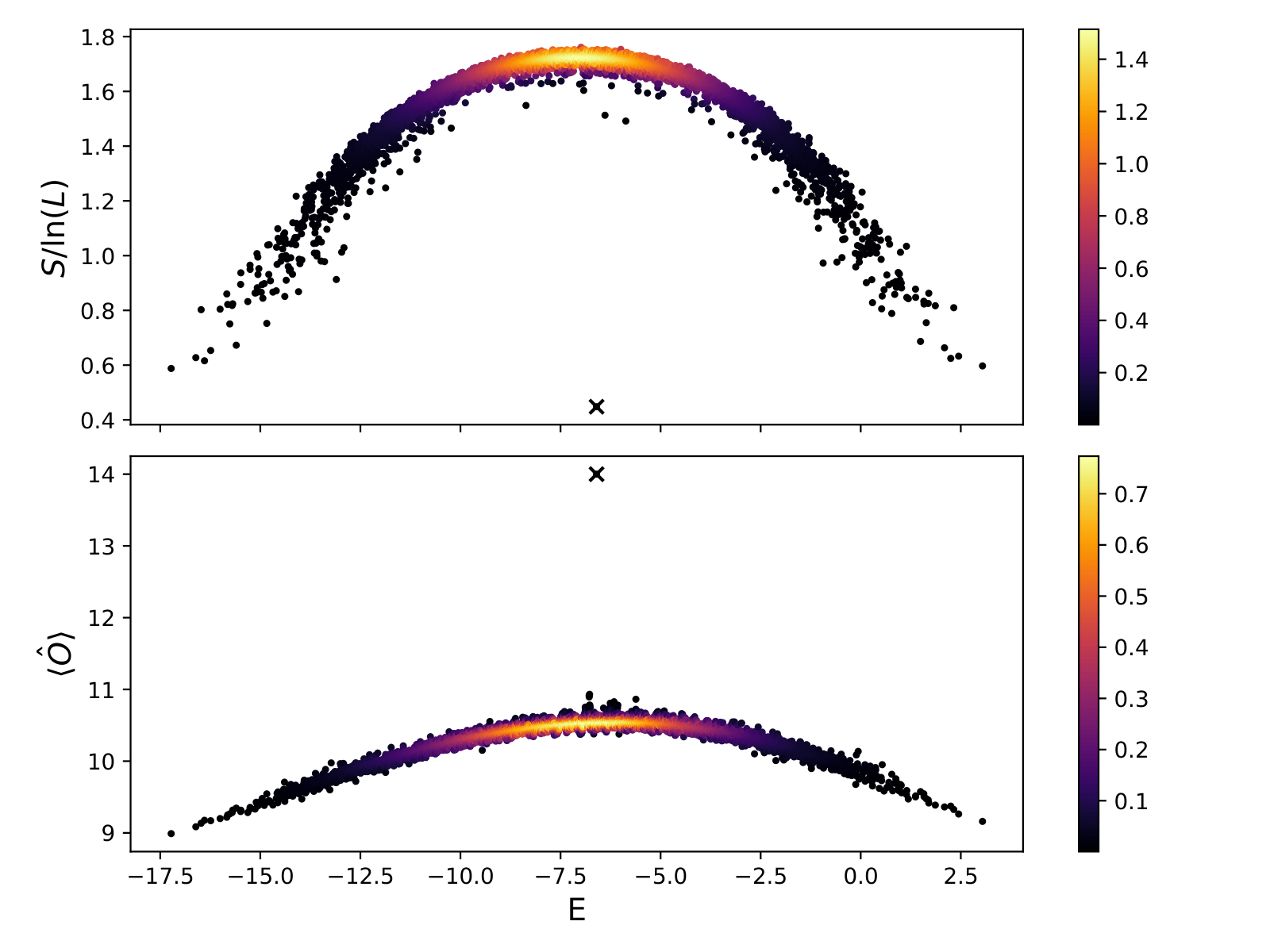}
        \caption[Short caption]{%
            Diagonal ETH and bipartite entanglement entropy for the model in Eq.~\eqref{eq:Hamiltonian-Spin1-XY} with uniform parameters $J_1=1$, $J_3=0.1$, $J_5=J_7=J_9=J_{11}= J_{13}=0$ $D=0.1$, $h=1$ for system size $L=14$, magnetization $m=-8$, and state parity $I=-1$ under chain inversion.%
        }
        \label{fig:eth_entropy_short}
    \end{subfigure}
    \hfill
    \begin{subfigure}[t]{0.49\linewidth}
        \centering
        \includegraphics[width=\linewidth]{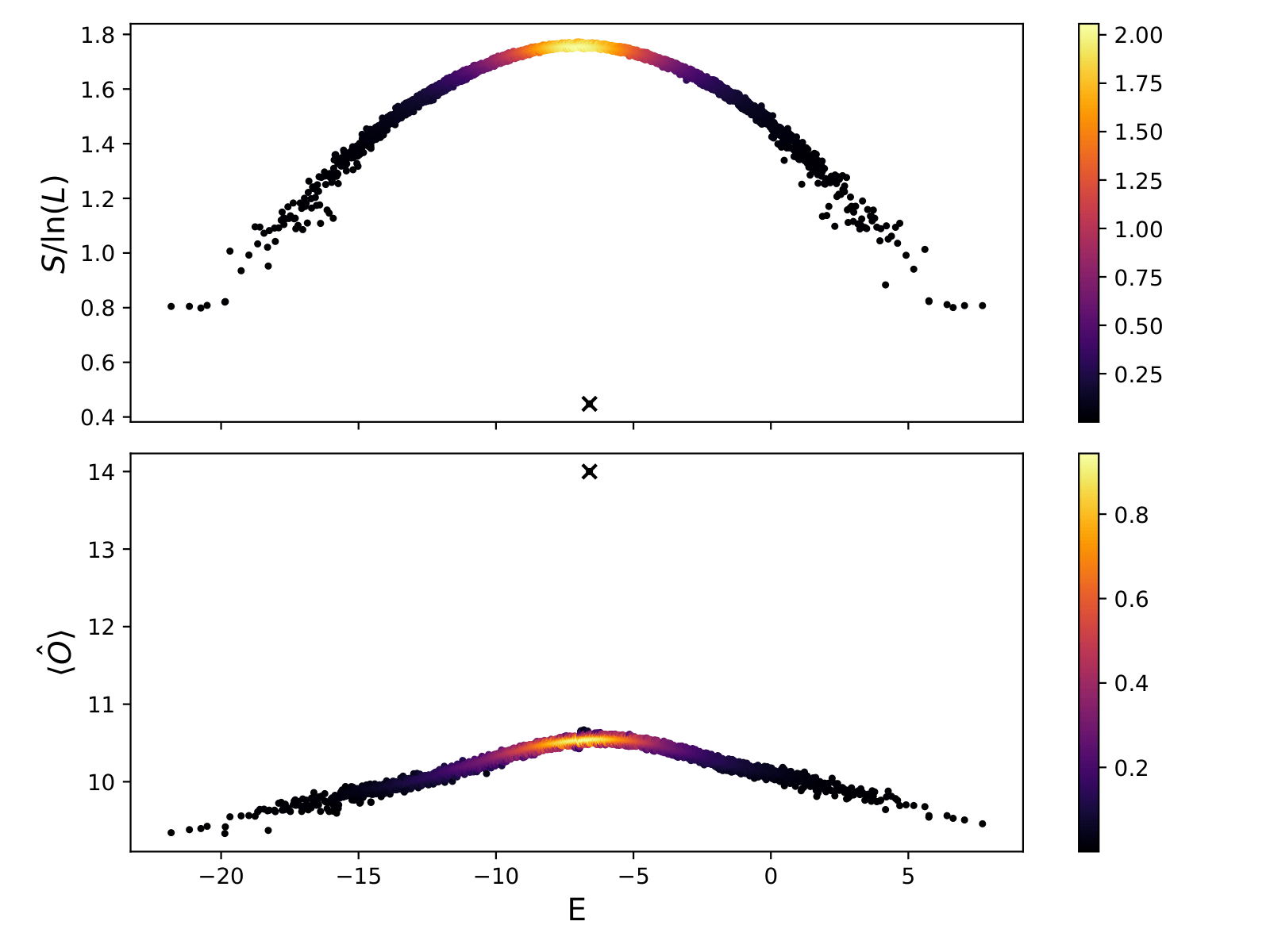}
        \caption[Long caption]{%
            Diagonal ETH and bipartite entanglement entropy for the long-range interacting model introduced, with the parameters $J_1=1$, $J_3=0.1$, $J_5=0.2$, $J_7=0.5$, $J_9=0.4$, $J_{11}=0.6$, $J_{13}=0.9$, $D=0.1$, $h=1$, system size $L=14$, magnetization $m=-8$, and state parity $I=-1$ under chain inversion.%
        }
        \label{fig:eth_entropy_longrange}
    \end{subfigure}

    \caption{
    Diagonal ETH for the extensive local observable $\hat{O}=\sum_{i} (S^{z}_{i})^{2}$ and bipartite entanglement entropy spectrum for the half-chain. The outlier data point ($\times$) near the center of the energy spectrum in each of the sub-figures indicates the single scar state in the sector. Note that with the addition of long-range terms, the diagonal ETH, and the entanglement entropy values of scars are unchanged.}
\end{figure*}

\subsection{Scars, spectrum-generating algebra, and ETH diagnostics}\label{sec:2_SGA}

The disorder-free model with no long-range interactions is known to host quantum many-body scars (QMBS) as was first shown by Schecter and Iadecola~\cite{PhysRevLett.123.147201}. However, as we will show below, the same scars persist even when disorder and long-range interactions are introduced as in Eq.~\eqref{eq:Hamiltonian-Spin1-XY} . We first 
define the bimagnon operator
\begin{equation}
  Q^{\dagger} \;=\; \sum_{i=1}^{L} (-1)^{i}\,(S_i^{+})^{2}, 
  \qquad Q=(Q^\dagger)^{\dagger}.
  \label{eq:Qdag-def}
\end{equation}
Let $|\Omega\rangle=|-1,-1,\dots,-1 \rangle$ the fully polarized ``vacuum'' and construct the tower
\begin{equation}
  |S_n\rangle \;=\; \mathcal N_n\, (Q^\dagger)^n | \Omega\rangle, \qquad n=0,1,\dots,L.
  \label{eq:scar-tower}
\end{equation}
where $\mathcal{N}_{n}$ is the normalization for the $n^{th}$ scar state in the tower.

These bimagnon operators form an emergent SU(2) algebra ~\cite{PhysRevLett.123.147201} i.e., 
\begin{equation}
  \left([\,H,\,Q^\dagger\,] -2 hQ^{\dagger}\right)\;\mathcal{W} \;=0,
\end{equation}
here, the subspace $\mathcal{W}=\mathrm{span}\{|S_n\rangle \, , \, \forall \, n =0, 1, ...L\}$ is closed under \(H\), and the tower forms an equally spaced ``ladder'' with spacing \(2h\):
\begin{equation}
  H |S_n\rangle \;=\; (E_0 + 2hn)\,\ |S_n\rangle,
  \qquad |S_{n+1}\rangle\propto Q^\dagger |S_n\rangle.
  \label{eq:ladder}
\end{equation}
The elements \(\{Q^\dagger,Q,Q^z \equiv H\}\) close an \(\mathfrak{su}(2)\) algebra only within the $\mathcal{W}$ subspace, and the $|S_{n}\rangle$ are \emph{exact} eigenstates for OBC\footnote{The uniform-field term \(h\sum_i S_i^z\) is what fixes the ladder spacing \(2h\); the single-ion term \(D\sum_i (S_i^z)^2\) commutes with the ladder and merely shifts all energies uniformly; see Appendix~\ref{app:bonds} for details.}. Any perturbation that preserves the scar subspace equivalently implies the existence of an SU(2) SGA, although the generators may need to be redefined. For illustration, we refer the reader to Appendix~\ref {app: SGA and commutant}. Throughout the text, we refer to the SGA as a symmetry, although it is not an exact symmetry of the Hamiltonian but only exists in a subspace of the Hilbert space.

The model in Eq.~\eqref{eq:Hamiltonian-Spin1-XY} satisfies ETH weakly, as can be seen from the diagonal ETH and entanglement entropy spectrum (see Fig.~\eqref{fig:eth_entropy_short}).
Within a fixed magnetization $m$ sector, the member of the scar tower that resides in the sector violates ETH. The numerical data in Fig.~\eqref{fig:eth_entropy_short} shows that such states exhibit anomalously low bipartite entanglement entropy and violate the diagonal ETH for the observable, $\hat{O} = \sum_{i} (S^{z}_{i})^{2}$. Since the analytic structure of scars is known in this case, an explicit verification of these numerical observations is known and is provided in the supplemental material.

\begin{figure*}[!t]
    \centering
    \includegraphics[width=\linewidth]{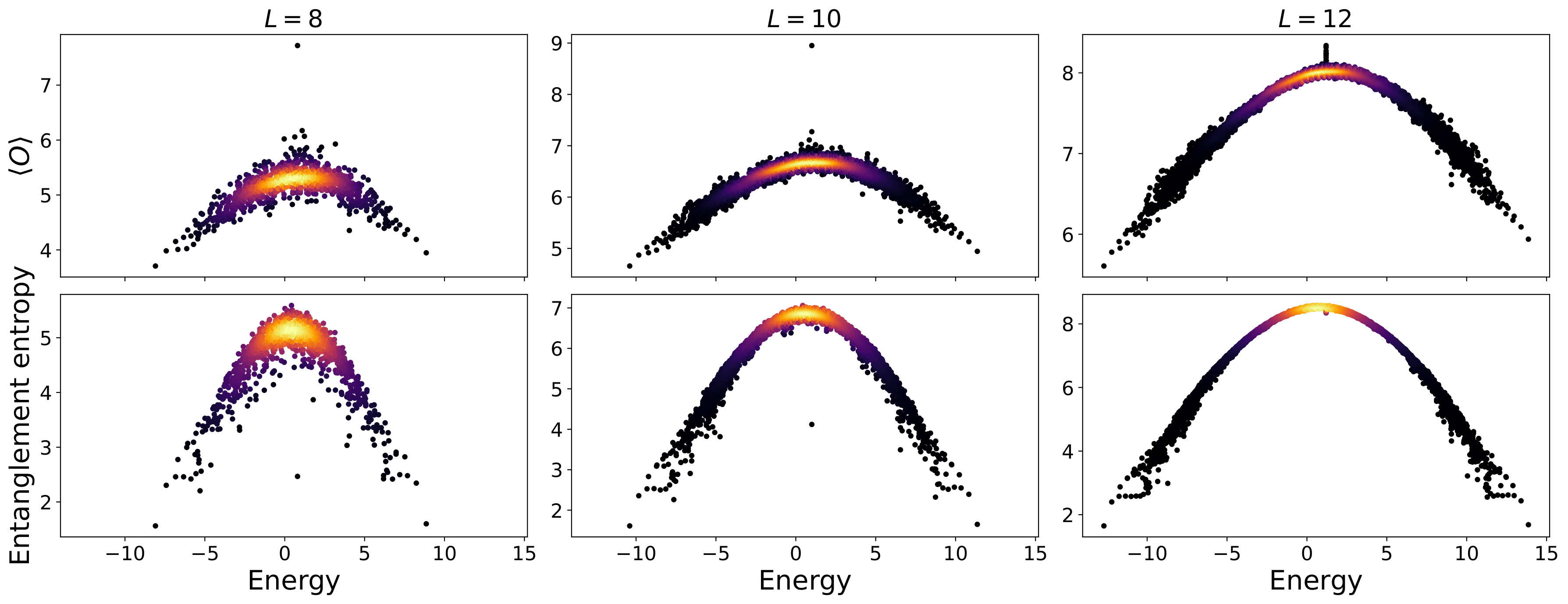}
    \caption{The plot of diagonal ETH and entropy for a fixed perturbation strength $\epsilon=0.05$ of onsite disorder. The putative scar (the point that is isolated) is found to merge with the ergodic spectrum as system size is increased, which shows this is not a true scar state despite its outlier character at smaller system sizes.}
    \label{fig:disorder_plot}
\end{figure*}

It is easily seen that  the disorder and long-range interactions introduced in Eq.~\eqref{eq:Hamiltonian-Spin1-XY} do not affect the scars. For bond disorders in $J_{1, i},$ the scar states constructed from Eq.~\eqref{eq:Qdag-def} and Eq.~\eqref{eq:ladder} are unchanged. This is on account of the existence of a commutant algebra~\cite{PhysRevX.12.011050, mohapatra2025unravelingadditionalquantummanybody}:
\begin{align}\label{eq:commutants}
    \left[T_{i, i+a}, (Q^{\dagger})^{n}|\Omega\rangle \langle \Omega| (Q)^{n}\right] & = 0,  \nonumber \\
    \left[(S^{z}_{i})^{2}, (Q^{\dagger})^{n}|\Omega\rangle \langle \Omega| (Q)^{n}\right] & = 0, \nonumber \\
    \left[\sum_i S^{z}_{i}, (Q^{\dagger})^{n}|\Omega\rangle \langle \Omega| (Q)^{n}\right] & = 0.
\end{align}
The commutant algebra is a set of operators that commute with building blocks (usually the local terms) of the Hamiltonian. Here $T_{i, i+a} = S^{x}_{i}S^{x}_{i+a} + S^{y}_{i}S^{y}_{i+a}$ are the bond operators. The first identity above holds for $\forall a \in$ odd, and $\forall i \in \Lambda,$ where $\Lambda$ refers to the lattice. These identities are easily verified analytically since the scar wavefunctions are known, see Appendix~\ref{app:bonds} for details. Similar to bond disorder, on-site disorder in the anisotropic interaction $D$ also preserves the commutant algebra, which follows from the last of Eq.~\eqref{eq:commutants}. 

We also consider long-range interactions, i.e., non-trivial $J_{5}$ and longer-range couplings, and find that the scars still exist at an individual level. The effects of the existence of a commutant algebra can be also seen in the numerically evaluated entanglement entropy and diagonal ETH. Figure~\eqref{fig:eth_entropy_longrange} shows calculated values of the diagonal ETH for the observable $\hat{O}=\sum_{i}(S^{z}_{i})^{2}$ and bipartite entanglement spectrum in the presence of the long-range interaction terms. Evidently, the outlier points have anomalously low entanglement entropies and also violate diagonal ETH, suggesting they could be scars. Moreover, the value of the entanglement and diagonal matrix element doesn't change when long-range hoppings are introduced, which further points to the existence of an underlying commutant algebra which we have of course explicitly identified.

We now consider the effect of a type of on-site disorder, $\epsilon S^{Z}_{L/2}$, which is an inhomogeneous Zeeman interaction in the bulk of the chain. We verify diagonal ETH as before and calculate the entanglement entropy for a fixed disorder strength as shown in Fig.~\eqref{fig:disorder_plot}. We observe that with increasing the system size, the scar state steadily approaches thermal behavior. We understand this as originating from the breaking of the commutant algebra (which protects the scars) by the impurity term, hence eliminating the exact scar structure; and consequently, the putative scar states do not survive as genuine scars in the thermodynamic limit. In the literature, scars are commonly identified by locating low-entanglement states in the middle of the spectrum and verifying that these states violate the diagonal component of the ETH. Thus, our analysis also tells us that scar detection using such methods may not always be appropriate. In Sec.~\eqref{sec: spt} we will define an appropriate (twist) operator whose expectation value indicates the breaking of the commutant algebra, hence providing a new way to probe such putative scars, being exact or not.

\subsection{Commutant Hamiltonian}\label{sec: commutant hamiltonian}
We see that the emergent $SU(2)$ symmetry, or SGA, is not a global symmetry of the Hamiltonian considered here, i.e.,eq.~\eqref{eq:Hamiltonian-Spin1-XY}. Consider instead a Hamiltonian that possesses the SGA as a global symmetry, and for which the scars are the ground states: we call this the commutant Hamiltonian: 
\begin{equation}
    H_{C} = -\sum_{n, m} (Q^{\dagger})^{n}\rho (Q)^{m},
\end{equation}
$\rho = |\Omega\rangle \langle \Omega|$, where $|\Omega\rangle = |-1 -1 \cdots -1\rangle$. This Hamiltonian is highly nonlocal and gapped since it is a sum of projectors. The commutant Hamiltonian possesses an exponentially large number of conserved charges -- the elements of the bond algebra -- thus, the commutant Hamiltonian is integrable. We claim that these conserved charges are the reason for the symmetry protection of the scars that we discuss in Sec.~\eqref{sec: spt}. Note that this construction does not work for any thermal eigenstate since the projection operator for a thermal state, the only conserved charge will be a unique Hamiltonian, i.e., a single charge; this goes back to the existence of a unique parent Hamiltonian for thermal states~\cite{tarun_single_eigenstate_encode, Qi2019determininglocal}. 

We now consider the symmetries of the commutant Hamiltonian. Note that $\prod_{i} (S^{z}_{i})^{2}$ is a conserved charge with a value of $+1$ since the scars are composed of local states $|-1\rangle,|1\rangle$ . Futhermore, the commutant Hamiltonian also has a symmetry given by $\prod_{i} \bar{\sigma}^{x}_{i}$, where 
\begin{align}
    \bar{\sigma}^{x} = \begin{pmatrix}
        0 & 0 & 1 \\
        0 & 0 & 0 \\
        1 & 0 & 0
    \end{pmatrix}
\end{align}
This operation interchanges the operators $Q$ and $Q^{\dagger}$, and we call this the flip symmetry. For the scar tower, this transformation simply flips the tower. The second symmetry we see is that of the sublattice symmetry. This symmetry transformation exchanges the sub-lattices with a $\pi$ phase, i.e., 
\begin{align}
    O_{i_{A}} \otimes O_{i_{B}} \to - O_{i_{B}} \otimes O_{i_{A}}  
\end{align}
where $O_{i_{x}}$ is a operator acting on the site $i_{x}$, where $x$ labels the sublattice $A, B$. Note that under this transformation the $Q$ and $Q^{\dagger}$ are invariant,
\begin{align}
    Q^{\dagger} = \sum_{i_{A}} \left((S^{+}_{i_{A}})^{2} - (S^{+}_{i_{A}+1})^{2}\right) \to \\
    \nonumber \sum_{i_{A}} \left(-(S^{+}_{i_{A}+1})^{2} + (S^{+}_{i_{A}})^{2}\right) = Q^{\dagger}     
\end{align}
The same goes for $Q$ the operator, since $Q$ and $Q^{\dagger}$ are invariant; hence, this is a symmetry of the commutant Hamiltonian. We further note that both of these transformations are $Z_{2}$. Note that $\left(\prod_{i}\bar{\sigma}^{x}_{i}\right)^{2} = \prod_{i}(S^{z}_{i})^{2}$, hence the $Z_{2}$ nature is apparent only when we are in the subspace with $\prod_{i}(S^{z}_{i})^{2} = +1$  charge sector.

If one adds certain perturbations to the spin 1 XY Hamiltonian eq.~\eqref{eq:Hamiltonian-Spin1-XY} that don't respect the above symmetries, the scar subspace is broken. In this paper we will consider the perturbations of different types that respectively respect or break one or more $Z_{2}$ symmetries. The first among them is $V_{1} = S^{z}_{L/2}$ which does not respect any of the two $Z_2$ symmetries, as this impurity term is not invariant under the sublattice transformation; hence, the scars should not exist. Next, we analyze the perturbation $V_{2} = \sum_{i} S^{x}_{i}$ which is not invariant under sublattice symmetry and also violates the conserved charge $\prod_{i} (S^{z}_{i})^{2}=+1$. The next perturbation we consider is $V_{3} = \sum_{i} (S ^{x}_{i})^{2}$: this perturbation is invariant under the flip symmetry but not invariant under the sublattice symmetry. Finally, we consider the perturbation $V_{4} = \sum_{i}(-1)^{i} (S^{x}_{i})^{2}$, which is invariant under the above symmetry transformations and also has the conserved charge of $\prod_{i}(S^{z}_{i})^{2} = +1$; hence, we expect that this transformation preserves the scars. That this is indeed the case is shown Appendix~\ref{app:numerical digaonlization for pert}.

\section{SPt nature of scars}\label{sec: spt}
In this section, we demonstrate that the scar subspace is a symmetry-protected trivial (SPt) state preserved by an SGA, while individual scars are protected by the commutant algebras~\cite{PhysRevX.14.041069}. SPt characterization is conventionally used in the context of the ground states of certain Hamiltonians; for example, in the spin-1 Haldane chain\cite{PhysRevB.81.064439} with an anisotropy term, i.e., $\sum_{i} (S^{z}_{i})^{2}$. The symmetry protection we refer to here is more general, i.e., going beyond ground states, and including the entire scar subspace which lies in the bulk of the spectrum.

\begin{figure}
    \centering
    \includegraphics[width=0.9\linewidth, trim = 0 0 0 0, clip]{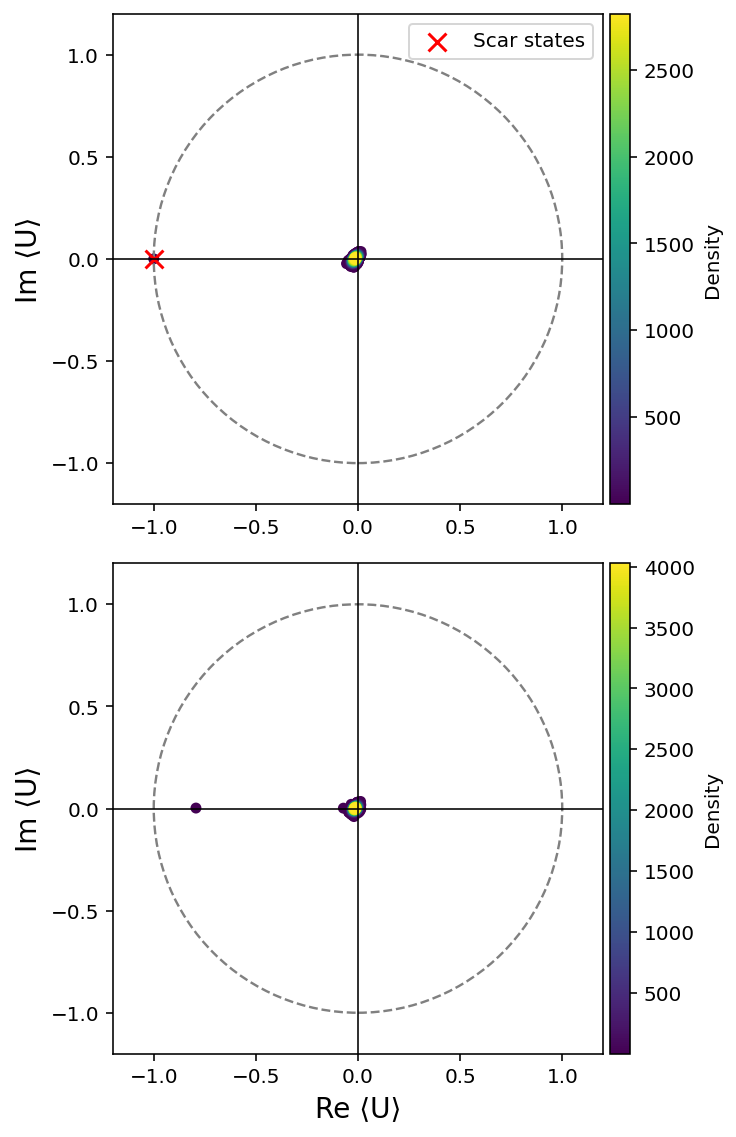}
    \caption{The expectation value of the twist operator $U(0)$ defined in eq.~\eqref{eq:twist_op} for the middle one-third of the spectrum, with system size $L=12$, and magnetization $m=0$ sector. The first subfigure is with no perturbation, and the second figure is when the perturbation is $\epsilon V = 0.01S^{z}_{L/2}$, and the isolated point from the cluster is the scar state with no perturbation, but now it is not as it does not lie at $-1$.}
    \label{fig:twist_expt}
\end{figure}

The hidden antiferromagnetic order in the well-known spin-1 Haldane chain was first studied by Kennedy and Tasaki~\cite{Kennedy1992} using string order parameters, and subsequently through the Lieb-Schultz-Mattis (LSM) twist operators by Nakamura and Todo~\cite{PhysRevLett.89.077204}, who showed that the expectation value of the twist operator can also detect the hidden order. We take this latter route to detect the SPt character of the scar subspace. 

An SPT state does not break any global symmetries and is distinct from a trivial product state only as long as a specific set of symmetries is preserved. If the protecting symmetry is explicitly broken, the topological distinction vanishes, and the state can be adiabatically connected to a trivial product state. In our case, a similar principle is apparent; namely, the scar subspace exists while certain symmetries, which in our case is the SGA, are preserved. Another symmetry that preserves the scar wavefunctions at an individual level rather than at the subspace level is the much stronger commutant symmetry: as soon as the perturbation violates the commutant symmetry, there is a mixing between the scar states, since the individual scars are the commutants, and the commutant symmetry is broken. We will characterize this explicit symmetry protection by defining an appropriate twist operator below Eq.~\eqref{eq:twist_op}.  To motivate the twist operator, we refer to Sec.~\ref{sec: commutant hamiltonian}, where we introduced the conserved  charges $(S^{z}_{i})^{2}$. We make an analogy with the spin-1 Haldane chain where one has $U(1)$ symmetry, and to uncover the hidden order one looks at the twist of the conserved charge. In the same vein, we consider the twist generated by the conserved charge of our commutant Hamiltonian, namely $\sum_{i} (S^{z}_{i})^{2}.$ To supplement this idea, we also numerically study the expectation value of the string order parameter~\cite{Kennedy1992} and its modifications (see Appendix~\ref{app:stringorder} for details), essentially replacing the $S_{z}$ operator with $(S^{z})^{2}$ in the spirit that now our conserved charge is associated with $(S^{z})^{2}$. The string order parameter conventionally is defined as
\begin{align}
    O^{z}_{\text{string}} = \lim_{|i-k|\to \infty} \omega \left(S^{z}_{i} \, \prod_{j=i+1}^{k-1}e^{i\pi S^{z}_{j}} \, S^{z}_{k}\right).     
\end{align}
\begin{figure}
    \centering
    \includegraphics[width=0.9\linewidth, trim = 0 0 0 0, clip]{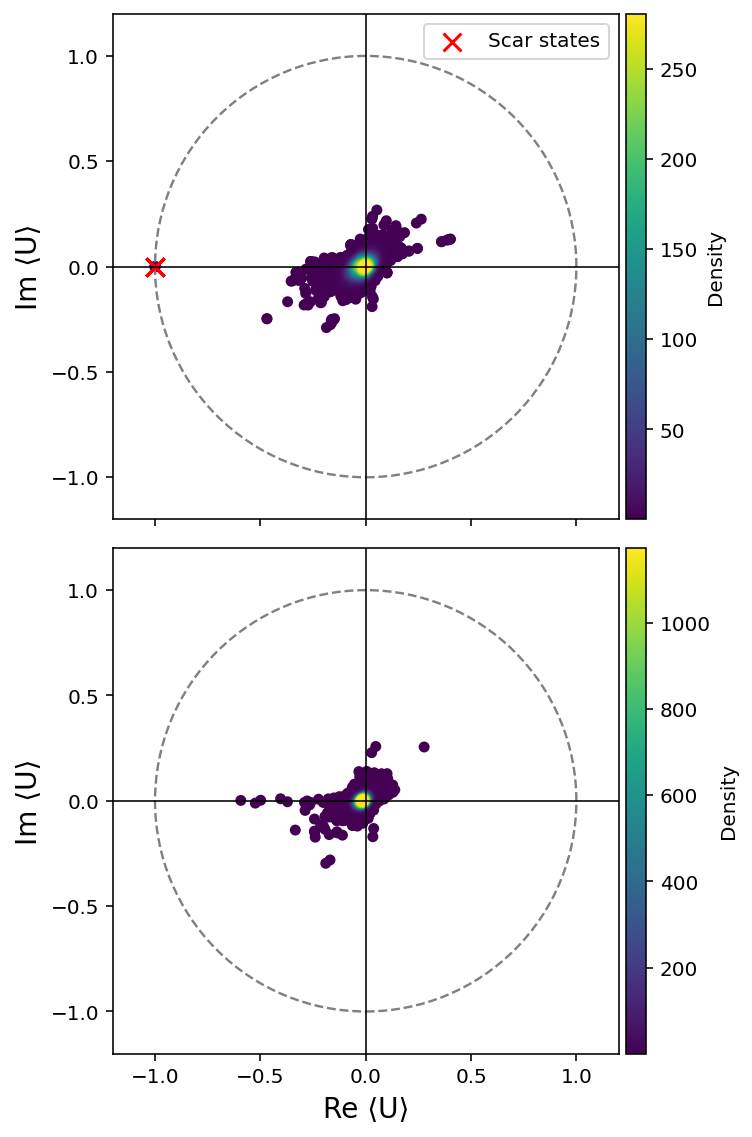}
    \caption{The expectation value of the twist operator $U(0)$ defined in eq.~\eqref{eq:twist_op} for the middle one-third of the spectrum, with system size $L=10$. The top figure is when there is no perturbation; hence, scars are exact in this regime, while the bottom figure is when the perturbation is $\epsilon V = 0.1\sum_{i}S^{x}_{i}$, which breaks the SGA, hence the scars are not exact indicated by the departure from the value of $-1$.}
    \label{fig:twist_expt_sumx}
\end{figure}

$\omega$ usually denotes averaging with respect to the ground state, but here we lift that to include any state in the spectrum. We find that the string order parameter with $(S^{z})^{2}$ on boundaries, which are actually $(S^{z})^{2}$ correlations when Kennedy-Tasaki transformation~\cite{Kennedy1992} is performed, yields an expectation value of $-1$. In light of the above discussion, we introduce a modified Lieb-Schultz-Mattis (LSM) twist operator~\cite{PhysRevLett.89.077204} by replacing $S^{z}$ in the LSM twist operator originally obtained for the spin-1 Haldane chain with $(S^{z})^{2}$ as follows:
\begin{equation}\label{eq:twist_op}
U(\theta)
=
\exp\!\left[
\displaystyle
\frac{2\pi i}{L} \sum_{j} j \left(S^{z}_{j}\right)^{2}
+
\frac{i \theta }{2} \sum_{j} S^{z}_{j}
\right].
\end{equation}

 It is easily seen that the expectation value of the twist operator in the scar states eq.~\eqref{eq:scar-tower} (for even system sizes) is given by
 \begin{equation}
     \langle U(\theta)\rangle_{n} = \langle S_{n}| U(\theta)|S_{n}\rangle =(-1)exp \left[\dfrac{i\theta(-L+2n)}{2}\right].
 \end{equation}
 Therefore, for the one-parameter family of twist operators $U(\theta)$, the scars lie on the unit circle in the complex plane of $\langle U(\theta) \rangle$, and for the special case of $\theta=0$(this will be of our interest for SGA breaking), all the scars give $\langle U(0)\rangle = -1$. On the other hand, for all the ergodic states, the one-parameter family of twist operators always clusters around $\langle U(\theta)\rangle=0$ as shown in Fig.~\eqref{fig:twist_expt}. We observe that for small system sizes, the spread of the cluster is larger, but as the system size is increased, the spread of the cluster around $\langle U(\theta)\rangle=0$ progressively shrinks. Hence, we posit that the twist operator acts as a non-local order parameter characterizing the SPt order in the thermodynamic limit.
 
In order to identify the symmetries that protect the SPt character of the scar states, we ask what symmetries protect the scar states. There are two different symmetries protecting the scars. At an individual scar level, the protection comes from the emergent SGA together with the $U(1)$ global symmetry of the Hamiltonian. At the scar subspace level, the protection is provided only by the SGA. When SGA is broken, the scar subspace ceases to exist. Examples of SGA breaking perturbations we have considered include $V_{2}=\sum_{i}S^{x}_{i}$ (this also breaks the $U(1)$ symmetry), and $V_{1}=S^{z}_{L/2}$ (that preserves the $U(1)$ symmetry). $\langle U(0)\rangle$ acts as a twist operator that detects the twist in the bulk and provides a diagnostic of the SPt phase protected by this symmetry: 
\[
\langle U(0) \rangle =
\begin{cases}
-1 \ \text{  for scar states}, \\[6pt]
0  \ \text{  for ergodic states}.
\end{cases}
\]
Figure~\eqref{fig:twist_expt}, and Figure~\eqref{fig:twist_expt_sumx}, respectively show the distribution of $\langle U(0)\rangle$ under the perturbations $V_{1}$ and $ V_{2}.$  Any departure from $\langle U(0)\rangle = -1$ implies mixing with the ergodic subspace and breaking of the SGA. We also observe similar results in the entanglement spectra, see Appendix~\ref{app:numerical digaonlization for pert}, Fig.~\eqref{fig: app, extensive non scar preserving}. 
Another implication of our LSM twist order parameter is that the SGA endows the scar subspace with a non-trivial twist that is detected by $\langle U(0)\rangle,$ and this symmetry is independent of the original $U(1)$ symmetry of the Hamiltonian.

\begin{figure}
    \centering
    \includegraphics[width=0.9\linewidth, trim = 0 0 0 20, clip]{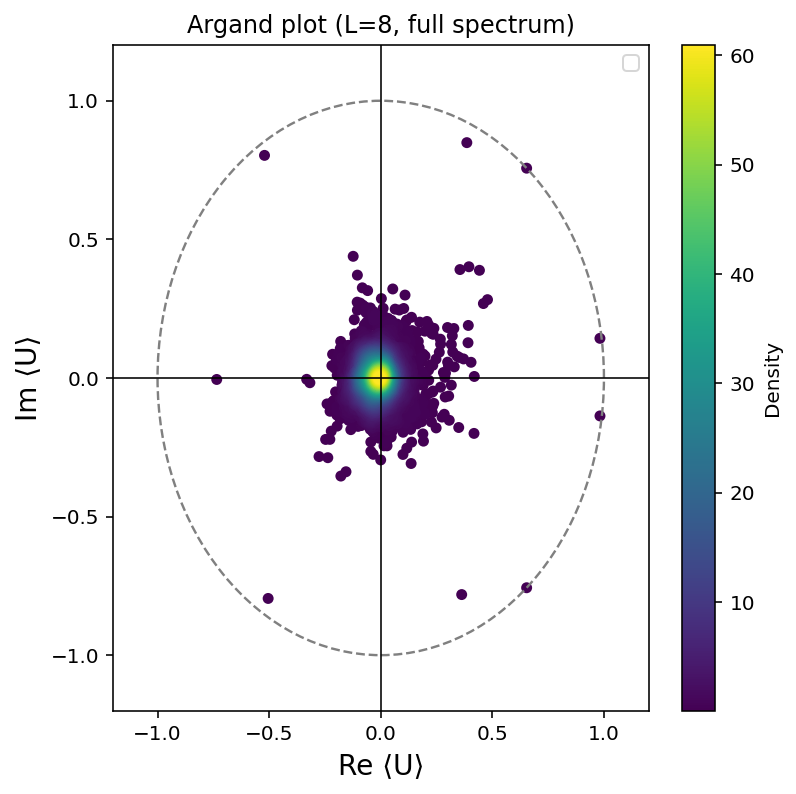}
    \caption{Twist operator $\langle U(\theta)\rangle$ for whole spectrum for systems size $L=8$, and perturbation  $\epsilon=0.05$. The outermost points forming a deformed circle are the putative scars that one can think of as scars numerically, see Fig.~\eqref{fig:disorder_plot}, but for larger system sizes, it mixes with the ETH prediction. Also, the leftmost points close to the unit circle are the states that were scars close to the spectral edges, and since the spectral edges do not have a high density of states, one cannot detect the deviation from the unit circle confidently.}
    \label{fig:twist_l8_full_z}
\end{figure}

We now examine if the commutants endow the scars with an SPt structure; for this we use the diagnostic $\langle U(\theta)\rangle$ with $\theta \neq  0$.  Consider the bond algebra ($\mathcal{B}$) consisting of $\{T_{i, i+1}, (S^{z}_{i})^{2}, \sum_{i} S^{z}_{i}\}$ and the lifting operator (Zeeman field); for details, see Ref.~\cite{PhysRevX.14.041069}. The commutants are the set $\{|S_{n}\rangle \langle S_{n}|\, \, \, \forall n=0, ..., L\}$ along with the conserved $U(1)$ charges of the Hamiltonian. If one doesn't add the lifting operator to the bond algebra, then the commutants are the set $\{|S_{n}\rangle \langle S_{m}|\}$, i.e., the degenerate scar subspace, along with the conserved $U(1)$ charges from the Hamiltonian. Under these conditions, the SGA and commutant algebra are the same. Since we are interested in scars forming a tower, we work with the bond algebra $\mathcal{B}$ in accord with the conjecture $\mathrm{III}$ of Ref.~\cite{PhysRevX.14.041069}.
To detect the commutant breaking, i.e., mixing of scars while the scar subspace is preserved, we use the twist operator $U(\theta)$ with $\theta \neq 0,$ 
\[
\langle U(\theta) \rangle =
\begin{cases}
-e^{i M_n \theta / 2} & \text{for scar states}, \\
0 & \text{for ergodic states},
\end{cases}
\]
This can occur in cases when the scar states mix with each other, but under such scenarios the commutants break, i.e. $|S_{n}\rangle \langle S_{n}| $ are not commutants, and this cannot be detected by the twist operator $U(0)$, as can be seen by expanding
\begin{align}
    \nonumber |\tilde{S}_{m}\rangle = \sum_{n} c_{mn} |S_{n}\rangle
\end{align}
and calculating the twist operator $U(0)$ expectation value:
\begin{align}
    \nonumber \langle U(0)\rangle_{m} &= (-1) \sum_{n} |c_{mn}|^{2} =-1.
\end{align}
Thus we need $\theta\neq 0$ in order to detect intra-scar subspace mixing, or in other words, commutant breaking. For $\theta\neq 0$ the twist operators for the new scars take the expectation value 
\begin{align}
    \langle U(\theta)\rangle_{m} &= (-1) \sum_{n} |c_{mn}|^{2}e^{i\theta M_n/2}.
\end{align}
The extra phases from the $\theta$ term cause dephasing; thus, $|\langle U(\theta)\rangle| < 1.$ Figure~\eqref{fig:twist_l8_full_z} shows an example of the movement $|\langle U(\theta)\rangle|$ when the model is subjected to commutant-breaking perturbations.  

Detection of the commutant breaking is thus a two-step process: one needs to identify the scars using the twist operator $U(0)$, followed by measurement of the twist operator with $\theta \neq 0$.  These twist operators, if they can be measured experimentally, could be a sensitive probe to search for scars because of the SPt character they have.

Having established the SPt character through the twist operator, we now identify the symmetry that protects the scar subspace. SPt states usually studied are protected by some global symmetry of the Hamiltonian; in contrast, our SGA, while describing an emergent $SU(2)$ symmetry, is not a global property of the Hamiltonian but a property of the scar subspace. However, the emergent $SU(2)$ symmetry is a global symmetry of another Hamiltonian constructed out of the commutant set. The important observation we want to make here is that the element of the bond algebra that define our hamiltonian each represent local conserved quantities of the commutant Hamiltonian defined above 

Conventionally, in order to characterize the SPt phases, one looks at the symmetry group that protects the states and uses its cohomology classification~\cite{PhysRevB.87.155114} to determine the number of distinct phases that can arise from it. In our case, we can't find the symmetry group that protects the scar subspace, so a classification of SPt phases is not possible in our case. A similar work along these lines has been done recently~\cite{matsui2025symmetryprotectedtopologicalscarsubspaces} in the context of the AKLT model, where it is argued that the entire scar subspace inherits the SPt character from the AKLT ground state that is also an SPt, evident from their calculation of the string order parameter. A crucial distinction between Ref.~\cite{matsui2025symmetryprotectedtopologicalscarsubspaces} and our work is that in our case, the SPt character is not inherited from the ground state, as our scar subspace is not tied to the ground state. Instead, the root state in our case $|-1 -1 \cdots -1\rangle$, which generates the scar tower, may or may not be the ground state of the system we consider. One difference between our work and the recent work~\cite{matsui2025symmetryprotectedtopologicalscarsubspaces} is that we consider a general form of perturbations which may or may not preserve the SGA generators, rather than generalized lifts (see eq.(74) of \cite{matsui2025symmetryprotectedtopologicalscarsubspaces}) which preserve the SGA generators, i.e. $Q, Q^{\dagger}$. In our case, we consider perturbations that can modify the SGA, e.g. $V_{4} = \sum_{i}(-1)^{i} (S^{x}_{i})^{2}$, in which case one still has the scars and SGA but the generators are now redefined, for a discussion; see Appendix~\ref{app: SGA and commutant}.

Although our symmetry protection is different from the symmetry protection considered in the recent work on the AKLT model~\cite{matsui2025symmetryprotectedtopologicalscarsubspaces}, there are similarities too. The similarity can be seen by looking at the commutant Hamiltonian and their symmetries, which are $Z_{2} \times Z_{2}$, this symmetry group is usually encountered in SPt. One of the $Z_{2}$ charges is the same as in the case of the recent work~\cite{matsui2025symmetryprotectedtopologicalscarsubspaces}, which is the $Z_{2}$ symmetry associated with the flipping of the tower, while the other $Z_{2}$ in our case is different from the one they considered. A cohomology classification of this symmetry group yields 2 phases, one of which is trivial, which we associate to the ergodic part of the spectrum, and the other being an SPt, which is associated with the scars.

\begin{figure}
    \centering
    \includegraphics[width=0.9\linewidth]{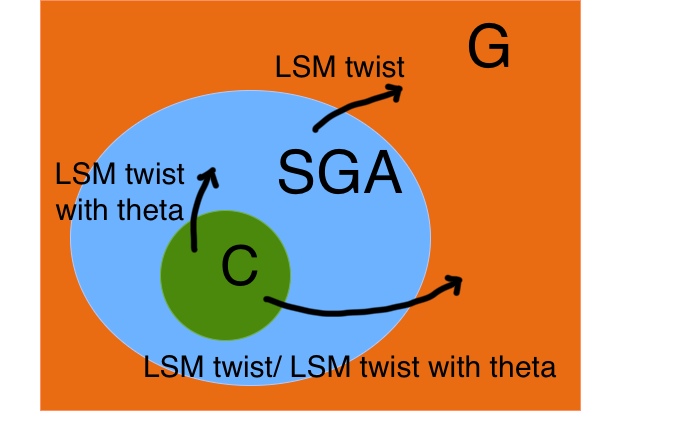}
    \caption{Schematic describing the twist operator to be used to detect the type of symmetry. Here, G is the general space of operators, SGA is the space of operators that preserve the scar subspace, and C is the space of operators that preserve the commutants or individual scar wavefunctions.}
    \label{fig:cartoon_spt}
\end{figure}

To summarize this section, we have two types of symmetries that protect the scars (a) at a subspace level and (b) at an individual level. The SGA, which can be detected using the twist operator with $\theta=0$ is associated with the entire scar subspace. The symmetry that protects the scars at an individual level can be probed through the twist operator with $\theta\neq 0$. This is illustrated schematically in Fig.~\eqref{fig:cartoon_spt}.
We would like to highlight that using a twist operator to detect a scar is much better than numerical diagonalization and finding entanglement entropy. 

\begin{figure}
    \centering
    \includegraphics[width=0.8\linewidth]{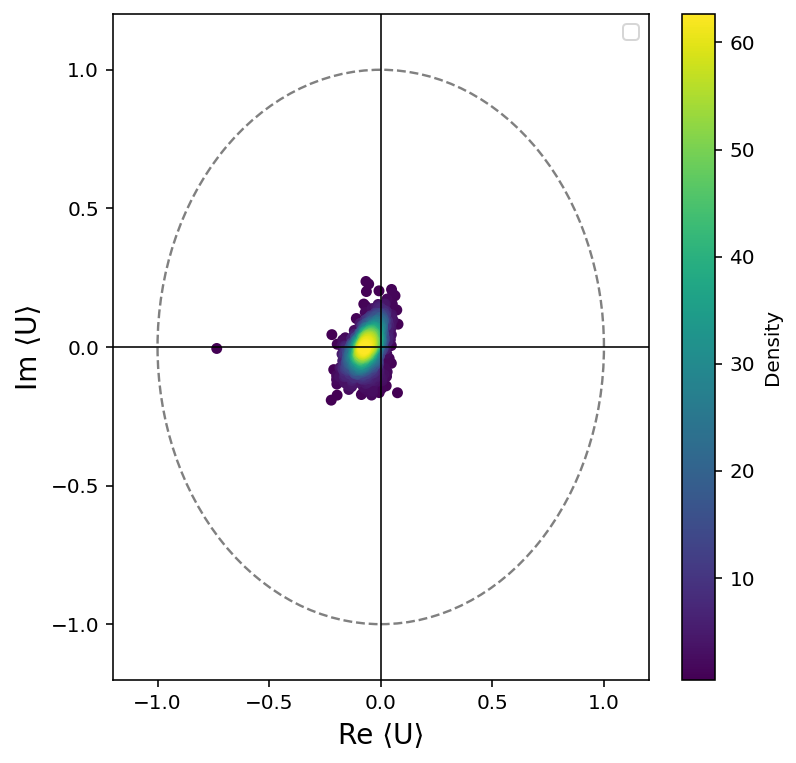}
    \caption{$\langle U(0) \rangle$ for the whole spectrum, for $L=8$, magnetization $m=0$, and perturbation $V_{1}=0.05 \, S^{z}_{L/2}$}
    \label{fig:L8_pertz_m0}
\end{figure}

As an aside, we would like to mention here that the twist operators that we used above to establish the SPt character are also more sensitive probes for detecting the scar states in comparison to the entanglement entropy that we studied earlier. By considering the perturbation $V_{1}=\epsilon \, S^{z}_{L/2}$ whose entanglement spectrum is shown in Fig.~\eqref{fig:disorder_plot} for $L=8$, from the figure, one may think that it is a scar, but in reality it is not, as is evident from the entanglement spectrum for $L=12$. The twist operator can detect the deviation from the unit circle for this perturbation, confirming that it is not a scar, as shown in Fig.~\eqref{fig:L8_pertz_m0}. This shows that measuring the twist of a suspected scar is a much more reliable way to find out if they are really scars or not.

\section{Scar Stability to perturbations}\label{sec:qfi}

We attempt now an understanding of the stability of the scars under perturbations to the Hamiltonian, turning to the Loschmidt echo as a diagnostic tool. 

At short times, the Loschmidt Echo is related to the Quantum Fisher Information (QFI), and the latter is well known in quantum metrology and quantum information as a witness for multipartite entanglement. Quantum Fisher information (QFI) is a crucial tool for estimating unknown parameters $\theta$ on which a particular state may depend. When the state $\rho(\theta)$ depends on $\theta$, the mean-square error of any \emph{locally unbiased} estimator obeys the quantum Cramér–Rao bound
\begin{equation}\label{eq:Cramer rao bound}
    (\Delta \theta)^{2} \ge \frac{1}{F[\rho,\mathcal{A}]},
\end{equation}
where for \emph{unitary} parameter encoding generated by \(\mathcal{A}\),
\(\rho(\theta)=e^{i\mathcal{A}\theta}\rho_{0}e^{-i\mathcal{A}\theta}\),
the QFI ($F[\rho, \mathcal{A}]$) admits the spectral (Helstrom) form~\cite{Helstrom1969}
\begin{equation}\label{eq:QFI for lu}
    F[\rho,\mathcal{A}]
    = 2 \sum_{p_i+p_j\neq 0}
      \frac{(p_i-p_j)^2}{p_i+p_j}\,|\langle i|\mathcal{A}|j\rangle|^{2}
    \;\le\; 4\,(\Delta \mathcal{A})^{2}.
\end{equation}
Here \(\rho=\sum_{i} \,  p_i \, |i\rangle\!\langle i|\) and \((\Delta \mathcal{A})^{2}\) is the variance of \(\mathcal{A}\) in \(\rho\). 
The inequality becomes an equality for \emph{pure} states, i.e.\ \(F=4\,(\Delta \mathcal{A})^{2}\), which is the case here. More generally, for arbitrary (not necessarily unitary) parameter dependence, the QFI is defined via the symmetric logarithmic derivative (SLD) \( \mathrm{L}\), introduced through
\[
    \frac{\partial \rho}{\partial \theta}
    = \frac{1}{2}\big(\rho\,\mathrm{L}+\mathrm{L}\,\rho\big),
\]
in which case
\begin{equation}\label{eq: QFI for general protocol}
    F\!\left[\rho,\frac{\partial \rho}{\partial \theta}\right]
    = \mathrm{Tr}\!\left(\rho\,\mathrm{L}^{2}\right).
\end{equation}
This formulation is closely connected to quantum statistical distance and the problem of distinguishing nearby density operators~\cite{PhysRevD.23.357,PhysRevLett.72.3439}; see also the reviews~\cite{Toth_2014,Liu_2020}.

The QFI per particle (QFI density) provides a diagnostic of multipartite entanglement. In an $N$-qubit $k$-producible state  ($k$-producible state is a state that can be written as a tensor product of at most $k$ entangled particles, for instance, the particles numbering up to $k$ form blocks within each of which the particles are entangled, and the state is simply a tensor product of such blocks), the  QFI density is bounded by $k$. This would mean that a state that is not $k$-producible must have at least $k+1$ multipartite entanglement. In this sense, QFI density acts as a witness of multi-partite entanglement; one has the upper bound~\cite{PhysRevA.85.022321,PhysRevA.85.022322}
\begin{align}
    F[\rho,\mathcal{A}]
    \;\le\; s\,k^{2} + (N - s k)^{2},
\end{align}
where \(s=\lfloor N/k \rfloor\). 
In the special case where \(N\) is divisible by \(k\) (i.e.\ \(N=s k\)), this bound reduces to
\begin{equation}
    f =\dfrac{F[\rho,\mathcal{A}]}{N} \;\le\; k.
\end{equation}
Values \(f\) exceeding the appropriate \(k\)-producible bound, certify genuine \((k{+}1)\)-partite entanglement that is accessible through the operator $\mathcal{A}$.
We will show in Sec.~\eqref{sec: qfistudy} that for scar states the QFI density scales as $f\sim N,$ signifying genuine $N$-partite entanglement. This is to be compared with thermal states which are not genuinely $N$-partite entangled, as has been demonstrated numerically~\cite{PhysRevLett.129.020601, PhysRevLett.124.040605} for the PXP model --  where as thermal states have the QFI density scaling $f \sim O(1).$

We also emphasize that QFI is experimentally accessible, enabling direct benchmarking of these predictions. One can measure QFI using randomized measurements~\cite{PhysRevResearch.3.043122}, or one can rely on certain inequalities on spin squeezing parameters that bound QFI~\cite{PRXQuantum.5.030338, Yu2022}. 

\subsection{Loschmidt echo and dynamical stability}\label{sec:loschmidtecho}

To quantify the dynamical sensitivity of an initial state $|\psi_{0}\rangle$ to a static perturbation $V$, we consider the Loschmidt amplitude
\begin{equation}
  m(t)
  = \langle \psi_{0}| e^{+i(H_0+V)t} e^{-i H_0 t} | \psi_{0} \rangle,
\end{equation}
and its modulus squared, the Loschmidt echo (LE),
\begin{equation}\label{eq:LE_def}
  M(t) = |m(t)|^2.
\end{equation}
Thus $M(t)$ is the fidelity between the states evolved under the forward Hamiltonian $H_0$ and the backward (perturbed) Hamiltonian $H_2=H_0+V$, for our case, eq.~\eqref{eq:Hamiltonian-Spin1-XY} is $H_0$. 
If $|\psi_{0}\rangle$ is a simultaneous eigenstate of $H_0$ and $V$, then $m(t)=e^{i v t}$ and $M(t)=1$ for all $t$, i.e. perfect dynamical stability. 
If $|\psi_{0}\rangle$ is not an eigenstate of $V$ but an eigenstate of $H_0$, then $m(t)=\langle e^{i V t}\rangle_{\psi_0}$ and $M(t)\le 1$ due to dephasing in the $V$ eigenbasis. The decay of the Loschmidt echo quantifies coherent dephasing under a unitary perturbation. Expanding at short times, one finds that the leading decay is determined by the quantum Fisher information, which therefore defines the characteristic dephasing timescale.

We will focus on the short-time behavior of the Loschmidt echo, which measures the "speed" with which two copies of an initial state under unitary evolution, respectively governed by two slightly differing Hamiltonians, drift away from each other, thus providing a measure of the stability of the initial state under a perturbation. This aligns with earlier studies of the Loschmidt echo~\cite{TomazProsen_2002, GORIN200633} regarding the stability of states under perturbation.

\begin{figure}
    \centering
    \includegraphics[width=\linewidth]{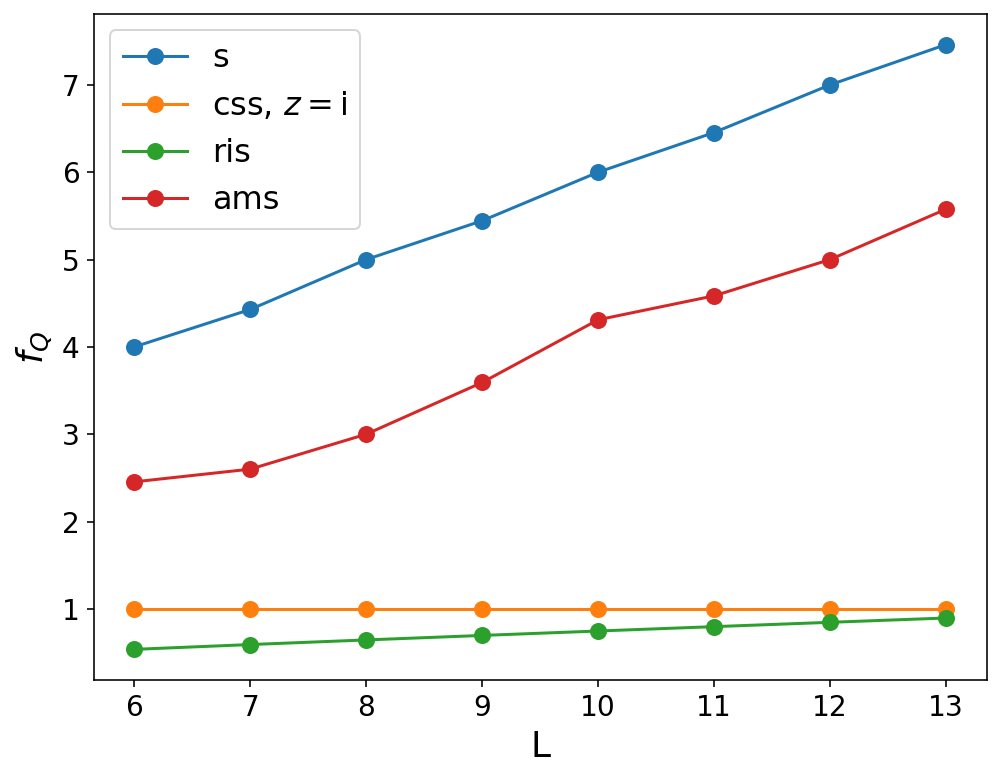}
    \caption{This plot shows the QFI density ($f$) for the perturbation $V=\sum_{i} (-1)^{i}(S^{x}_{i})^{2}$ for various different states, s=Scar, css=Coherent States, ris=Random Initial state, ams=Asymptotic Many Body scars. The scars lie in the middle of the spectrum, and ams are deformations of the scars from the middle with k=$\pi/2$.}
    \label{fig:QFI_scaling}
\end{figure}

The long-time behavior of the Loschmidt echo is governed by the whole spectrum and often decays to saturation for finite-size systems regardless of the initial state; hence, the long-time behavior is less informative as a stability diagnostic. Hence, the short-time behavior of the Loschmidt echo can be used as a stability diagnostic.

For time-independent perturbations, which are our focus here, the short-time expansion is well known and depends on the Quantum Fisher Information (QFI) (see, for eg.~\cite{LIU2014167}) as
\begin{equation}\label{eq:LE_short_time}
    M(t) = 1 - \dfrac{t^{2}}{4}F[V, |\psi_{0}\rangle] +O(t^{3})
\end{equation}
Equation~\eqref{eq:LE_short_time} shows that the initial curvature of $M(t)$ is set by the fluctuations of the perturbation $V$ in the initial state. This turns out to be true even for mixed states. Here, we take the perturbation V to be $V=\sum_{i}(-1)^{i}(S^{x}_{i})^{2}$, which doesn't break the SGA hence we still have scars, but the scars mix since this doesn't preserve $U(1)$ symmetry.

The dynamic stability of different states is characterized by their QFI as shown by Eq.~\eqref{eq:LE_short_time}. Figure~\eqref{fig:QFI_scaling} shows numerically calculated QFI finite-size scaling data for the following states:
\begin{enumerate}
    \item Scar states, which form the basis of the emergent SU(2) spin $L$ representation, for which QFI $F_{Q} \sim L^{2}$
    \item Coherent states: given the spin $L$ representation, we can create spin coherent states (see, e.g., ~\cite{JMRadcliffe_1971}, acting with $ |z\rangle= \,e^{z \, S^{-}}|S^{z}_{tot}=-L\rangle$). We show numerically that QFI scales $F_{Q}\sim L$  the same as thermal states, shown in Fig.~\eqref{fig:QFI_scaling}.
    \item Asymptotic Many body scars, introduced in~\cite{PhysRevLett.131.190401}, are approximate eigenstates and become true eigenstates in the thermodynamic limit, whose QFI scales $F\sim L^{2}$ as shown in Fig.~\eqref{fig:QFI_scaling} but is less than that of exact scars. 
    \item Random initial states: Since Berry’s conjecture~\cite{1977RSPTA.287..237B} suggests that chaotic eigenstates resemble Gaussian random variables in the semiclassical limit, we use random initial states to investigate QFI scaling. This approach provides a theoretical window into the eigenstate structure of non-integrable systems.
\end{enumerate}
We use these states to assess how much time it takes to reach a fidelity cutoff. From eq.\eqref{eq:LE_short_time} we see that, 
\begin{align}
    t_{*} \sim \dfrac{1}{\sqrt{F[Q, |\psi_{0}\rangle]}}
\end{align}
It therefore follows that both the scar states and the asymptotic quantum many-body scars are highly unstable, as the relevant timescale is inversely proportional to the QFI. This behavior is consistent with the interpretation of extensive QFI density as indicating enhanced sensitivity to perturbations. In contrast, coherent states and random initial states exhibit significantly longer timescales to reach the same fidelity threshold, which may be viewed as a minimal fidelity requirement for state preparation in quantum protocols.

We would also like to point out that the Asymptotic many body scars have a QFI scaling similar to scarred states, which hasn't been reported in the literature to the best of our knowledge.

\subsection{Super extensive scaling of QFI}\label{sec: qfistudy}

 In a related recent work, the QFI has been studied for the PXP model~\cite{PhysRevLett.129.020601} where a near-superextensive scaling was reported for numerically extracted putative scar states: specifically, $F \sim N^{2}$ (or equivalently $f \sim N$) has been observed for system sizes up to $N\sim28$ beyond which the scaling exponent deviates to a fitted value below $2$. This is owing to the fact that the scars in the PXP model are not exact, and there is some hybridization with the ergodic states. Nevertheless, the SU(2) symmetry can be made exact by adding suitable deformations~\cite{PhysRevLett.122.220603}, whereupon the $N^2$ scaling for QFI is expected to persist in the thermodynamic limit. Here, we intend to study QFI for the case of the Spin 1 XY model introduced above and show analytically that the super-extensive QFI exists for scars, and is linked to the existence of the SGA.
 
\begin{figure}[t]
    \centering
    \includegraphics[width=1.1\linewidth, trim=0 0 0 41, clip]{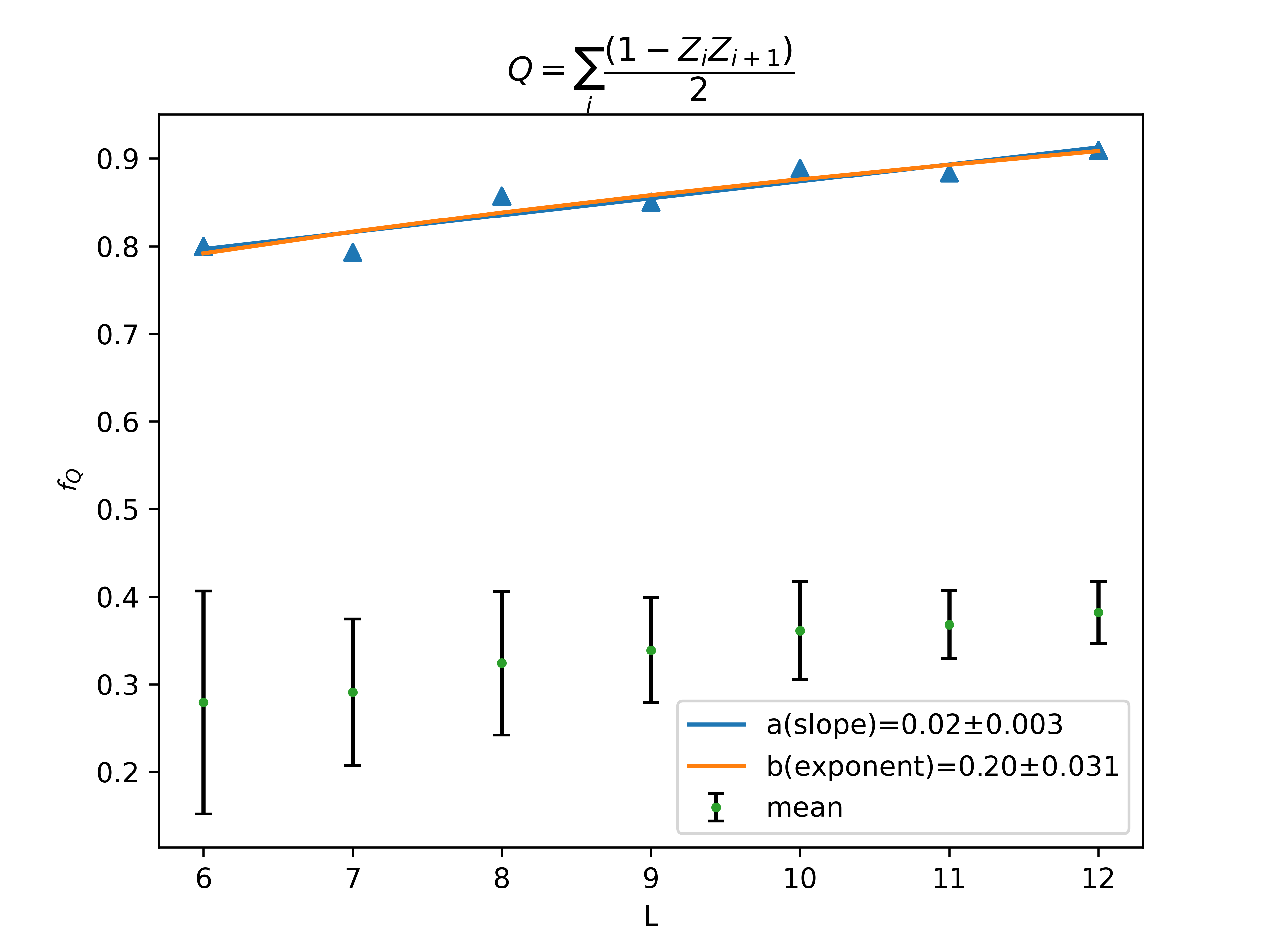}
    \caption{The QFI density $f_{Q}$ scaling in the $m=0$ sector without chain inversion symmetry resolved, with parameters $Jz=1$, $J3=0.1$, $D=0.1.$ The blue line is the linear fit given to $aL+d$, the orange line is a power-law fit to $cL^{b}.$ The blue triangles are the data from exact diagonalization, the green points denote the mean QFI density for the entire spectrum in the given magnetization sector, and the black error bars indicate the variance within that data. For the range of system sizes considered, the expected linear scaling is not unambiguously observable.}
    \label{fig:QFIscaling}
\end{figure}

\begin{figure*}[t]
    \centering

    \begin{subfigure}{0.32\linewidth}
        \centering
        \includegraphics[width=\linewidth]{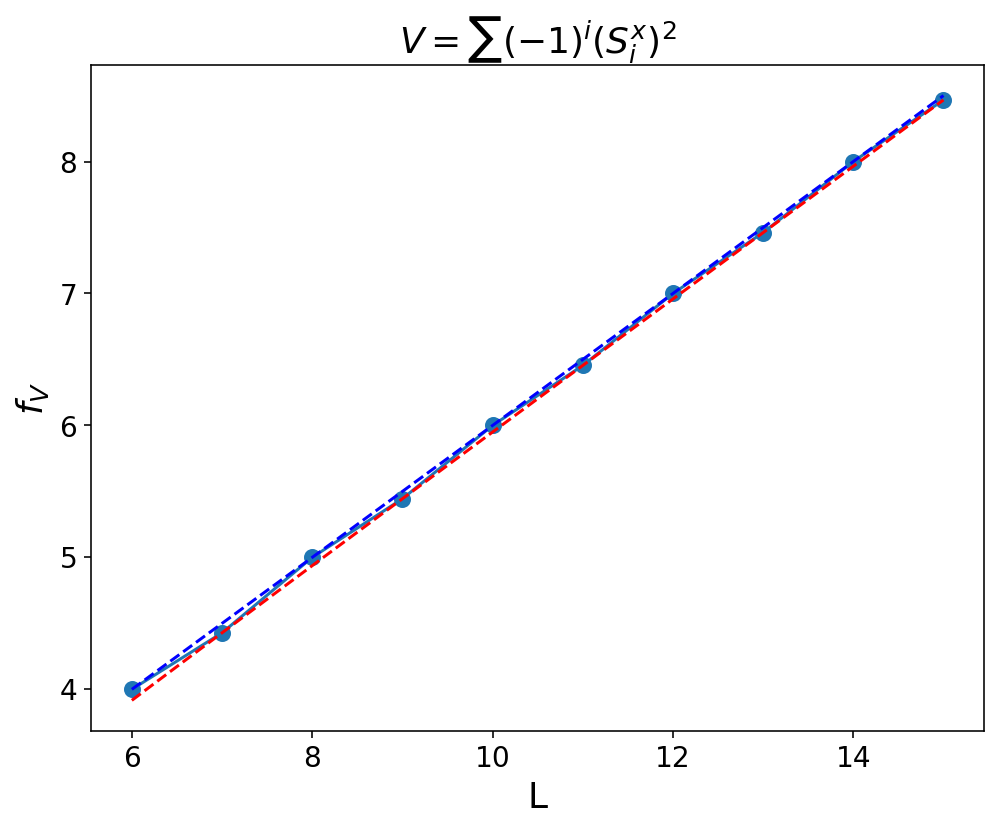}
        \caption{Extensive and scar preserving}
        \label{fig:sub_a}
    \end{subfigure}
    \hfill
    \begin{subfigure}{0.32\linewidth}
        \centering
        \includegraphics[width=\linewidth]{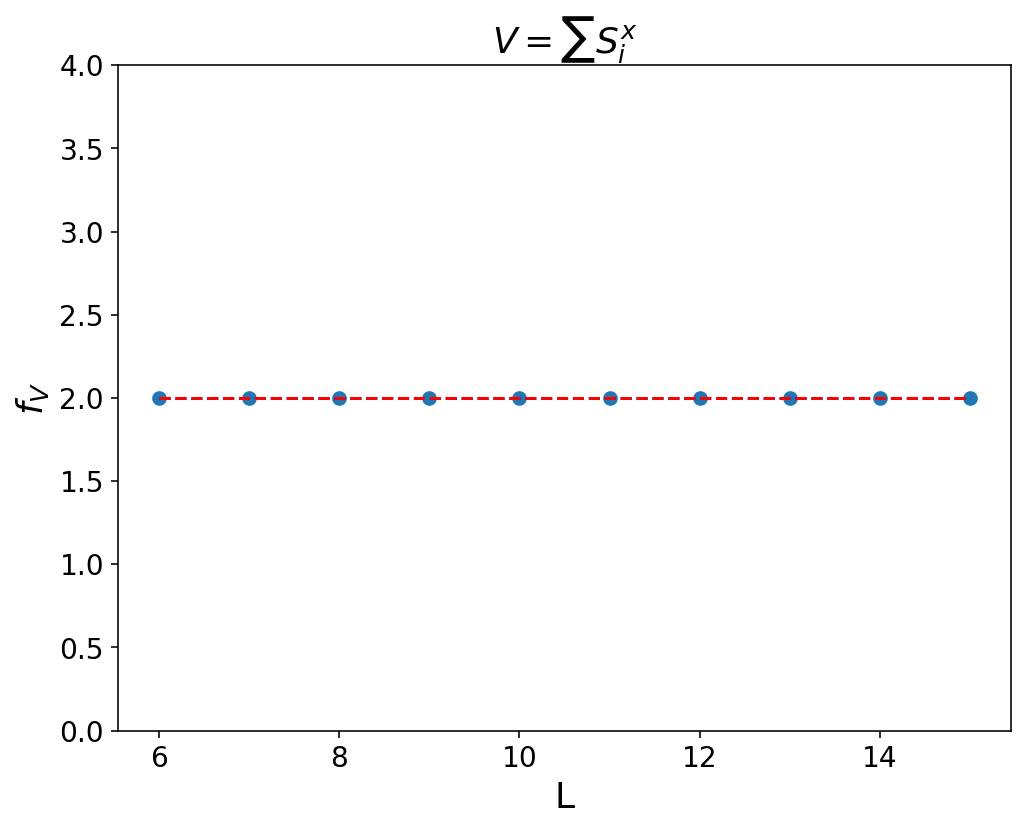}
        \caption{Extensive and non-scar preserving}
        \label{fig:sub_b}
    \end{subfigure}
    \hfill
    \begin{subfigure}{0.32\linewidth}
        \centering
        \includegraphics[width=\linewidth]{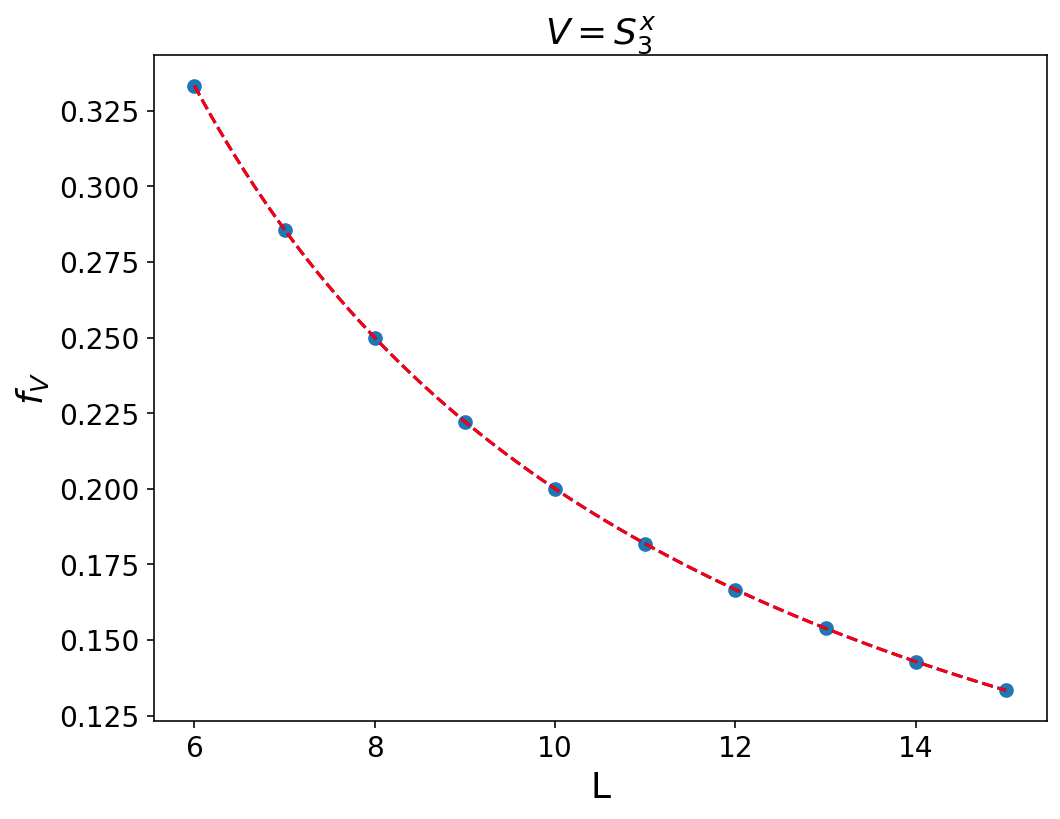}
        \caption{Intensive}
        \label{fig:sub_c}
    \end{subfigure}

    \caption{The 3 different scaling regimes of the different types of operators are mentioned above the figures. The blue dots are the numerical data obtained by creating the scars using the SGA and then finding the QFI, this task scales better than exact diagonalization, hence one can go to larger system sizes. The dashed lines are the analytical results, which can be found in the supplemental material; the red and blue dashed lines are for odd and even system sizes, respectively. In subfigures (b), (c), the distinction between odd and even system sizes does not exist; hence, only red indicates the analytical expression.}
    \label{fig:scaling regimes of observables}
\end{figure*}

Consider now the finite-size scaling of QFI density for scars and thermal states, respectively, where analytic expressions for the scar wavefunctions are available. 

Figure~\eqref{fig:QFIscaling} shows numerically calculated finite-size scaling of QFI density for $N\leq 12.$ It is difficult to ascertain with the limited range of system sizes we have if QFI density obeys a linear law. Fortunately, we know the scar wave functions and can obtain the scaling analytically, which was no possible in the earlier work~\cite{PhysRevLett.129.020601} because the pxp scars are known to not have a analytical expression for the wavefunction. We first calculate the $k$-body reduced density matrix of scars. For a sanity check, we obtained the scaling of the entanglement entropy (see Appendix~\ref{app:density_matrix_calculation} for details), which matches with the well known sub extensive scaling for entanglement entropy.

We then analytically calculate the QFI density for some operators and confirm that the QFI density scales as $f \sim N$. Nevertheless, the above super-extensive scaling of QFI does not hold for every operator. We observe three different scaling regimes of the operators: (i) Class $\mathrm{I}$ operators: Operators that preserve the scar subspace, i.e., the emergent SU(2) algebra, follow an extensive QFI density scaling, $f\sim N;$ (ii) Class $\mathrm{II}$ operators: Extensive operators that break the scar subspace show a size-independent scaling, i.e., $f \sim O(1)$; (iii) Class $\mathrm{III}$ operators: Intensive operators that break the scar subspace show a scaling $f \sim O(1/N).$ These different scaling regimes for the QFI are shown in Fig.~\eqref{fig:scaling regimes of observables}. Details of the analytical calculation of the QFI for these three regimes are provided in the Appendix~\ref{app:QFI for operators}.

While the above distinction between the three operator classes is meaningful, it is important to note a subtle qualification. 
Let $O_1$ belong to class $\mathrm{I}$ (whose quantum Fisher information (QFI) exhibits $N^2$ scaling), and let $O_2$ belong to class $\mathrm{II}$. 
Consider now the composite operator.
\begin{equation}
    O = O_1 + O_2.
\end{equation}
In the thermodynamic limit, the QFI density associated with $O$ can still scale as $N^2$, since the superextensive contribution arising from $O_1$ dominates. 
However, the presence of $O_2$ breaks the invariant scar subspace. 
Such mixed operators, therefore, do not fall cleanly into the classification described above.

One can instead construct a basis of strictly local operators and show that any operator $O$ formed from this basis with 2-site translational invariance, both (i) retains the $N^2$ QFI scaling and (ii) preserves the scar subspace. 
Remarkably, this distinguished basis coincides with the set of local operators appearing as generators of the spectrum-generating algebra (SGA). 
The single-site building blocks of this construction are listed in Eq.~\eqref{eq: basis for fundamental operators}. 
We refer to operators assembled from these elements as \emph{SGA-local operators}. 
When attention is restricted to this class, the above classification becomes well defined and reveals a close connection between the scaling of the QFI and perturbations that break the SGA structure.
\begin{equation}\label{eq: basis for fundamental operators}
        \underbrace{\begin{pmatrix}
        0 & 0 & 1 \\
        0 & 0 & 0 \\
        1 & 0 & 0
    \end{pmatrix}}_{\Large{\Bar{\sigma}^{x}}},
        \underbrace{\begin{pmatrix}
        0 & 0 & -i \\
        0 & 0 & 0 \\
        i & 0 & 0
    \end{pmatrix}}_{\Bar{\sigma}^{y}},
        \underbrace{\begin{pmatrix}
        1 & 0 & 0 \\
        0 & 0 & 0 \\
        0 & 0 & -1
    \end{pmatrix}}_{\Bar{\sigma}^{z}},
        \underbrace{\begin{pmatrix}
        1 & 0 & 0 \\
        0 & 0 & 0 \\
        0 & 0 & 1
    \end{pmatrix}}_{\Bar{\mathds{1}}}
            ,
        \underbrace{\begin{pmatrix}
        0 & 0 & 0 \\
        0 & 1 & 0 \\
        0 & 0 & 0
    \end{pmatrix}}_{\gamma^{z}}
\end{equation}

One can show that any operators of the form $\hat{O} = \sum_{i} (-1)^{i} \left( c_{x}\Bar{\sigma}^{x}_{i} + c_{y}\Bar{\sigma}^{y}_{i}+ c_{z}\Bar{\sigma}^{z}_{i}+ c_{1}\Bar{\mathds{1}}_{i}+ c_{\gamma}\gamma^{z}_{i}\right)$ will have the QFI scaling as $\sim N^{2}$ and also preserve the scar subspace. This form generalizes to any non-Hermitian perturbations, hence hinting at the possibility of the presence of scars in open quantum systems.

\section{Summary and Discussion}\label{sec:conclusion}

We studied Quantum Many-Body Scars (QMBS) in the spin 1XY model under OBC with different types of disorder and long-range interactions, hence confirming the existence of an underlying commutant algebra that gives rise to the scars. We constructed different probes to detect the scars and, more importantly, the breaking of SGA and commutant symmetries, as measured by an appropriate Lieb-Schultz-Mattis (LSM) type twist operator and the Quantum Fisher Information (QFI). We find that the scars subspace has an SPt character. We trace this SPt character of the scar subspace to a hidden $Z_{2}\times Z_{2}$ symmetry associated with a sub-lattice symmetry and also a flipping symmetry of local terms $|-1\rangle, |1\rangle$, these symmetries are the global symmetries of a different model, which we call commutant Hamiltonian. These symmetries and the bond disorder can be explained by considering the Hamiltonian made from the commutants, dubbed as the commutant Hamiltonian. We find that this $ Z_{2} \times Z_{2}$ symmetry exists only within the sector of conserved charge of the commutant Hamiltonian, namely $\prod_{i} (S^{z}_{i})^{2}=+1$ the sector. We create a Lieb-Schultz-Mattis (LSM) type twist operator, which captures this SPt character via a nontrivial twist through the bulk. It is to our surprise that our twist operator is more sensitive at detecting putative scars as compared to exact diagonalization and the entanglement entropy plot, as illustrated in the text.

To study about the stability of these scars, we also use Loschmidt echo as a probe to the dynamical stability of the scars, to various perturbations, and find that it is related to the Quantum Fisher Information (QFI). We find analytically, that the super-extensive scaling of QFI can be used for classification of perturbations, which in principle can also be non-Hermitian, as scar-preserving or not. The extensive QFI indicates that scars are useful from a metrological point of view as compared to the thermal states, as also shown recently for the case of spin 1 Dzyaloshinksii-Moriya Model (DMI)~\cite{Shane_Dooley_Quantum_sensing_scars_2021}.

It would be interesting to see if one can get the same SPt character by starting from the Liouvillian superoperator~\cite{PRXQuantum.5.040330, PhysRevB.107.224312} defined as
\begin{align}
    \hat{\mathcal{L}}_{\alpha} = \hat{h}_{\alpha}\otimes \mathds{1} - \mathds{1} \otimes \hat{h}_{\alpha}
\end{align}
where $\hat{h}_{\alpha}$ are the local terms of the Hamiltonian. Within this formalism, the scars become the frustration-free zero modes of the local super hamiltonian $\mathcal{P}=\sum_{\alpha}\hat{\mathcal{L}}^{\dagger}_{\alpha}\hat{\mathcal{L}}_{\alpha}$, which is also the ground state since, $\mathcal{P}$ is a positive definite operator. This could in principle explain that the scar states, which are high in the energy spectra, can still be SPt.

The possible future directions for this work would be to explore whether the SPt character of the scars exists for other models too; we suspect this to be true, as our construction of the commutant Hamiltonian relies on the fact that the scars are the eigenstates of non-commuting local operators, which make them non-thermal as argued recently in~\cite{mohapatra2025unravelingadditionalquantummanybody}. Furthermore, it would be interesting to see if a similar SPt characteristic is present in towers of scars in different models and if one can understand this twist operator in terms of the properties of the commutants, as commutants are the ground state of the Liouvillian, and if one can rigorously derive the twist operator from some fundamental principles and symmetries of the Liouvillian. It would be interesting to study if the QFI scaling for scars originating from different group actions in the scar subspace and to examine whether scars with no structure also have certain characteristic QFI scaling, hence their relevance to the metrology. Recently, QFI has been related to the spectral function of the ETH~\cite{PhysRevLett.124.040605, Hauke2016}, and also has been extended to include QFI for various different protocols~\cite{iniguez2023quantumfisherinformationdifferent}, the implication of the QFI scaling would be a different behavior in the transport properties in the scars and thermal states, which we think could explain the recently seen superdiffusive transport~\cite{transportinSpin1xy}. 

\begin{acknowledgments}
The authors acknowledge support of the Department
of Atomic Energy, Government of India, under Project
Identification No. RTI 4002, and the Department of Theoretical Physics, TIFR, for computational resources. The authors would also like to thank Marco Manya and Avijit Maity for the helpful and insightful discussions. The numerical calculations were performed using the Python library QuSpin~\cite{10.21468/SciPostPhys.2.1.003} for spin systems. 
\end{acknowledgments}

\bibliographystyle{apsrev4-2}
\bibliography{main}  

@article{PhysRevLett.123.147201,
  title = {Weak Ergodicity Breaking and Quantum Many-Body Scars in Spin-1 $XY$ Magnets},
  author = {Schecter, Michael and Iadecola, Thomas},
  journal = {Phys. Rev. Lett.},
  volume = {123},
  issue = {14},
  pages = {147201},
  numpages = {5},
  year = {2019},
  month = {Oct},
  publisher = {American Physical Society},
  doi = {10.1103/PhysRevLett.123.147201},
  url = {https://link.aps.org/doi/10.1103/PhysRevLett.123.147201}
}

@article{AtsuhiroKitazawa_2003,
doi = {10.1088/0305-4470/36/23/104},
url = {https://dx.doi.org/10.1088/0305-4470/36/23/104},
year = {2003},
month = {may},
publisher = {},
volume = {36},
number = {23},
pages = {L351},
author = {Atsuhiro Kitazawa and Keigo Hijii and Kiyohide Nomura},
title = {An SU(2) symmetry of the one-dimensional spin-1 XY model},
journal = {Journal of Physics A: Mathematical and General}
}

@article{PhysRevB.102.085140,
  title = {$\ensuremath{\eta}$-pairing in Hubbard models: From spectrum generating algebras to quantum many-body scars},
  author = {Moudgalya, Sanjay and Regnault, Nicolas and Bernevig, B. Andrei},
  journal = {Phys. Rev. B},
  volume = {102},
  issue = {8},
  pages = {085140},
  numpages = {12},
  year = {2020},
  month = {Aug},
  publisher = {American Physical Society},
  doi = {10.1103/PhysRevB.102.085140},
  url = {https://link.aps.org/doi/10.1103/PhysRevB.102.085140}
}

@article{PhysRevB.101.024306,
  title = {Quantum many-body scar states with emergent kinetic constraints and finite-entanglement revivals},
  author = {Iadecola, Thomas and Schecter, Michael},
  journal = {Phys. Rev. B},
  volume = {101},
  issue = {2},
  pages = {024306},
  numpages = {14},
  year = {2020},
  month = {Jan},
  publisher = {American Physical Society},
  doi = {10.1103/PhysRevB.101.024306},
  url = {https://link.aps.org/doi/10.1103/PhysRevB.101.024306}
}

@article{PhysRevB.101.195131,
  title = {Unified structure for exact towers of scar states in the Affleck-Kennedy-Lieb-Tasaki and other models},
  author = {Mark, Daniel K. and Lin, Cheng-Ju and Motrunich, Olexei I.},
  journal = {Phys. Rev. B},
  volume = {101},
  issue = {19},
  pages = {195131},
  numpages = {17},
  year = {2020},
  month = {May},
  publisher = {American Physical Society},
  doi = {10.1103/PhysRevB.101.195131},
  url = {https://link.aps.org/doi/10.1103/PhysRevB.101.195131}
}

@article{PhysRevB.98.235155,
  title = {Exact excited states of nonintegrable models},
  author = {Moudgalya, Sanjay and Rachel, Stephan and Bernevig, B. Andrei and Regnault, Nicolas},
  journal = {Phys. Rev. B},
  volume = {98},
  issue = {23},
  pages = {235155},
  numpages = {31},
  year = {2018},
  month = {Dec},
  publisher = {American Physical Society},
  doi = {10.1103/PhysRevB.98.235155},
  url = {https://link.aps.org/doi/10.1103/PhysRevB.98.235155}
}

@article{PhysRevB.98.235156,
  title = {Entanglement of exact excited states of Affleck-Kennedy-Lieb-Tasaki models: Exact results, many-body scars, and violation of the strong eigenstate thermalization hypothesis},
  author = {Moudgalya, Sanjay and Regnault, Nicolas and Bernevig, B. Andrei},
  journal = {Phys. Rev. B},
  volume = {98},
  issue = {23},
  pages = {235156},
  numpages = {43},
  year = {2018},
  month = {Dec},
  publisher = {American Physical Society},
  doi = {10.1103/PhysRevB.98.235156},
  url = {https://link.aps.org/doi/10.1103/PhysRevB.98.235156}
}

@article{PhysRevLett.122.220603,
  title = {Emergent SU(2) Dynamics and Perfect Quantum Many-Body Scars},
  author = {Choi, Soonwon and Turner, Christopher J. and Pichler, Hannes and Ho, Wen Wei and Michailidis, Alexios A. and Papi\ifmmode \acute{c}\else \'{c}\fi{}, Zlatko and Serbyn, Maksym and Lukin, Mikhail D. and Abanin, Dmitry A.},
  journal = {Phys. Rev. Lett.},
  volume = {122},
  issue = {22},
  pages = {220603},
  numpages = {6},
  year = {2019},
  month = {Jun},
  publisher = {American Physical Society},
  doi = {10.1103/PhysRevLett.122.220603},
  url = {https://link.aps.org/doi/10.1103/PhysRevLett.122.220603}
}

@article{PhysRevB.98.155134,
  title = {Quantum scarred eigenstates in a Rydberg atom chain: Entanglement, breakdown of thermalization, and stability to perturbations},
  author = {Turner, C. J. and Michailidis, A. A. and Abanin, D. A. and Serbyn, M. and Papi\ifmmode \acute{c}\else \'{c}\fi{}, Z.},
  journal = {Phys. Rev. B},
  volume = {98},
  issue = {15},
  pages = {155134},
  numpages = {23},
  year = {2018},
  month = {Oct},
  publisher = {American Physical Society},
  doi = {10.1103/PhysRevB.98.155134},
  url = {https://link.aps.org/doi/10.1103/PhysRevB.98.155134}
}

@article{Toth_2014,
doi = {10.1088/1751-8113/47/42/424006},
url = {https://dx.doi.org/10.1088/1751-8113/47/42/424006},
year = {2014},
month = {oct},
publisher = {IOP Publishing},
volume = {47},
number = {42},
pages = {424006},
author = {Tóth, Géza and Apellaniz, Iagoba},
title = {Quantum metrology from a quantum information science perspective},
journal = {Journal of Physics A: Mathematical and Theoretical}
}

@article{PhysRevD.23.357,
  title = {Statistical distance and Hilbert space},
  author = {Wootters, W. K.},
  journal = {Phys. Rev. D},
  volume = {23},
  issue = {2},
  pages = {357--362},
  numpages = {0},
  year = {1981},
  month = {Jan},
  publisher = {American Physical Society},
  doi = {10.1103/PhysRevD.23.357},
  url = {https://link.aps.org/doi/10.1103/PhysRevD.23.357}
}

@article{PhysRevLett.72.3439,
  title = {Statistical distance and the geometry of quantum states},
  author = {Braunstein, Samuel L. and Caves, Carlton M.},
  journal = {Phys. Rev. Lett.},
  volume = {72},
  issue = {22},
  pages = {3439--3443},
  numpages = {0},
  year = {1994},
  month = {May},
  publisher = {American Physical Society},
  doi = {10.1103/PhysRevLett.72.3439},
  url = {https://link.aps.org/doi/10.1103/PhysRevLett.72.3439}
}

@article{Liu_2020,
doi = {10.1088/1751-8121/ab5d4d},
url = {https://dx.doi.org/10.1088/1751-8121/ab5d4d},
year = {2019},
month = {dec},
publisher = {IOP Publishing},
volume = {53},
number = {2},
pages = {023001},
author = {Liu, Jing and Yuan, Haidong and Lu, Xiao-Ming and Wang, Xiaoguang},
title = {Quantum Fisher information matrix and multiparameter estimation},
journal = {Journal of Physics A: Mathematical and Theoretical}
}

@article{PhysRevX.12.011050,
  title = {Hilbert Space Fragmentation and Commutant Algebras},
  author = {Moudgalya, Sanjay and Motrunich, Olexei I.},
  journal = {Phys. Rev. X},
  volume = {12},
  issue = {1},
  pages = {011050},
  numpages = {44},
  year = {2022},
  month = {Mar},
  publisher = {American Physical Society},
  doi = {10.1103/PhysRevX.12.011050},
  url = {https://link.aps.org/doi/10.1103/PhysRevX.12.011050}
}

@article{PhysRevB.102.075132,
  title = {$\ensuremath{\eta}$-pairing states as true scars in an extended Hubbard model},
  author = {Mark, Daniel K. and Motrunich, Olexei I.},
  journal = {Phys. Rev. B},
  volume = {102},
  issue = {7},
  pages = {075132},
  numpages = {18},
  year = {2020},
  month = {Aug},
  publisher = {American Physical Society},
  doi = {10.1103/PhysRevB.102.075132},
  url = {https://link.aps.org/doi/10.1103/PhysRevB.102.075132}
}

@article{PhysRevA.85.022321,
  title = {Fisher information and multiparticle entanglement},
  author = {Hyllus, Philipp and Laskowski, Wies\l{}aw and Krischek, Roland and Schwemmer, Christian and Wieczorek, Witlef and Weinfurter, Harald and Pezz\'e, Luca and Smerzi, Augusto},
  journal = {Phys. Rev. A},
  volume = {85},
  issue = {2},
  pages = {022321},
  numpages = {10},
  year = {2012},
  month = {Feb},
  publisher = {American Physical Society},
  doi = {10.1103/PhysRevA.85.022321},
  url = {https://link.aps.org/doi/10.1103/PhysRevA.85.022321}
}

@article{PhysRevA.85.022322,
  title = {Multipartite entanglement and high-precision metrology},
  author = {T\'oth, G\'eza},
  journal = {Phys. Rev. A},
  volume = {85},
  issue = {2},
  pages = {022322},
  numpages = {8},
  year = {2012},
  month = {Feb},
  publisher = {American Physical Society},
  doi = {10.1103/PhysRevA.85.022322},
  url = {https://link.aps.org/doi/10.1103/PhysRevA.85.022322}
}

@article{PhysRevLett.124.040605,
  title = {Multipartite Entanglement Structure in the Eigenstate Thermalization Hypothesis},
  author = {Brenes, Marlon and Pappalardi, Silvia and Goold, John and Silva, Alessandro},
  journal = {Phys. Rev. Lett.},
  volume = {124},
  issue = {4},
  pages = {040605},
  numpages = {6},
  year = {2020},
  month = {Jan},
  publisher = {American Physical Society},
  doi = {10.1103/PhysRevLett.124.040605},
  url = {https://link.aps.org/doi/10.1103/PhysRevLett.124.040605}
}

@article{PhysRevLett.129.020601,
  title = {Extensive Multipartite Entanglement from su(2) Quantum Many-Body Scars},
  author = {Desaules, Jean-Yves and Pietracaprina, Francesca and Papi\ifmmode \acute{c}\else \'{c}\fi{}, Zlatko and Goold, John and Pappalardi, Silvia},
  journal = {Phys. Rev. Lett.},
  volume = {129},
  issue = {2},
  pages = {020601},
  numpages = {7},
  year = {2022},
  month = {Jul},
  publisher = {American Physical Society},
  doi = {10.1103/PhysRevLett.129.020601},
  url = {https://link.aps.org/doi/10.1103/PhysRevLett.129.020601}
}

@article{PhysRevB.75.155111,
  title = {Localization of interacting fermions at high temperature},
  author = {Oganesyan, Vadim and Huse, David A.},
  journal = {Phys. Rev. B},
  volume = {75},
  issue = {15},
  pages = {155111},
  numpages = {5},
  year = {2007},
  month = {Apr},
  publisher = {American Physical Society},
  doi = {10.1103/PhysRevB.75.155111},
  url = {https://link.aps.org/doi/10.1103/PhysRevB.75.155111}
}

@article{PhysRevLett.110.084101,
  title = {Distribution of the Ratio of Consecutive Level Spacings in Random Matrix Ensembles},
  author = {Atas, Y. Y. and Bogomolny, E. and Giraud, O. and Roux, G.},
  journal = {Phys. Rev. Lett.},
  volume = {110},
  issue = {8},
  pages = {084101},
  numpages = {5},
  year = {2013},
  month = {Feb},
  publisher = {American Physical Society},
  doi = {10.1103/PhysRevLett.110.084101},
  url = {https://link.aps.org/doi/10.1103/PhysRevLett.110.084101}
}

@Article{Helstrom1969,
author={Helstrom, Carl W.},
title={Quantum detection and estimation theory},
journal={Journal of Statistical Physics},
year={1969},
month={Jun},
day={01},
volume={1},
number={2},
pages={231-252},
issn={1572-9613},
doi={10.1007/BF01007479},
url={https://doi.org/10.1007/BF01007479}
}

@article{PhysRevLett.131.190401,
  title = {Asymptotic Quantum Many-Body Scars},
  author = {Gotta, Lorenzo and Moudgalya, Sanjay and Mazza, Leonardo},
  journal = {Phys. Rev. Lett.},
  volume = {131},
  issue = {19},
  pages = {190401},
  numpages = {7},
  year = {2023},
  month = {Nov},
  publisher = {American Physical Society},
  doi = {10.1103/PhysRevLett.131.190401},
  url = {https://link.aps.org/doi/10.1103/PhysRevLett.131.190401}
}

@article{PhysRevLett.125.230602,
  title = {Many-Body Scars as a Group Invariant Sector of Hilbert Space},
  author = {Pakrouski, K. and Pallegar, P. N. and Popov, F. K. and Klebanov, I. R.},
  journal = {Phys. Rev. Lett.},
  volume = {125},
  issue = {23},
  pages = {230602},
  numpages = {6},
  year = {2020},
  month = {Dec},
  publisher = {American Physical Society},
  doi = {10.1103/PhysRevLett.125.230602},
  url = {https://link.aps.org/doi/10.1103/PhysRevLett.125.230602}
}

@article{PhysRevLett.126.120604,
  title = {Quasisymmetry Groups and Many-Body Scar Dynamics},
  author = {Ren, Jie and Liang, Chenguang and Fang, Chen},
  journal = {Phys. Rev. Lett.},
  volume = {126},
  issue = {12},
  pages = {120604},
  numpages = {6},
  year = {2021},
  month = {Mar},
  publisher = {American Physical Society},
  doi = {10.1103/PhysRevLett.126.120604},
  url = {https://link.aps.org/doi/10.1103/PhysRevLett.126.120604}
}

@article{LIU2014167,
title = {Fidelity susceptibility and quantum Fisher information for density operators with arbitrary ranks},
journal = {Physica A: Statistical Mechanics and its Applications},
volume = {410},
pages = {167-173},
year = {2014},
issn = {0378-4371},
doi = {https://doi.org/10.1016/j.physa.2014.05.028},
url = {https://www.sciencedirect.com/science/article/pii/S0378437114003926},
author = {Jing Liu and Heng-Na Xiong and Fei Song and Xiaoguang Wang}
}

@article{TomazProsen_2002,
doi = {10.1088/0305-4470/35/6/309},
url = {https://dx.doi.org/10.1088/0305-4470/35/6/309},
year = {2002},
month = {feb},
publisher = {},
volume = {35},
number = {6},
pages = {1455},
author = {Tomaz Prosen and Marko Znidaric},
title = {Stability of quantum motion and correlation decay},
journal = {Journal of Physics A: Mathematical and General}
}

@article{GORIN200633,
title = {Dynamics of Loschmidt echoes and fidelity decay},
journal = {Physics Reports},
volume = {435},
number = {2},
pages = {33-156},
year = {2006},
issn = {0370-1573},
doi = {https://doi.org/10.1016/j.physrep.2006.09.003},
url = {https://www.sciencedirect.com/science/article/pii/S0370157306003310},
author = {Thomas Gorin and Tomaž Prosen and Thomas H. Seligman and Marko Žnidarič}
}

@misc{iniguez2023quantumfisherinformationdifferent,
      title={Quantum Fisher Information for Different States and Processes in Quantum Chaotic Systems}, 
      author={Fernando Iniguez and Mark Srednicki},
      year={2023},
      eprint={2304.01657},
      archivePrefix={arXiv},
      primaryClass={cond-mat.stat-mech},
      url={https://arxiv.org/abs/2304.01657}, 
}

@article{PhysRevE.50.888,
  title = {Chaos and quantum thermalization},
  author = {Srednicki, Mark},
  journal = {Phys. Rev. E},
  volume = {50},
  issue = {2},
  pages = {888--901},
  numpages = {0},
  year = {1994},
  month = {Aug},
  publisher = {American Physical Society},
  doi = {10.1103/PhysRevE.50.888},
  url = {https://link.aps.org/doi/10.1103/PhysRevE.50.888}
}

@article{MarkSrednicki_1999,
doi = {10.1088/0305-4470/32/7/007},
url = {https://dx.doi.org/10.1088/0305-4470/32/7/007},
year = {1999},
month = {feb},
publisher = {},
volume = {32},
number = {7},
pages = {1163},
author = {Mark Srednicki},
title = {The approach to thermal equilibrium in quantized chaotic systems},
journal = {Journal of Physics A: Mathematical and General}
}

@article{MarkSrednicki_1996,
doi = {10.1088/0305-4470/29/4/003},
url = {https://dx.doi.org/10.1088/0305-4470/29/4/003},
year = {1996},
month = {feb},
publisher = {},
volume = {29},
number = {4},
pages = {L75},
author = {Mark Srednicki},
title = {Thermal fluctuations in quantized chaotic systems},
journal = {Journal of Physics A: Mathematical and General}
}

@article{RevModPhys.91.021001,
  title = {Colloquium: Many-body localization, thermalization, and entanglement},
  author = {Abanin, Dmitry A. and Altman, Ehud and Bloch, Immanuel and Serbyn, Maksym},
  journal = {Rev. Mod. Phys.},
  volume = {91},
  issue = {2},
  pages = {021001},
  numpages = {26},
  year = {2019},
  month = {May},
  publisher = {American Physical Society},
  doi = {10.1103/RevModPhys.91.021001},
  url = {https://link.aps.org/doi/10.1103/RevModPhys.91.021001}
}

@article{Calabrese_2016,
doi = {10.1088/1742-5468/2016/06/064001},
url = {https://dx.doi.org/10.1088/1742-5468/2016/06/064001},
year = {2016},
month = {jun},
publisher = {IOP Publishing and SISSA},
volume = {2016},
number = {6},
pages = {064001},
author = {Calabrese, Pasquale and Essler, Fabian H L and Mussardo, Giuseppe},
title = {Introduction to ‘Quantum Integrability in Out of Equilibrium Systems’},
journal = {Journal of Statistical Mechanics: Theory and Experiment}
}

@article{Moudgalya_2022,
doi = {10.1088/1361-6633/ac73a0},
url = {https://dx.doi.org/10.1088/1361-6633/ac73a0},
year = {2022},
month = {jul},
publisher = {IOP Publishing},
volume = {85},
number = {8},
pages = {086501},
author = {Moudgalya, Sanjay and Bernevig, B Andrei and Regnault, Nicolas},
title = {Quantum many-body scars and Hilbert space fragmentation: a review of exact results},
journal = {Reports on Progress in Physics}
}

@article{PhysRevX.10.011047,
  title = {Ergodicity Breaking Arising from Hilbert Space Fragmentation in Dipole-Conserving Hamiltonians},
  author = {Sala, Pablo and Rakovszky, Tibor and Verresen, Ruben and Knap, Michael and Pollmann, Frank},
  journal = {Phys. Rev. X},
  volume = {10},
  issue = {1},
  pages = {011047},
  numpages = {19},
  year = {2020},
  month = {Feb},
  publisher = {American Physical Society},
  doi = {10.1103/PhysRevX.10.011047},
  url = {https://link.aps.org/doi/10.1103/PhysRevX.10.011047}
}

@article{PhysRevB.101.174204,
  title = {Localization from Hilbert space shattering: From theory to physical realizations},
  author = {Khemani, Vedika and Hermele, Michael and Nandkishore, Rahul},
  journal = {Phys. Rev. B},
  volume = {101},
  issue = {17},
  pages = {174204},
  numpages = {17},
  year = {2020},
  month = {May},
  publisher = {American Physical Society},
  doi = {10.1103/PhysRevB.101.174204},
  url = {https://link.aps.org/doi/10.1103/PhysRevB.101.174204}
}

@Article{Bernien2017,
author={Bernien, Hannes
and Schwartz, Sylvain
and Keesling, Alexander
and Levine, Harry
and Omran, Ahmed
and Pichler, Hannes
and Choi, Soonwon
and Zibrov, Alexander S.
and Endres, Manuel
and Greiner, Markus
and Vuleti{\'{c}}, Vladan
and Lukin, Mikhail D.},
title={Probing many-body dynamics on a 51-atom quantum simulator},
journal={Nature},
year={2017},
month={Nov},
day={01},
volume={551},
number={7682},
pages={579-584},
issn={1476-4687},
doi={10.1038/nature24622},
url={https://doi.org/10.1038/nature24622}
}

@article{PhysRevLett.119.030601,
  title = {Systematic Construction of Counterexamples to the Eigenstate Thermalization Hypothesis},
  author = {Shiraishi, Naoto and Mori, Takashi},
  journal = {Phys. Rev. Lett.},
  volume = {119},
  issue = {3},
  pages = {030601},
  numpages = {6},
  year = {2017},
  month = {Jul},
  publisher = {American Physical Society},
  doi = {10.1103/PhysRevLett.119.030601},
  url = {https://link.aps.org/doi/10.1103/PhysRevLett.119.030601}
}

@article{PhysRevResearch.2.043305,
  title = {From tunnels to towers: Quantum scars from Lie algebras and $q$-deformed Lie algebras},
  author = {O'Dea, Nicholas and Burnell, Fiona and Chandran, Anushya and Khemani, Vedika},
  journal = {Phys. Rev. Res.},
  volume = {2},
  issue = {4},
  pages = {043305},
  numpages = {30},
  year = {2020},
  month = {Dec},
  publisher = {American Physical Society},
  doi = {10.1103/PhysRevResearch.2.043305},
  url = {https://link.aps.org/doi/10.1103/PhysRevResearch.2.043305}
}

@article{PhysRevResearch.3.043122,
  title = {Experimental estimation of the quantum Fisher information from randomized measurements},
  author = {Yu, Min and Li, Dongxiao and Wang, Jingcheng and Chu, Yaoming and Yang, Pengcheng and Gong, Musang and Goldman, Nathan and Cai, Jianming},
  journal = {Phys. Rev. Res.},
  volume = {3},
  issue = {4},
  pages = {043122},
  numpages = {9},
  year = {2021},
  month = {Nov},
  publisher = {American Physical Society},
  doi = {10.1103/PhysRevResearch.3.043122},
  url = {https://link.aps.org/doi/10.1103/PhysRevResearch.3.043122}
}

@article{PRXQuantum.5.030338,
  title = {Robust Estimation of the Quantum Fisher Information on a Quantum Processor},
  author = {Vitale, Vittorio and Rath, Aniket and Jurcevic, Petar and Elben, Andreas and Branciard, Cyril and Vermersch, Beno\^{\i}t},
  journal = {PRX Quantum},
  volume = {5},
  issue = {3},
  pages = {030338},
  numpages = {27},
  year = {2024},
  month = {Aug},
  publisher = {American Physical Society},
  doi = {10.1103/PRXQuantum.5.030338},
  url = {https://link.aps.org/doi/10.1103/PRXQuantum.5.030338}
}

@Article{Yu2022,
author={Yu, Min
and Liu, Yu
and Yang, Pengcheng
and Gong, Musang
and Cao, Qingyun
and Zhang, Shaoliang
and Liu, Haibin
and Heyl, Markus
and Ozawa, Tomoki
and Goldman, Nathan
and Cai, Jianming},
title={Quantum Fisher information measurement and verification of the quantum Cram{\'e}r--Rao bound in a solid-state qubit},
journal={npj Quantum Information},
year={2022},
month={May},
day={12},
volume={8},
number={1},
pages={56},
issn={2056-6387},
doi={10.1038/s41534-022-00547-x},
url={https://doi.org/10.1038/s41534-022-00547-x}
}

@article{PhysRevB.101.174308,
  title = {Quantum many-body scars from virtual entangled pairs},
  author = {Chattopadhyay, Sambuddha and Pichler, Hannes and Lukin, Mikhail D. and Ho, Wen Wei},
  journal = {Phys. Rev. B},
  volume = {101},
  issue = {17},
  pages = {174308},
  numpages = {14},
  year = {2020},
  month = {May},
  publisher = {American Physical Society},
  doi = {10.1103/PhysRevB.101.174308},
  url = {https://link.aps.org/doi/10.1103/PhysRevB.101.174308}
}

@article{PRXQuantum.5.040330,
  title = {Symmetries as Ground States of Local Superoperators: Hydrodynamic Implications},
  author = {Moudgalya, Sanjay and Motrunich, Olexei I.},
  journal = {PRX Quantum},
  volume = {5},
  issue = {4},
  pages = {040330},
  numpages = {41},
  year = {2024},
  month = {Nov},
  publisher = {American Physical Society},
  doi = {10.1103/PRXQuantum.5.040330},
  url = {https://link.aps.org/doi/10.1103/PRXQuantum.5.040330}
}

@article{JMRadcliffe_1971,
doi = {10.1088/0305-4470/4/3/009},
url = {https://doi.org/10.1088/0305-4470/4/3/009},
year = {1971},
month = {may},
publisher = {},
volume = {4},
number = {3},
pages = {313},
author = {J M Radcliffe},
title = {Some properties of coherent spin states},
journal = {Journal of Physics A: General Physics}
}

@article{PhysRevB.107.224312,
  title = {Numerical methods for detecting symmetries and commutant algebras},
  author = {Moudgalya, Sanjay and Motrunich, Olexei I.},
  journal = {Phys. Rev. B},
  volume = {107},
  issue = {22},
  pages = {224312},
  numpages = {19},
  year = {2023},
  month = {Jun},
  publisher = {American Physical Society},
  doi = {10.1103/PhysRevB.107.224312},
  url = {https://link.aps.org/doi/10.1103/PhysRevB.107.224312}
}

@article{PhysRevX.14.041069,
  title = {Exhaustive Characterization of Quantum Many-Body Scars Using Commutant Algebras},
  author = {Moudgalya, Sanjay and Motrunich, Olexei I.},
  journal = {Phys. Rev. X},
  volume = {14},
  issue = {4},
  pages = {041069},
  numpages = {49},
  year = {2024},
  month = {Dec},
  publisher = {American Physical Society},
  doi = {10.1103/PhysRevX.14.041069},
  url = {https://link.aps.org/doi/10.1103/PhysRevX.14.041069}
}

@misc{matsui2025symmetryprotectedtopologicalscarsubspaces,
      title={Symmetry-protected topological scar subspaces}, 
      author={Chihiro Matsui and Thomas Quella and Naoto Tsuji},
      year={2025},
      eprint={2512.11216},
      archivePrefix={arXiv},
      primaryClass={cond-mat.str-el},
      url={https://arxiv.org/abs/2512.11216}, 
}

@ARTICLE{1977RSPTA.287..237B,
       author = {{Berry}, M.~V.},
        title = "{Semi-Classical Mechanics in Phase Space: A Study of Wigner's Function}",
      journal = {Philosophical Transactions of the Royal Society of London Series A},
         year = 1977,
        month = oct,
       volume = {287},
       number = {1343},
        pages = {237-271},
          doi = {10.1098/rsta.1977.0145},
       adsurl = {https://ui.adsabs.harvard.edu/abs/1977RSPTA.287..237B}
}

@Article{10.21468/SciPostPhys.2.1.003,
	title={{QuSpin: a Python package for dynamics and exact diagonalisation of quantum many body systems part I: spin chains}},
	author={Phillip Weinberg and Marin Bukov},
	journal={SciPost Phys.},
	volume={2},
	pages={003},
	year={2017},
	publisher={SciPost},
	doi={10.21468/SciPostPhys.2.1.003},
	url={https://scipost.org/10.21468/SciPostPhys.2.1.003},
}

@Article{Kennedy1992,
author={Kennedy, Tom
and Tasaki, Hal},
title={Hidden symmetry breaking and the Haldane phase inS=1 quantum spin chains},
journal={Communications in Mathematical Physics},
year={1992},
month={Jul},
day={01},
volume={147},
number={3},
pages={431-484},
issn={1432-0916},
doi={10.1007/BF02097239},
url={https://doi.org/10.1007/BF02097239}
}

@article{PhysRevLett.89.077204,
  title = {Order Parameter to Characterize Valence-Bond-Solid States in Quantum Spin Chains},
  author = {Nakamura, Masaaki and Todo, Synge},
  journal = {Phys. Rev. Lett.},
  volume = {89},
  issue = {7},
  pages = {077204},
  numpages = {4},
  year = {2002},
  month = {Jul},
  publisher = {American Physical Society},
  doi = {10.1103/PhysRevLett.89.077204},
  url = {https://link.aps.org/doi/10.1103/PhysRevLett.89.077204}
}

@article{NAKAMURA20031000,
title = {Identification of topologically different valence bond states in spin ladders},
journal = {Physica B: Condensed Matter},
volume = {329-333},
pages = {1000-1001},
year = {2003},
note = {Proceedings of the 23rd International Conference on Low Temperature Physics},
issn = {0921-4526},
doi = {https://doi.org/10.1016/S0921-4526(02)02180-4},
url = {https://www.sciencedirect.com/science/article/pii/S0921452602021804},
author = {Masaaki Nakamura}
}

@article{transportinSpin1xy,
  title = {Transport in a system with a tower of quantum many-body scars},
  author = {Morettini, Gianluca and Capizzi, Luca and Fagotti, Maurizio and Mazza, Leonardo},
  journal = {Phys. Rev. B},
  volume = {112},
  issue = {13},
  pages = {134314},
  numpages = {12},
  year = {2025},
  month = {Oct},
  publisher = {American Physical Society},
  doi = {10.1103/821h-8yjz},
  url = {https://link.aps.org/doi/10.1103/821h-8yjz}
}

@Article{Hauke2016,
author={Hauke, Philipp
and Heyl, Markus
and Tagliacozzo, Luca
and Zoller, Peter},
title={Measuring multipartite entanglement through dynamic susceptibilities},
journal={Nature Physics},
year={2016},
month={Aug},
day={01},
volume={12},
number={8},
pages={778-782},
issn={1745-2481},
doi={10.1038/nphys3700},
url={https://doi.org/10.1038/nphys3700}
}

@article{PhysRevB.81.064439,
  title = {Entanglement spectrum of a topological phase in one dimension},
  author = {Pollmann, Frank and Turner, Ari M. and Berg, Erez and Oshikawa, Masaki},
  journal = {Phys. Rev. B},
  volume = {81},
  issue = {6},
  pages = {064439},
  numpages = {10},
  year = {2010},
  month = {Feb},
  publisher = {American Physical Society},
  doi = {10.1103/PhysRevB.81.064439},
  url = {https://link.aps.org/doi/10.1103/PhysRevB.81.064439}
}

@article{PhysRevB.87.155114,
  title = {Symmetry protected topological orders and the group cohomology of their symmetry group},
  author = {Chen, Xie and Gu, Zheng-Cheng and Liu, Zheng-Xin and Wen, Xiao-Gang},
  journal = {Phys. Rev. B},
  volume = {87},
  issue = {15},
  pages = {155114},
  numpages = {48},
  year = {2013},
  month = {Apr},
  publisher = {American Physical Society},
  doi = {10.1103/PhysRevB.87.155114},
  url = {https://link.aps.org/doi/10.1103/PhysRevB.87.155114}
}

@article{Shane_Dooley_Quantum_sensing_scars_2021,
  title = {Robust Quantum Sensing in Strongly Interacting Systems with Many-Body Scars},
  author = {Dooley, Shane},
  journal = {PRX Quantum},
  volume = {2},
  issue = {2},
  pages = {020330},
  numpages = {12},
  year = {2021},
  month = {May},
  publisher = {American Physical Society},
  doi = {10.1103/PRXQuantum.2.020330},
  url = {https://link.aps.org/doi/10.1103/PRXQuantum.2.020330}
}

@article{Qi2019determininglocal,
  doi = {10.22331/q-2019-07-08-159},
  url = {https://doi.org/10.22331/q-2019-07-08-159},
  title = {Determining a local {H}amiltonian from a single eigenstate},
  author = {Qi, Xiao-Liang and Ranard, Daniel},
  journal = {{Quantum}},
  issn = {2521-327X},
  publisher = {{Verein zur F{\"{o}}rderung des Open Access Publizierens in den Quantenwissenschaften}},
  volume = {3},
  pages = {159},
  month = jul,
  year = {2019}
}

@article{tarun_single_eigenstate_encode,
  title = {Does a Single Eigenstate Encode the Full Hamiltonian?},
  author = {Garrison, James R. and Grover, Tarun},
  journal = {Phys. Rev. X},
  volume = {8},
  issue = {2},
  pages = {021026},
  numpages = {24},
  year = {2018},
  month = {Apr},
  publisher = {American Physical Society},
  doi = {10.1103/PhysRevX.8.021026},
  url = {https://link.aps.org/doi/10.1103/PhysRevX.8.021026}
}

@misc{mohapatra2025unravelingadditionalquantummanybody,
      title={Unraveling additional quantum many-body scars of the spin-$1$ $XY$ model with Fock-space cages and commutant algebras}, 
      author={Sashikanta Mohapatra and Sanjay Moudgalya and Ajit C. Balram},
      year={2025},
      eprint={2511.14878},
      archivePrefix={arXiv},
      primaryClass={cond-mat.str-el},
      url={https://arxiv.org/abs/2511.14878}, 
}

@article{Omiya_embedding_PXP,
  title = {Quantum many-body scars in bipartite Rydberg arrays originating from hidden projector embedding},
  author = {Omiya, Keita and M\"uller, Markus},
  journal = {Phys. Rev. A},
  volume = {107},
  issue = {2},
  pages = {023318},
  numpages = {20},
  year = {2023},
  month = {Feb},
  publisher = {American Physical Society},
  doi = {10.1103/PhysRevA.107.023318},
  url = {https://link.aps.org/doi/10.1103/PhysRevA.107.023318}
}

@article{Fractionalization_pavesway,
  title = {Fractionalization paves the way to local projector embeddings of quantum many-body scars},
  author = {Omiya, Keita and M\"uller, Markus},
  journal = {Phys. Rev. B},
  volume = {108},
  issue = {5},
  pages = {054412},
  numpages = {18},
  year = {2023},
  month = {Aug},
  publisher = {American Physical Society},
  doi = {10.1103/PhysRevB.108.054412},
  url = {https://link.aps.org/doi/10.1103/PhysRevB.108.054412}
}

\clearpage
\onecolumngrid
\appendix

\section{Showing that the bonds form the Commutant algebra}\label{app:bonds}

We define the scar tower as
\begin{align}
   |S_n\rangle &= \dfrac{\mathcal{N}(n)}{n!}\,(Q^{\dagger})^{n}|\Omega\rangle , \\
   |\Omega\rangle &= |-1,-1,-1,\ldots,-1\rangle , \\
   Q^{\dagger} &= \sum_{i}(-1)^{i}(S^{+}_{i})^{2}.
\end{align}

 where we have introduced the factor of $n!$ in the definition, as the scars would have states of the form, $|-1 \, 1\, -1\, 1 \dots\rangle$ and the $|1\rangle$'s
 are permutation symmetric, as they can arise from $n!$ permutations of the $(S^{+}_{j})^{2}$ operator; for eg.  $(Q^{\dagger})^{2}|-1 -1 -1\rangle$ would have a term $|1 \, 1 -1\rangle$ that can come first by the action of $(S^{+}_{1})^{2}$ and then $(S^{+}_{2})^{2}$ or vice versa.
 
\subsection{Notation}

Since the product states appearing in $|S_{n}\rangle$ are configurations with exactly $n$ sites in the $|+1\rangle$ state, we introduce the shorthand
\begin{equation}
   |j_{1}j_{2}\cdots j_{n}\rangle
   \;\equiv\;
   |-1\cdots \underbrace{1}_{j_{1}}\cdots
   \underbrace{1}_{j_{n}}\cdots -1\rangle
   ,
\end{equation}
Acting with $(Q^\dagger)^{n}$ on $|\Omega\rangle = |-1 -1 \dots -1\rangle$ generates each such configuration $n!$ times due to the permutation symmetry of the bimagnon creation operators. 
The state $|j_{1}j_{2}\cdots j_{n}\rangle$ depends on the \textit{unordered set}
$\{j_{1},j_{2},\ldots,j_{n}\}$ of distinct sites.

The inner product in this basis is
\begin{equation}\label{eq:innerproduct}
   \langle j'_{1}\cdots j'_{m}\,|\,j_{1}\cdots j_{n}\rangle
   =\,\delta_{mn}\,
   \delta_{\{j'\},\{j\}},
\end{equation}
where $\delta_{mn}$ follows from the conservation of total magnetization due to the global $U(1)$ symmetry, and $\delta_{\{j'\},\{j\}}$ enforces equality of the two \textit{unordered sets} of sites. 

\subsection{Normalization}

Expressing the scar states in the above notation, we obtain
\begin{align}
   |S_{n}\rangle
   &= \mathcal{N}(n)\sum_{\{j\}}
      (-1)^{\sum_{k=1}^{n}j_{k}}
      |j_{1}\cdots j_{n}\rangle , \\[4pt]
   \langle S_{n}|S_{n}\rangle
   &= (\mathcal{N}(n))^{2}\sum_{\{j\}}1 \nonumber\\
   &= (\mathcal{N}(n))^{2}\binom{L}{n},
\end{align}
where the sum $\sum_{\{j\}}$ runs over all $\binom{L}{n}$ distinct subsets of $n$ sites. And also, the starting definition is the one which is intuitive, as after the action of $(Q^{\dagger})^{n}$ same state can arise in $n!$ different ways, and the prefactor of $1/n!$ actually corrects this overcounting of the same state.

Imposing the normalization condition $\langle S_{n}|S_{n}\rangle=1$, we find
\begin{equation}
   \mathcal{N}(n)=\sqrt{\frac{n!\,(L-n)!}{L!}}.
\end{equation}

\subsection{Commutation relations}

From SU(2),
\begin{align}
   [S^{+}_{i},S^{-}_{j}] &= 2\delta_{ij}S^{z}_{j}, \quad
   [S^{z}_{i},S^{\pm}_{j}] = \pm \delta_{ij}S^{\pm}_{i}, \\
   [S^{-}_{i},(S^{+}_{j})^{2}] &= -2\delta_{ij}\{S^{z}_{i},S^{+}_{i}\}, \quad
   [S^{z}_{i},(S^{+}_{j})^{2}] = 2\delta_{ij}(S^{+}_{i})^{2}
\end{align}

Consider the hopping term $T_{i,i+1}=S^{+}_{i}S^{-}_{i+1}+{\rm h.c.}$:
\begin{align}
   [T_{i,i+1},Q^{\dagger}]
   &=-2\sum_{j}(-1)^{j}\Big[
       \delta_{i+1,j}S^{+}_{i}\{S^{z}_{i+1},S^{+}_{i+1}\} \nonumber\\
   &\hspace{3.0cm}
       +\delta_{i,j}\{S^{z}_{i},S^{+}_{i}\}S^{+}_{i+1}
     \Big].
\end{align}

Acting on the vacuum $|\Omega\rangle$ gives
\begin{align}
    \nonumber\implies [T_{i,i+1}, Q^{\dagger}]|\Omega\rangle &= -2\Big[(-1)^{i+1}S^{+}_{i}\{S^{z}_{i+1}, S^{+}_{i+1}\}
    \nonumber \\ &\hspace{0.5cm}
    + (-1)^{i}\{S^{z}_{i}, S^{+}_{i}\}S^{+}_{i+1}\Big]|-1, -1\rangle_{i,i+1} \\
    \nonumber\implies [T_{i,i+1}, Q^{\dagger}]|\Omega\rangle &= 2\left[(-1)^{i+1} |00\rangle_{i,i+1}+(-1)^{i} |00\rangle_{i,i+1}\right] \\
    \implies [T_{i,i+1},Q^{\dagger}]|\Omega\rangle &= 0
\end{align}
We note that we can do this for any $T_{i, j}$ provided that $(i-j) \text{mod(2)} = 1$ we can still have this construction and create the first scar in the tower. \newline

Next we need to see if this commutation relation holds for higher members in the scar tower. To see that, Consider the expression, 
\begin{align}
    \nonumber[T_{i,i+1}, Q^{\dagger}] (Q^{\dagger})^{m}|\Omega\rangle
\end{align}
Note that the commutator acts on $(Q^{\dagger})^{m}|\Omega\rangle$ which consists of only $1$'s and $-1$'s so in all for the commutator above only 4 configurations are possible that can act on by the commutator which are $|1,1\rangle, |-1,1\rangle, |1,-1\rangle, |-1,-1\rangle$, One checks
\begin{equation}
   [T_{i,i+1},Q^{\dagger}]
   \begin{cases}
      |-1,-1\rangle &\mapsto 0, \\[4pt]
      |1,-1\rangle &\mapsto 0, \\[4pt]
      |-1,1\rangle &\mapsto 0, \\[4pt]
      |1,1\rangle &\mapsto 0,
   \end{cases}
\end{equation}
so that
\begin{align}\label{app,eq: commutator with bond}
   \nonumber[T_{i,i+1}, Q^{\dagger}] (Q^{\dagger})^{m}|\Omega\rangle &= 0 \\
   \implies [T_{i,i+1},(Q^{\dagger})^{m+1}]|\Omega\rangle &= 0,
   \qquad \forall m .
\end{align}
The last expression is a result of applying $T_{i, i+1}$ on the vacuum state ,i.e., $|\Omega\rangle= |-1 -1 \cdots -1\rangle$, which results in $T_{i, i+1}|\Omega\rangle =0$. The same reasoning applies for $T_{i,i+a}$ with $a$ odd:
\begin{equation}
   [T_{i,i+a},(Q^{\dagger})^{m}]|\Omega\rangle=0,
   \qquad \forall m,\; a\in{\rm odd}.
\end{equation}
Note, that the above equation defines our commutant algebra.

\subsection{Anisotropy term}

For the $D$ term,
\begin{align}
   [(S^{z}_{i})^{2},(S^{+}_{j})^{2}]|\Omega\rangle
   &= (S^{z}_{i})^{2}(S^{+}_{i})^{2}|-1\rangle_{i}
     -(S^{+}_{i})^{2}(S^{z}_{i})^{2}|-1\rangle_{i} \nonumber\\
   &= 0.
\end{align}
Thus,
\begin{equation}
   [(S^{z}_{i})^{2},Q^{\dagger}](Q^{\dagger})^{m}|\Omega\rangle=0.
\end{equation}
This holds true, cause the states in the scar tower are composed of $1$'s and $-1$'s and this commutator gives zero for both cases.
The scar states are therefore frustration-free eigenstates of
$\sum_{i}(S^{z}_{i})^{2}$.

\subsection{Uniform field term}

Finally,
\begin{align}
   \sum_{j}(-1)^{j}[S^{z}_{i},(S^{+}_{j})^{2}]
   &=2\sum_{j}(-1)^{j}\delta_{ij}(S^{+}_{j})^{2}, \\
   [S^{z}_{i},Q^{\dagger}] &= 2(-1)^{i}(S^{+}_{i})^{2}, \\
   \Big[\sum_{i}S^{z}_{i},Q^{\dagger}\Big] &= 2Q^{\dagger}.
\end{align}

\subsection{Spectrum generating algebra}

Combining results,
\begin{equation}\label{eq:SGA}
   \left([H,Q^{\dagger}]-2h\,Q^{\dagger}\right)\,\mathcal{W}=0 ,
\end{equation}
where $\mathcal{W}$ is the subspace spanned by the scar tower.  
Equation~\eqref{eq:SGA} is the spectrum generating algebra (SGA), this structure has been found in several models showing QMBS~\cite{ PhysRevB.101.024306,PhysRevB.98.235155, PhysRevB.98.235156,PhysRevLett.122.220603} and in several formalisms like~\cite{PhysRevB.101.195131} and is responsible for dynamical signatures like Fidelity revival ~\cite{PhysRevB.98.155134, PhysRevLett.123.147201}.\newline

From Eq.~\eqref{eq:SGA}, the eigenenergies of the scarred states follow as
\begin{equation}
   H|S_{n}\rangle = E_{n}\,|S_{n}\rangle ,
   \qquad E_{n} = E_{0}+2hn .
\end{equation}
Where $E_{0}$ is the energy of the vacuum state i.e. $|\Omega\rangle$. Also note that ~\eqref{app,eq: commutator with bond} tells that the scar states belong to the commutant algebra~\cite{PhysRevX.12.011050} generated by the bond terms.

\begin{figure}
    \centering
    \includegraphics[width=0.5\linewidth, trim = 0 0 0 20, clip]{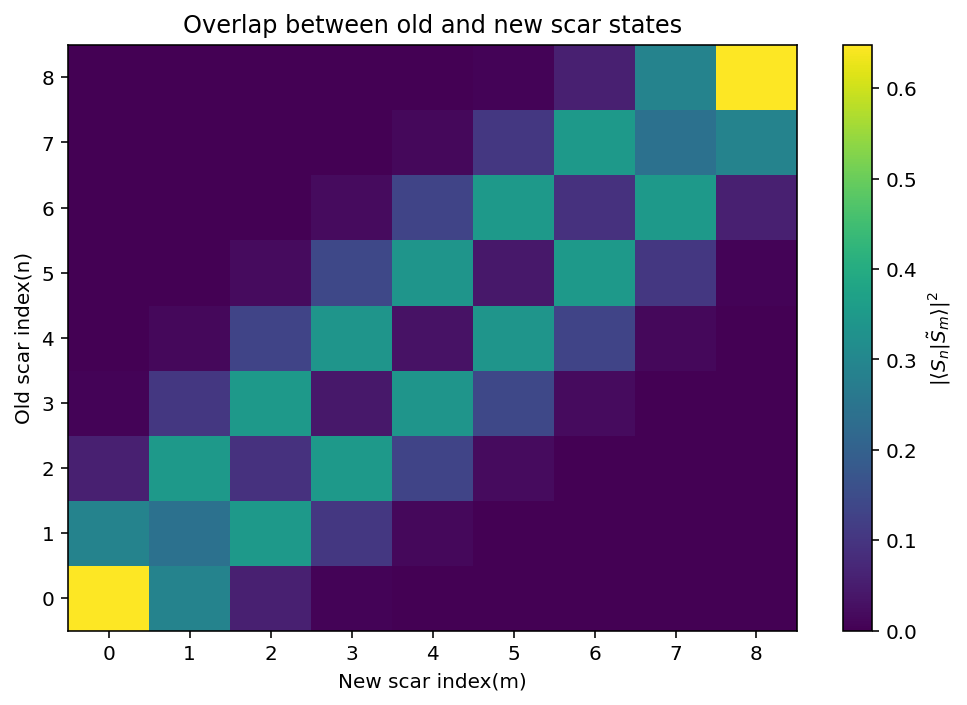}
    \caption{The overlap between old scars with index $n$ in the y-axis with no perturbation, and new scars with the $V$ perturbation, that doesn't break the SGA, the index $m$ along the x-axis}
    \label{fig:scar_overlap}
\end{figure}

\section{Difference between an SGA and a Commutant algebra}\label{app: SGA and commutant}
We define the bond algebra as $\mathcal{B}=\{ \{T_{i, i+a}, \,\, \forall \,\,(a\%2)=1\}, \,\,, (S^{z}_{i})^{2}, \,\, \sum_{i}S^{z}_{i}\}$, and for this the commutant algebra constitutes the scars,i.e., $\mathcal{C}=\{|S_{n}\rangle\langle S_{n}| \,\, \forall \,n=0, ... , L\}$. This implies that any mixing of scars due to some perturbation would break the commutant algebra, but on the other hand, if the perturbation $V$ is such that it mixes the scars in such a way that the scar subspace is preserved, it implies a redefinition of the basis in the scar subspace. To illustrate this, we consider the perturbation $V=\sum_{j} (-1)^{j} (S^{x}_{j})^{2}$, which can be decomposed as $V = Q/2 + Q^{\dagger}/2 + \mathds{1}/2$ for details the reader is referred to the supplemental material. It can be shown that within the scar subspace, 
\begin{align}
    [H_{0}+V, Q^{\dagger}] = 2hQ^{\dagger} +c
\end{align}
here, $H_{0}$ is the Hamiltonian that preserves the commutant algebra, the constant c would simply couple the nearby scars and hence would become equivalent to a tight-binding model in the energy eigenspace. To construct the SGA, one can take linear combinations of $|\tilde{S}_{n}\rangle=\sum_{m} c_{nm}|S_{m}\rangle $ as shown in Fig.~\eqref{fig:scar_overlap} and hence get an SGA redefined in terms of the powers of the original SGA generators.

\section{Numerical investigation of scars under different perturbations}\label{app:numerical digaonlization for pert}

In the main text, we discuss certain perturbations that break or preserve the scar subspace. In this appendix, we provide numerical evidence of the above scar subspace preservation or destruction, as shown in Fig.~\eqref{fig:app, extensive scar preserving}, Fig.~\eqref{fig: app, extensive non scar preserving}, Fig.~\eqref{fig: app, intensive}. 

\begin{figure}[h]
    \centering
    \includegraphics[width=0.6\linewidth]{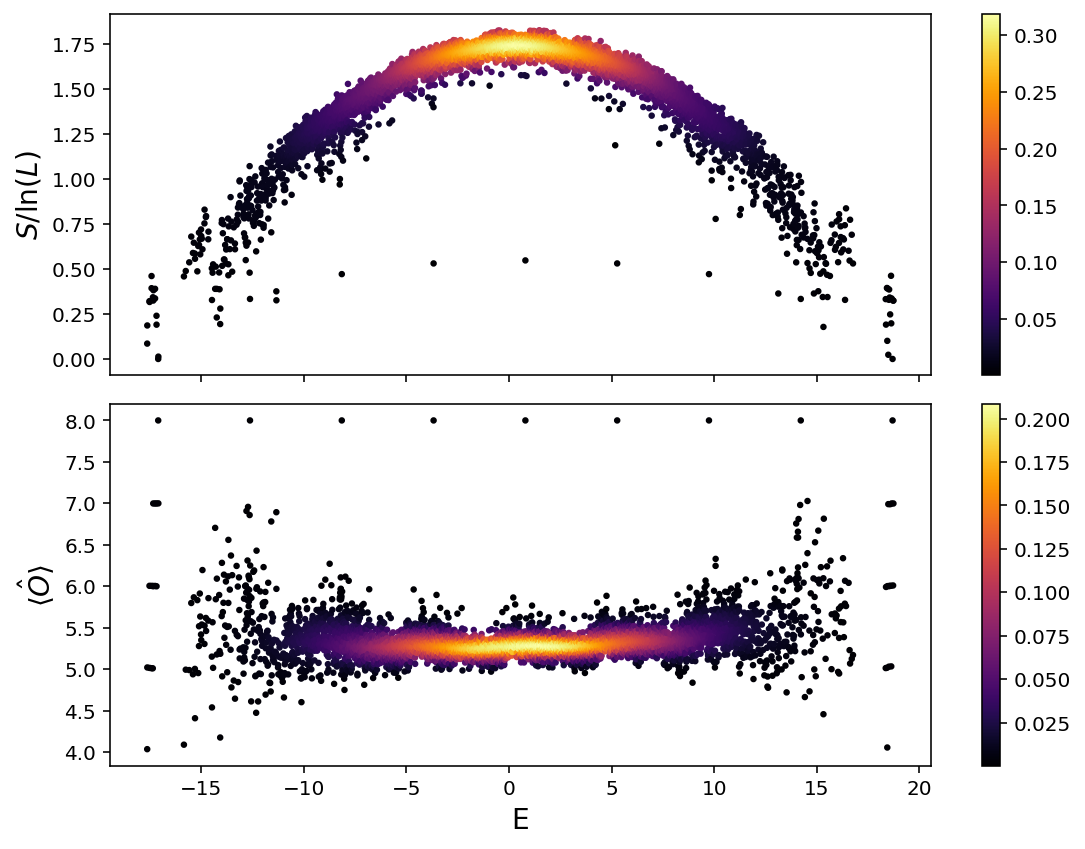}
    \caption{Adding the perturbation $V=\sum (-1)^{i}(S^{x}_{i})^{2}$ breaks the $U(1)$ symmetry, hence we cannot go for large system sizes, but we can see that we have outlier points especially in the middle of the spectrum and also throughout the spectrum which are equally spaced, have low entanglement entropy and also violate the Diagonal ETH}
    \label{fig:app, extensive scar preserving}
\end{figure}

\begin{figure}[h]
    \centering
    \includegraphics[width=0.6\linewidth]{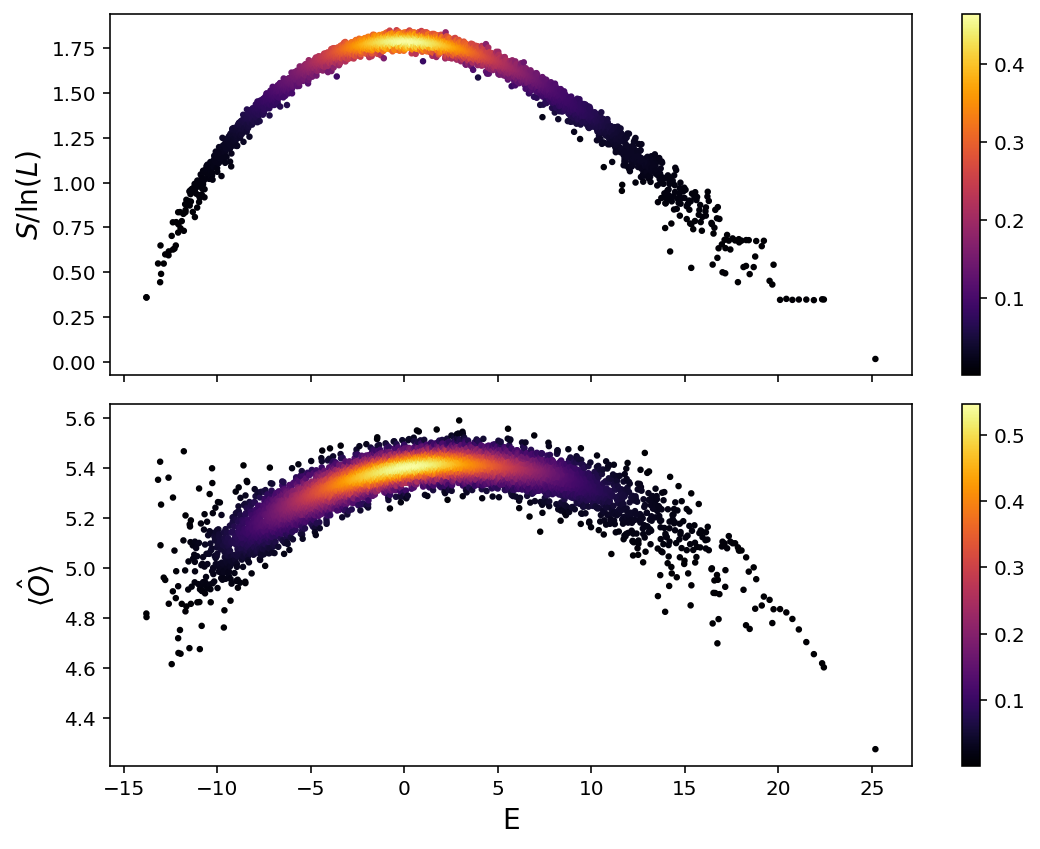}
    \caption{Adding the perturbation $V=\sum S^{x}_{i}$ breaks the $U(1)$ symmetry; hence, we cannot go for large system sizes, but we can see that there are no outlier points throughout the spectrum, and the system satisfies the Diagonal ETH nicely with no anomalously low entanglement eigenstates}
    \label{fig: app, extensive non scar preserving}
\end{figure}

\begin{figure}[h]
    \centering
    \includegraphics[width=0.6\linewidth]{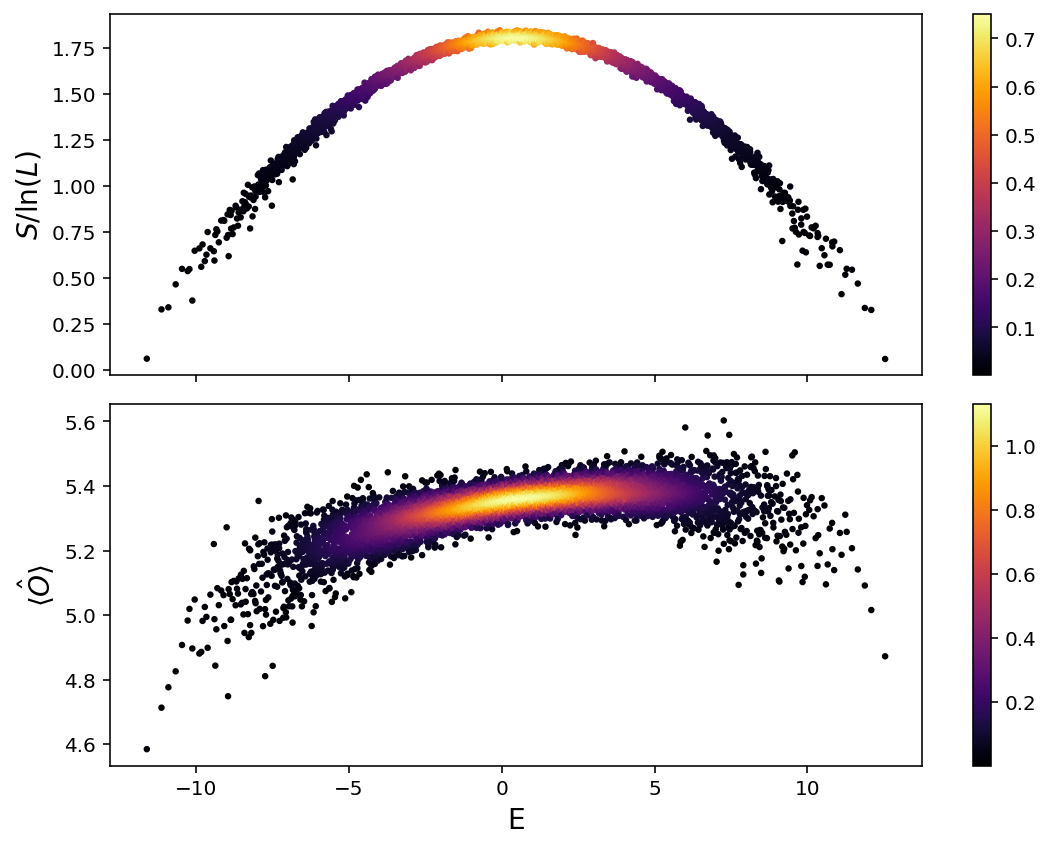}
    \caption{Adding the perturbation $V=S^{x}_{4}$ breaks the $U(1)$ symmetry; hence, we cannot go for large system sizes, but we can see that there are no outlier points throughout the spectrum, and the system satisfies the Diagonal ETH nicely with no anomalously low entanglement eigenstates}
    \label{fig: app, intensive}
\end{figure}

\section{Numerics for the string order parameter}\label{app:stringorder}
In this appendix, we examine the string order parameter~\cite{Kennedy1992}, which was investigated to uncover the hidden $Z_{2}\times Z_{2}$ symmetry in the spin-1 Haldane chain. It is known that the LSM twist operator~\cite{PhysRevLett.89.077204, NAKAMURA20031000} provides an alternative diagnostic of the SPt. Both the string order parameter and the LSM twist operator, therefore, have been widely used in spin-1 or spin 1/2 chains.

\begin{figure*}
    \centering
    \captionsetup{justification=centering}

    \begin{subfigure}{0.45\textwidth}
        \centering
        \includegraphics[width=1\linewidth]{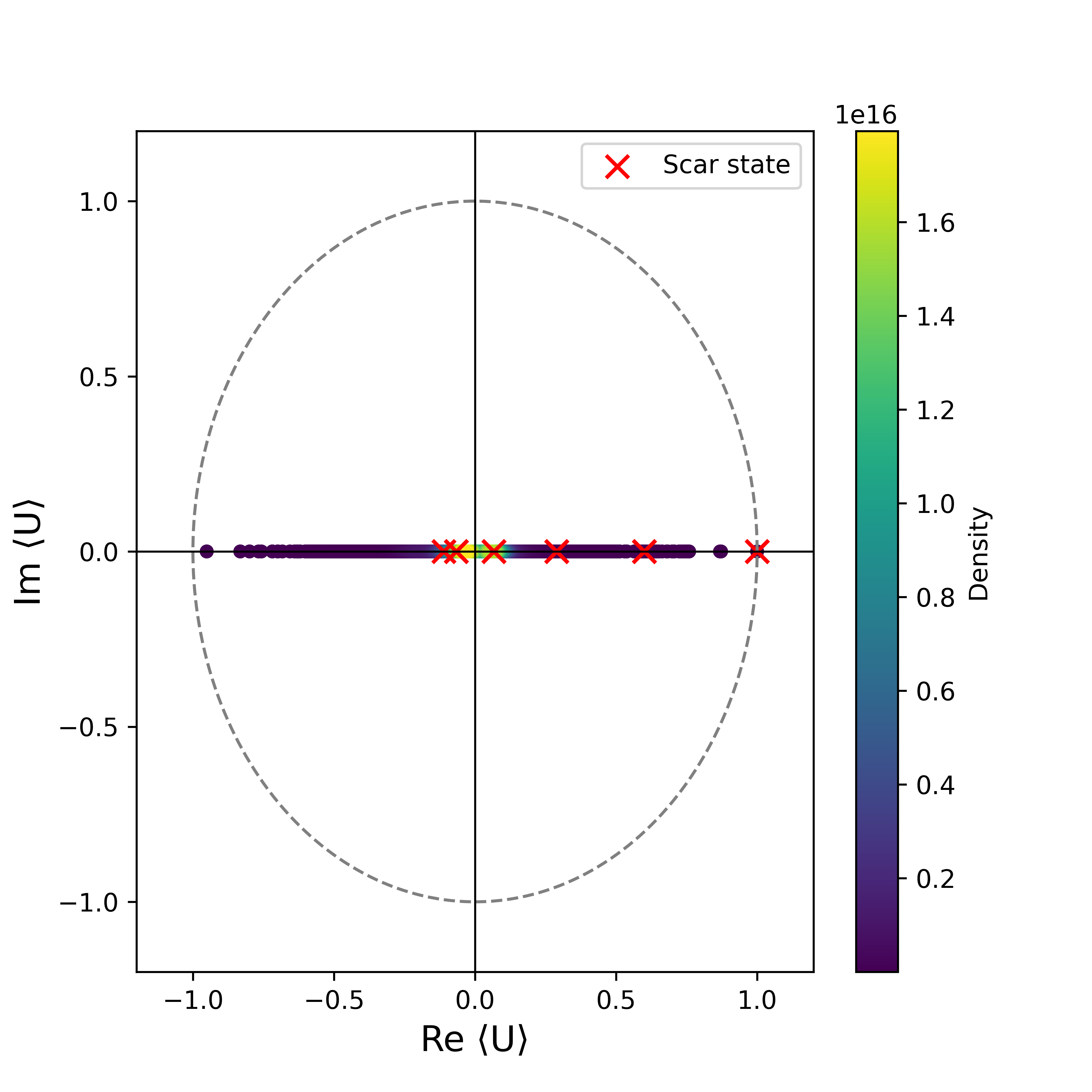}
        \caption{$\langle O_{1}\rangle = \langle  S^{z}_{1}\, \prod_{j=2}^{L-1}e^{i\pi S^{z}_{j}}S^{z}_{L} \rangle $}
        \label{fig:a}
    \end{subfigure}
    \hfill
    \begin{subfigure}{0.45\textwidth}
        \centering
        \includegraphics[width=1\linewidth]{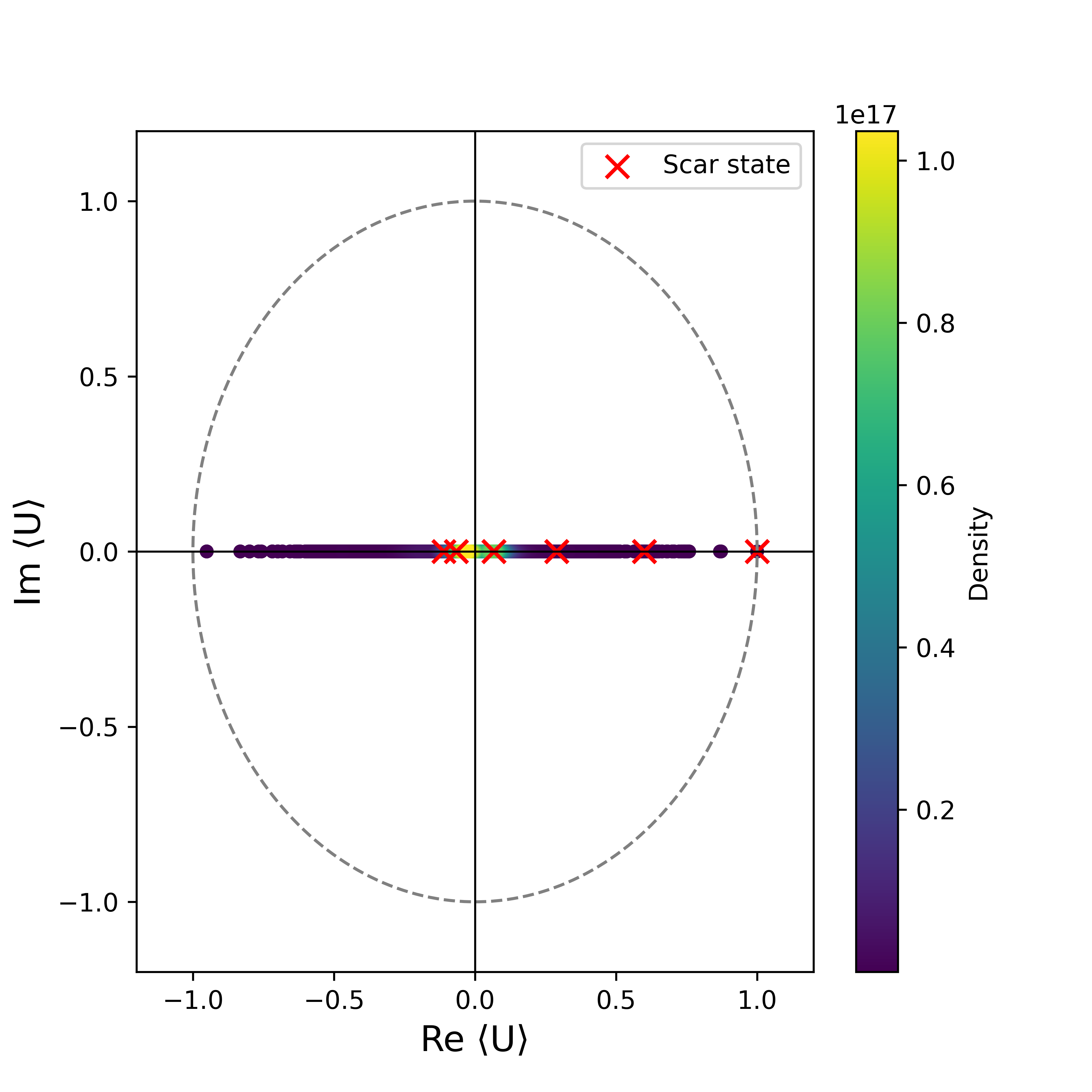}
        \caption{$\langle O_{2}\rangle = \langle S^{z}_{1}\, \prod_{j=2}^{L-1}e^{i\pi (S^{z}_{j})^{2}}S^{z}_{L} \rangle$}
        \label{fig:b}
    \end{subfigure}

    \vspace{0.2cm}

    \begin{subfigure}{0.45\textwidth}
        \centering
        \includegraphics[width=1\linewidth]{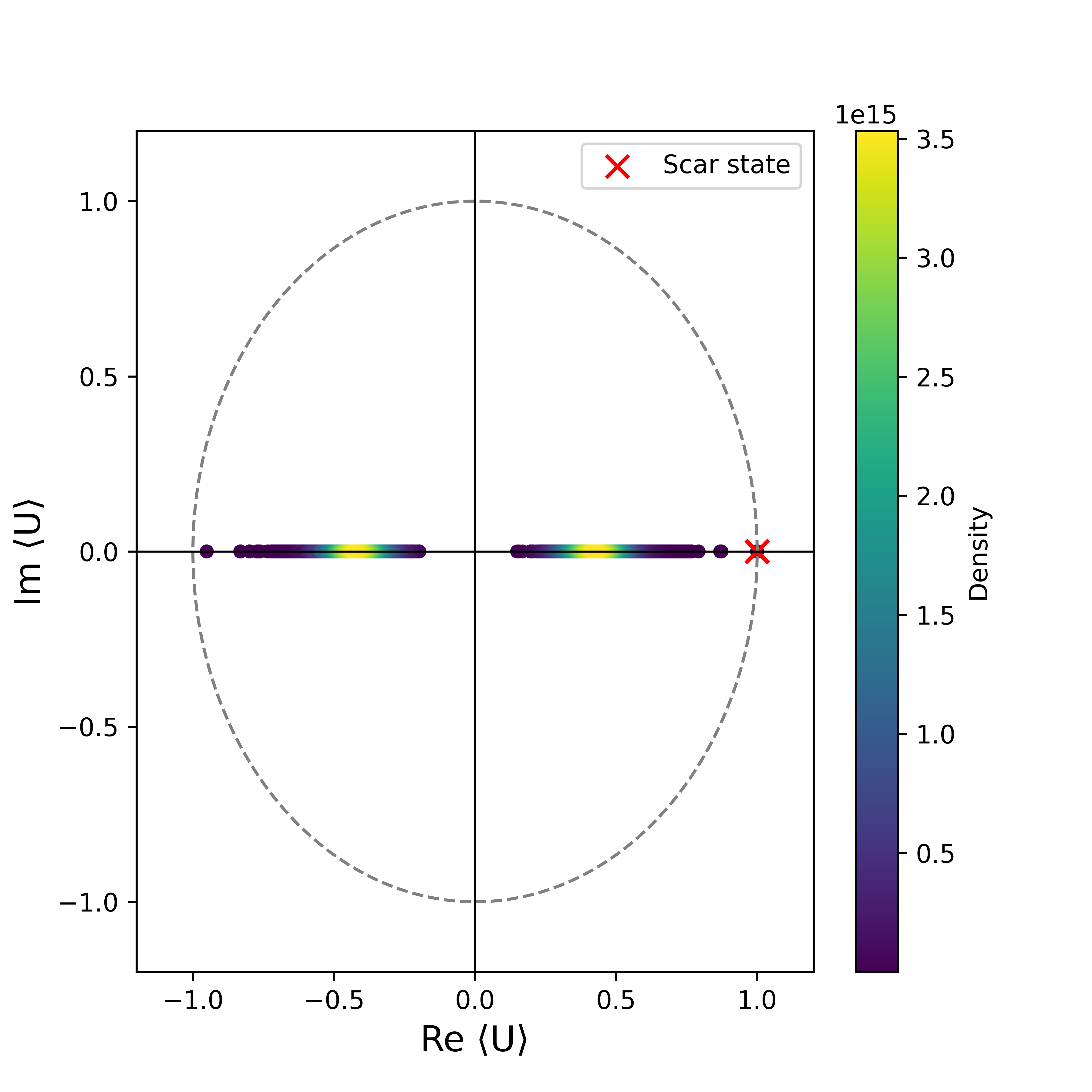}
        \caption{$\langle O_{3} \rangle = \langle (S^{z}_{1})^{2}\, \prod_{j=2}^{L-1}e^{i\pi S^{z}_{j}}(S^{z}_{L})^{2} \rangle $}
        \label{fig:c}
    \end{subfigure}
    \hfill
    \begin{subfigure}{0.45\textwidth}
        \centering
        \includegraphics[width=1\linewidth]{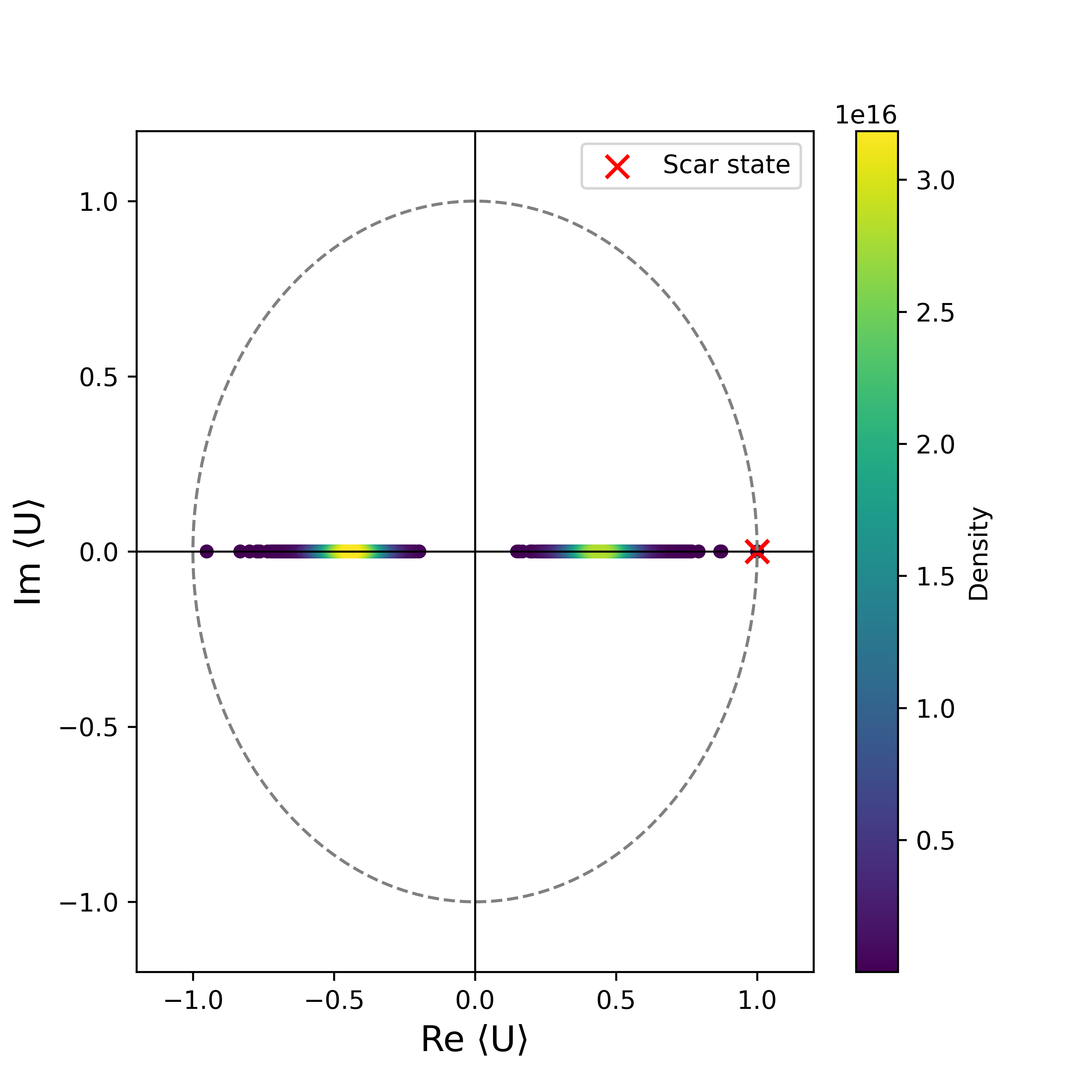}
        \caption{$\langle O_{4}\rangle  = \langle (S^{z}_{1})^{2}\, \prod_{j=2}^{L-1}e^{i\pi (S^{z}_{j})^{2}}(S^{z}_{L})^{2} \rangle$}
        \label{fig:d}
    \end{subfigure}

    \caption{The expectation value of the various string order parameters throughout the spectrum for system size $L=10$.}
    \label{fig:stringorders}
\end{figure*}

In the main text, we have presented the numerical data for the appropriate twist operator defined in Eq.~\eqref{eq:twist_op}. In this appendix, we aim to show a similar analysis for the string order parameter and hence try to motivate our twist operator. To this end, we consider the following 4 different types of string order parameter, 
\begin{align}
    O_{1} &= S^{z}_{1}\, \prod_{j=2}^{L-1}e^{i\pi S^{z}_{j}}S^{z}_{L}  \\
    O_{2} &= S^{z}_{1}\, \prod_{j=2}^{L-1}e^{i\pi (S^{z}_{j})^{2}}S^{z}_{L}   \\
    O_{3} &= (S^{z}_{1})^{2}\, \prod_{j=2}^{L-1}e^{i\pi S^{z}_{j}}(S^{z}_{L})^{2}   \\
    O_{4} &= (S^{z}_{1})^{2}\, \prod_{j=2}^{L-1}e^{i\pi (S^{z}_{j})^{2}}(S^{z}_{L})^{2}   
\end{align}
An empirical observation motivating this construction is that for states within the scar subspace, $\sum_{j}(S^{z}_{j})^{2}$ is conserved during the time evolution in contrast to generic ergodic states. This suggests one to look at the string order parameter in which the hidden order is characterized by a $\pm1$ with no zeros. In the original Kennedy-Tasaki (KT) transformation~\cite{Kennedy1992}, the hidden antiferromagnetic order is revealed by a non-local transformation that effectively removes the zeros in between of the $|1\rangle$'s or $|-1\rangle$'s. The idea here is similar, but to detect the order $|\pm 1\rangle$, we use $(S^{z})^{2}$ as it puts $|1\rangle$ and $|-1\rangle$ on the same footing, while excluding $|0\rangle$ from contributing to the order parameter.
Indeed, this reasoning seems to be true as shown in Fig.~\eqref{fig:stringorders}, all the scar states lie on $1$ when we consider the string orders given by $O_{3}, O_{4}$ whereas for the string order $O_{1}, O_{2}$, the scars are distributed and mixed among the ergodic states. We would like to highlight that $O_{3}, O_{4}$ are equivalent since,
\begin{align}
    e^{i\pi S^{z}_{j}} |m\rangle = -|m\rangle \quad \text{, m=$\pm1$}  \\
    e^{i\pi (S^{z}_{j})^{2}} |m\rangle = -|m\rangle \quad \text{, m=$\pm1$}
\end{align}
The observation that the string order parameters $O_3, O_4$ detect the order present in the scars motivates us to modify the original LSM twist operator used in the LSM argument by replacing $S^{z}$ with $(S^{z})^{2}$. Finally, note that the twist operators and string order parameters only make physical sense for ground states, and since scars are the ground states of the Liouvillian from the commutant~\cite{PhysRevX.14.041069} approach, we expect that these string order and the LSM twist operator makes physical sense for such states. For completeness, we also examine the string order parameter for ergodic states. Their expectation values do not show any clear signatures of order, consistent with their non-ground-state nature.

\section{Computing the reduced density matrices of Scars}\label{app:density_matrix_calculation}
We review here the notation to represent these scar states; this notation is motivated by the fact that the product states involved in the scar states are always of the form $|-1 ... \underbrace{1}_{j}... -1\rangle$ that $j$ tells about the location of $1$ that sits at the $j^{th}$ site.
\begin{align}
|j_{1}j_{2}\cdots j_{n}\rangle
   \;\equiv\; 
   |-1\cdots \underbrace{1}_{j_{1}}\cdots
   \underbrace{1}_{j_{n}}\cdots -1\rangle
\end{align}

It is important to realize that $|j_{1}, j_{2}\dots j_{n}\rangle$ is actually generated $n!$ times via the action of the bimagnon operator $Q^{\dagger}$ , since all the local excitations commute i.e. $(S^{+}_{j})^{2}$ commute. So the scar state when expressed using our notation is simply,
\begin{equation}
    |S_{n}\rangle = \dfrac{\mathcal{N}(n)}{n!} (Q^{\dagger})^{n}|\Omega\rangle = \mathcal{N}(n)\sum_{\{j\}} |j_{1}\, j_{2} \, \dots \, j_{n}\rangle
\end{equation}
Now, a very important property about these states is that the inner product in this notation is,
\begin{equation}\label{eq:innerproduct rule}
    \langle j'_{1} j'_{2} ... j'_{m}| j_{1} j_{2} ... j_{n}\rangle =  \delta_{mn} \delta_{\{j\}, \{j'\}}
\end{equation}
Here $\delta_{mn}$ arises because of the Global $U(1)$ symmetry i.e. conservation of magnetization, and the delta function is on the \textit{unordered set} $\{j\}$. Using this property and expressing scar states in the notation above,
\begin{align*}
    |S_{n}\rangle =& \mathcal{N}(n) \sum_{\{j\}}(-1)^{\sum_{k=1}^{n} j_{k}}|j_{1} j_{2} ...j_{n}\rangle  \\
    \implies \langle S_{n}| S_{n} \rangle=&(\mathcal{N}(n))^{2} \sum_{\{j\}, \{j'\}} (-1)^{\sum_{k=1}^{n} j'_{k}}(-1)^{\sum_{k=1}^{n} j_{k}} \langle j'_{1} j'_{2}... j'_{n}|j_{1} j_{2}... j_{n}\rangle \\
    \implies 1=&(\mathcal{N}(n))^{2}  \sum_{\{j\}} 1 \\
    \implies 1=& (\mathcal{N}(n))^{2}  \dfrac{L!}{(L-n)!n!} \\
    \implies \mathcal{N}(n)=&\sqrt{\dfrac{n!(L-n)!}{L!}}
\end{align*}
here, $\sum_{\{j\}}$ is simply a sum over all possible values of $j_{1}, j_{2}... j_{n}$ we can get i.e. all possible places for $n$ $1's$ in L sites which is just the combinatorial factor of L choose n.

We first do a local unitary transformation to remove  the extra phases of $(-1)^{i}$ can be removed by a suitable local unitary transformation given by $U=\prod_{j\in odd} e^{i\frac{\pi}{2} \sigma^{z}_{j}}$. We see that because $U$ is a spin 1 representation of $SO(3)$ which rotates $S^{x} \rightarrow S^{y}$ and $S^{y} \rightarrow -S^{x}$ this transformation implies that, $S^{+} \rightarrow (-i)S^{+}$ and $S^{-} \rightarrow i S^{-}$ therefore after squaring, the phase factor of $-1$ is removed from the odd sites, so now we are going to talk about scar in these rotated bases.
We start by writing the scar states in the notation introduced earlier.
\begin{align*}
    |S_{n}\rangle=\dfrac{\mathcal{N}(n)}{n!}\sum_{\{j\}}|j_{1} j_{2} ...j_{n}\rangle
\end{align*}

Because of the permutation symmetry of the scar states, the k-body reduced density matrix is the same for all partitions of the k-bodies, meaning tracing out any $(L-k)$ would give me the same result.\newline

\subsection{1-body reduced density matrix}
We first find the 1-body reduced density matrix, $\rho^{(1)}(n)$, where $n$ is the index of scar state. Here, we are tracing out the whole system except the first spin in the lattice.
\begin{align*}
    \rho(n)=(\mathcal{N}(n))^{2}\sum_{\{j\},\{j'\}} | j'_{1} j'_{2} ...j'_{n} \rangle \langle j_{1} j_{2} ...j_{n}| 
\end{align*}
    
\[
\begin{aligned}
\rho^{(1)}(n)
= \mathcal{N}(n)^{2}\Big[
& |-1\rangle\langle-1|\;
   \left(\sum_{\{j\},\{j'\}}
   \langle j'_1 j'_2\!\cdots\! j'_n \mid j_1 j_2\!\cdots\! j_n\rangle\right) + |1\rangle\langle-1|\;
   \left(\sum_{\{j\},\{j'\}}
   \langle j'_2\!\cdots\! j'_n \mid j_1 j_2\!\cdots\! j_n\rangle\right) \\
& + |-1\rangle\langle 1|\;
   \left(\sum_{\{j\},\{j'\}}
   \langle j'_1 j'_2\!\cdots\! j'_n \mid j_2\!\cdots\! j_n\rangle\right) + |1\rangle\langle 1|\;
   \left(\sum_{\{j\},\{j'\}}
   \langle j'_2\!\cdots\! j'_n \mid j_2\!\cdots\! j_n\rangle\right)
\Big].
\end{aligned}
\]

Now in the above expression, note that the terms $|1\rangle \langle -1|$ ,$|-1 \rangle \langle 1|$ have coefficients that correspond to the inner product of states that have an unequal number of $1$'s and therefore reside in different symmetry sectors and  result in a zero inner product just from~\eqref{eq:innerproduct rule} and now using the inner product rule ,
\begin{align}\label{eq:inner product calc for reduced matrices}
    \sum_{\{j\},\{j'\}} \langle j'_{1} j'_{2} ...j'_{n} | j_{1} j_{2} ...j_{m}\rangle=\delta_{m,n} \sum_{\{j\},\{j'\}} \delta_{\{j\},\{j'\}} = \delta_{m,n} \sum_{\{j\}} 1= \delta_{m,n} \Comb{L}{n} 
\end{align}

\begin{figure}[h!]\hspace{-2cm}
    \centering
    \includegraphics[width=0.6\linewidth]{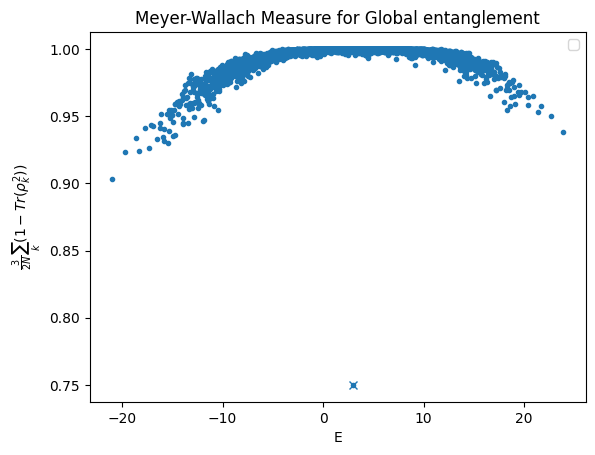}
    \caption{Meyer-Wallach measure for Spin 1XY model in the m=0 sector for L=10 system size}
    \label{fig:MW_measure}
\end{figure}

This is a very important relation, which we are going to use to get the elements of the reduced density matrices and in the end, all of the $\alpha$-Renyi entropy can be calculated analytically.\newline
Using~\eqref{eq:inner product calc for reduced matrices} on the expression for 1-body reduced density matrix, we see that all of those inner products can be rewritten as,
\begin{align*}
    \rho^{(1)}(n)=\dfrac{1}{\Comb{L}{n}}\left[ |-1\rangle \langle -1| \Comb{L-1}{n}   +  |1\rangle \langle 1|  \Comb{L-1}{n-1}   \right]
\end{align*}
Here, the normalisation as we mentioned as be seen from~\eqref{eq:inner product calc for reduced matrices}. Now, firstly, the terms like $|-1\rangle \langle 1|$ gave us zero cause in the inner product, we have the inner product of states of length $L-1$ and having a different number of $1$'s so it results in zero.  For the coefficient of $|-1\rangle \langle -1|$ we see that it is an inner product of states having length $L-1$ with $n$ $1$'s and hence from~\eqref{eq:inner product calc for reduced matrices} we get that result. Now for the term with $|1\rangle\langle1|$ , in this case we simply have a length of $L-1$ and with $n-1$ $1$'s which results in a combinatorial factor of $\Comb{L-1}{n-1}$
\begin{equation}\label{eq: 1 body reduced density matrix}
    \rho^{(1)}(n)=\dfrac{\Comb{L-1}{n}}{\Comb{L}{n}}|-1\rangle \langle -1|+ \dfrac{\Comb{L-1}{n-1}}{\Comb{L}{n}}|1\rangle \langle 1| = \dfrac{L-n}{L}|-1\rangle \langle -1|+ \dfrac{n}{L}|1\rangle \langle 1|
\end{equation}
To see if our result matches, we test with a few known cases, for eg. EPR pair, $|\phi^{+}\rangle=(|01\rangle +|10\rangle)/\sqrt{2}$ , which corresponds to $L=2$ and $n=1$. Putting these values, we see that indeed the 1-body reduced density matrix matches the reduced density matrix of the EPR pair. Next, we test on the state with $L=3$ and $n=1$ which is our familiar W state given as, 
\begin{align}\label{eq:W state}
    |W\rangle=\dfrac{1}{\sqrt{3}}\left(|001\rangle + |010\rangle+ |100\rangle\right)
\end{align}
Here $|0\rangle$ corresponds to $|-1\rangle$ in our scars and $|1\rangle$ corresponds to $|1\rangle$ in our scars. Now we know the 1-body reduced density matrix for the W state is given as, 
\begin{align*}
    \rho^{(1)}_{W}=\dfrac{2}{3}|0\rangle \langle 0| + \dfrac{1}{3} |1\rangle \langle 1|
\end{align*}
which matches exactly, we also calculate the Meyer-Wallach measure of entanglement, which is essentially the average $2$nd Rényi entropy per site. We can find this numerically as seen in fig(\ref{fig:MW_measure}) and we see that for the scar state having $N/2$ $1$'s this matches the analytic prediction, providing more support for our calculation.

\subsection{2-body reduced density matrix}
Now we get the 2-body reduced density matrix. We follow the same idea that the scar states only contain $-1$'s and $1$'s, and the scar states are permutation symmetric with respect to the sites. We start with the full density matrix for the scar state and then trace out all the sites except  the $1^{st}$ and $2^{nd}$ site,
\begin{align*}
    \rho(n)=(\mathcal{N}(n))^{2}\sum_{\{j\},\{j'\}} | j'_{1} j'_{2} ...j'_{n} \rangle \langle j_{1} j_{2} ...j_{n}|
\end{align*}
\[
\begin{aligned}
\rho^{(2)}(n)
&= \mathcal{N}(n)^{2}\Big[
   |-1,-1\rangle\langle-1,-1|\;\sum_{\{j\},\{j'\}}
     \langle j'_{1}\!\cdots\! j'_{n}\mid j_{1}\!\cdots\! j_{n}\rangle \\
&\quad + \,(|-1,1\rangle+|1,-1\rangle)
     (\langle-1,1|+\langle1,-1|)\;\sum_{\{j\},\{j'\}}
     \langle j'_{2}\!\cdots\! j'_{n}\mid j_{2}\!\cdots\! j_{n}\rangle \\
&\quad + \;|1,1\rangle\langle1,1|\;\sum_{\{j\},\{j'\}}
     \langle j'_{3}\!\cdots\! j'_{n}\mid j_{3}\!\cdots\! j_{n}\rangle
\Big].
\end{aligned}
\]

Now from here we again use~\eqref{eq:inner product calc for reduced matrices} to get the inner product, , and as mentioned prior, we have omitted the terms like $|-1,-1\rangle \langle -1, 1|$ simply because in the inner product while tracing, we would have the inner product of states having an unequal number of $1$'s, which would give zero.
\[
\begin{aligned}
\rho^{(2)}(n)
&= \mathcal{N}(n)^{2}\Big[
   |-1,-1\rangle\langle-1,-1|\,\Comb{L-2}{n}  \\
&\qquad +  (|-1,1\rangle+|1,-1\rangle)
   (\langle-1,1|+\langle1,-1|)\,
   \Comb{L-2}{n-1} \\
&\qquad + |1,1\rangle\langle1,1|\,
   \Comb{L-2}{n-2}
\Big]
\end{aligned}
\]

We would like to point out 2 interesting structures that arise from this, the first is that the density matrix is simply a direct sum of blocks, each having a different number of $1$'s this is a result of the $U(1)$ symmetry of the original Hamiltonian, and also, obviously, these blocks are solely characterized by states with no $0$ in the state, which is a result of bimagnon excitations. Moreover, because of the permutation symmetry of the scar states, these blocks have a special form, which is that all the matrix elements of these blocks are the same number.
\begin{align}
    F(a,2)\equiv\begin{pmatrix}
        a & a \\
        a & a
    \end{pmatrix}
\end{align}
Above is the block corresponding to one $1$ in the state, where $a=\frac{\Comb{L-2}{n-1}}{\Comb{L}{n}}$ for a 2-body reduced density matrix. We define this matrix as $F(a,m)$ where m is the dimension of the matrix and $a$ is the entry in the matrix. 
So, keeping in mind these points, we can find the density matrix for k sites, but before that, we write down the explicit form of the 2-body reduced density matrix. 
\begin{multline}\label{eq: 2body reduced density matrix}
    \rho^{(2)}(n)= \left[\dfrac{(L-n)(L-n-1)}{L(L-1)}|-1,-1\rangle \langle -1,-1|+ \dfrac{n(L-n)}{L(L-1)}(|1,-1\rangle+|-1,1\rangle)( \langle 1,-1|+\langle -1,1|
    + \dfrac{n(n-1)}{L(L-1)}|1,1\rangle \langle 1,1|\right]
\end{multline}
Now we verify it with W state~\eqref{eq:W state}, we know that the 2-body reduced density matrix for W state is,
\begin{equation*}
    \rho^{(2)}_{W}=\dfrac{1}{3}\left[|00\rangle \langle 00|+|01\rangle \langle 01|+|01\rangle \langle 10|+|10\rangle \langle 01|+|10\rangle \langle 10|\right]
\end{equation*}
by putting $L=3, n=1$ in~\eqref{eq: 2body reduced density matrix} we get the same result up to the identification of $|0\rangle$ with $|-1\rangle$ in scars and $|1\rangle$ with $|1\rangle$ in scars. We are going to represent the density matrix in the basis $|1,1\rangle, |-1,1\rangle, |1,-1\rangle, |-1,-1\rangle$ cause these scar states don't have any $|0\rangle$ in spin 1 representation, so the matrix looks like,

\begin{equation*}
    \rho^{(2)}=
    \begin{bmatrix}
     F\left(\frac{\binom{L-2}{n}}{\binom{L}{n}},1\right) & 0 & 0 \\
     0 & F\left(\frac{\binom{L-2}{n-1}}{\binom{L}{n}},2\right) & 0 \\
     0 & 0 & F\left(\frac{\binom{L-2}{n-2}}{\binom{L}{n}},1\right)
     \end{bmatrix}
\end{equation*}
\vspace{0.5cm}

This way of writing the reduced density matrix will help us to write the k-body reduced density matrix.

\subsection{k-body reduced density matrix}
\begin{figure}[t]
    \centering
    \begin{subfigure}[b]{0.45\textwidth}
        \centering
        \includegraphics[width=\textwidth]{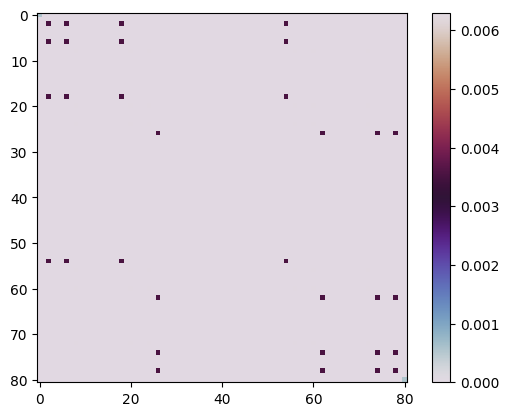}
        \caption{The 4- body reduced density matrix for the scarred state}
        \label{fig:scar rdm}
    \end{subfigure}
    \hfill
    \begin{subfigure}[b]{0.45\textwidth}
        \centering
        \includegraphics[width=\textwidth]{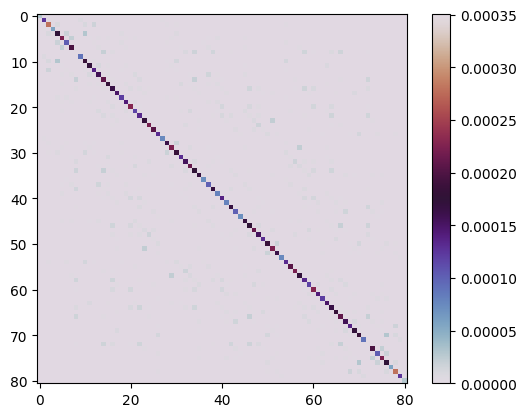}
        \caption{The 4- body reduced density matrix for the thermal state}
        \label{fig:thermal rdm}
    \end{subfigure}
    \caption{4-body reduced density matrix for Spin 1 XY model with L=10 sites}
    \label{fig:reduced_density_matrix_implot}
\end{figure}

Now using the arguments used and illustrated in the previous 2 sections, here we directly write the density matrix,
\begin{equation}
    \rho^{(k)}=
    \begin{bmatrix}
     F\left(\frac{\binom{L-k}{n}}{\binom{L}{n}},\binom{k}{0}\right) & & & &\\
        & F\left(\frac{\binom{L-k}{n-1}}{\binom{L}{n}},\binom{k}{1}\right) & & &\\
         & & F\left(\frac{\binom{L-k}{n-2}}{\binom{L}{n}},\binom{k}{2}\right) & & \\
         & & & \ddots & \\
         & & & & F\left(\frac{\binom{L-k}{n-k}}{\binom{L}{n}},\binom{k}{k}\right)
    \end{bmatrix}
\end{equation}
where each $F$ is a matrix which we defined earlier, the shape of the $F$ matrix is decided by the number of ways to distribute the $1$'s in k sites which, is given by $\Comb{k}{r}$ if there are $r$ $1$'s to be distributed. First we verify that this density matrix has a unit trace for a sanity check,
\begin{align*}
   \operatorname{Tr}(\rho^{(k)})&=\dfrac{1}{\binom{L}{n}}\sum_{r} \binom{L-k}{n-r}\binom{k}{r}\\
     \implies \operatorname{Tr}(\rho^{(k)})&=\dfrac{1}{\binom{L}{n}} \binom{L}{n}=1
\end{align*}
This identity is known as Vandermonde's identity. For another sanity check, we calculate the 4-body reduced density matrix for $L=10$ in $m=0$ magnetization sector, and plot the colour map of the reduced density matrix shown in fig.\eqref{fig:reduced_density_matrix_implot}, we indeed see that the expression is true. Note that the block structure is not apparent because of the way the computational basis is stored in the computer, i.e., by their equivalent integer representation of states, which is if the state is $|-1,0,1\rangle \rightarrow |0,1,2\rangle \rightarrow 5$ and the basis is arranged in increasing order of their representation values.

Although obtaining the full spectrum of the k-site reduced density matrix may be algebraically cumbersome, its block-diagonal structure makes it straightforward to compute products of $\rho^{(k)}$. This motivates working with Rényi entropies rather than the von Neumann entropy. The Rényi entropies only require evaluating traces of powers of the density matrix, Tr($\rho^{(k)}$), which can be computed efficiently due to the simple block structure. Consequently, the Rényi entropies provide an analytically tractable measure of entanglement in this setting. Firstly, note that the $F$ matrix has this nice property that, 
\begin{align*}
     &(F(a,m))^{2}=F(ma^{2},m) \\
     \implies &\operatorname{Tr}(F(a,m)^{2})=m^{2}a^{2}
\end{align*}
This enables us to write the square of the k-body reduced density matrix.
\begin{align*}
    (\rho^{(k)}(n))^{2}&=
    \begin{bmatrix}
         \left(F\left(\frac{\binom{L-k}{n}}{\binom{L}{n}},\binom{k}{0}\right)\right)^{2} & & & &\\
         & \left(F\left(\frac{\binom{L-k}{n-1}}{\binom{L}{n}},\binom{k}{1}\right)\right)^{2} & & &\\
         &  & \left(F\left(\frac{\binom{L-k}{n-2}}{\binom{L}{n}},\binom{k}{2}\right)\right)^{2} & & \\
         & & & \ddots & \\
         & & & & \left(F\left(\frac{\binom{L-k}{n-k}}{\binom{L}{n}},\binom{k}{k}\right)\right)^{2}
    \end{bmatrix}  \\
     \implies & \operatorname{Tr}((\rho^{(k)}(n))^{2})= \text{Sum of the traces of all } F \text{ matrices} \\
     \implies & \operatorname{Tr}((\rho^{(k)}(n))^{2})=\dfrac{1}{\left(\binom{L}{n}\right)^{2}}\sum_{r} \binom{L-k}{n-r}^{2} \binom{k}{r}^{2}
\end{align*}

From here in principal one can get an exact expression for the Renyi entropy, but we couldn't find any exact expression for such sum, so we are going to give an approximate value of Renyi entropy for the scar states which are in the middle of the spectrum, i.e. $n=L/2$, and we also focus on $k=L/2$ which is the half-chain Renyi entropy,i.e. $S_{2}$
\begin{align*}
    e^{-S_{2}}&=Tr((\rho^{(L/2)}(L/2))^{2})=\dfrac{1}{(\Comb{L}{L/2})^{2}}\sum_{r} (\Comb{L/2}{L/2-r})^{2} (\Comb{L/2}{r})^{2} \\
    e^{-S_{2}}&=\dfrac{1}{(\Comb{L}{L/2})^{2}}\sum_{r} (\Comb{L/2}{r})^{4}
\end{align*}

Define $r=aL/2$ where $a$ is an $O(1)$ number, and then now we will write the sum in terms of $a$ for large L formally $\lim_{L \rightarrow \infty} $,
\begin{align}\label{eq:2nd renyi entropy int form}
    e^{-S_{2}}=\dfrac{1}{(\Comb{L}{L/2})^{2}} \dfrac{L}{2}\int_{0}^{1} (\Comb{L/2}{aL/2})^{4} da
\end{align}
From here, we use Stirling's approximation, which is $n! \sim  \sqrt{2\pi n} n^{n} e^{-n}$, keeping in mind this holds for large $n$, we have to restrict the values of $a$ in the sense that it can't be very small or to be precise, it can't be $O(\frac{1}{L})$ , cause in that case one cannot apply Stirling's approximation to the integrand as $r$ is no longer a large quantity in the limit of large $L$. With this in mind we proceed,
\begin{align*}
    \Comb{L}{L/2}&=\dfrac{L!}{((L/2)!)^{2}} \sim \dfrac{\sqrt{2\pi L} \, L^{L} e^{-L}}{(\sqrt{\pi L} (L/2)^{L/2} e^{-L/2})^{2}} = \dfrac{\sqrt{2\pi L} \, L^{L} e^{-L}}{\pi L \, (L/2)^{L} e^{-L}} \\
    \implies \Comb{L}{L/2}&= \sqrt{\dfrac{2}{\pi \,L}} \, 2^{L} 
\end{align*}
\begin{align*}
    \Comb{L/2}{aL/2}&=\dfrac{(L/2)!}{(L/2a)!(L/2(1-a))!} \sim 
    \dfrac{\sqrt{\pi \, L}\, (L/2)^{L/2} e^{-L/2}}{\sqrt{\pi \, a\, L}\, (aL/2)^{aL/2} e^{-aL/2}\sqrt{\pi \, L \, (1-a) }\, (L(1-a)/2)^{(1-a)L/2} e^{-(1-a)L/2}}  \\
    \implies & \sqrt{\dfrac{1}{\pi \, L \, a(1-a)}} \dfrac{(L/2)^{L/2}}{(L/2)^{aL/2}(L/2)^{L/2(1-a)}} \dfrac{1}{a^{aL/2}(1-a)^{(1-a)L/2}}  \\
    \implies & \sqrt{\dfrac{1}{\pi \, L \, a(1-a)}} e^{{L/2}\left({-a\, \ln(a) - (1-a) \, \ln(1-a)}\right)}
\end{align*}
Using the above 2 formula's we got from Stirling's approximation in~\eqref{eq:2nd renyi entropy int form}
\begin{align*}
    e^{-S_{2}}&\approx\dfrac{\pi \, L}{2} \, 4^{-L} \dfrac{L}{2} \int_{0}^{1} \left( \dfrac{1}{\pi \, L \, a(1-a)}\right)^{2} e^{{2L}\left({-a\, \ln(a) - (1-a) \, \ln(1-a)}\right)} da\\
    \implies e^{-S_{2}}&\approx \dfrac{4^{-L}}{4 \, \pi } \int_{0}^{1}\dfrac{e^{{2L}\left({-a\, \ln(a) - (1-a) \, \ln(1-a)}\right)}}{(a(1-a))^{2}} da
\end{align*}
The integral of the above can be solved by using Laplace's method to approximate such integrals. Laplace's method is essentially the stationary phase approximation applied to the Euclidean integrand. Laplace's method tells us that we need to find the maximum of the exponential function, 
\begin{align}\label{eq: Laplace method}
    \int h(x) e^{M f(x)} dx\approx \sqrt{\dfrac{2\pi}{M |f''(x_{0})|}} h(x_{0})e^{Mf(x_{0})}
\end{align}
We find the maximum of the function $f(a)=-a\ln(a)-(1-a)\ln (1-a)$,
\begin{align*}
    f'(a)&=-\ln(a)-1+\ln(1-a)+1=\ln(1-a)-\ln(a)=0
    \implies a=\dfrac{1}{2}  \\
    f''(a)&=-\dfrac{1}{1-a}-\dfrac{1}{a} \implies \left|f''(\dfrac{1}{2})\right|=4
\end{align*}
From here we use the laplace's method into the 2nd renyi entropy calculation,
\begin{align}\label{eq: renyi entropy for mid scar}
    \nonumber e^{-S_{2}}&\approx \dfrac{4^{-L}}{4 \, \pi } \sqrt{\dfrac{2 \pi }{8L}} 4^{2} e^{{2L}\ln(2)} \\
    \nonumber \implies e^{-S_{2}}&\approx \dfrac{4^{-L}}{ \sqrt{\pi \, L} } 2 e^{{2L}\ln(2)}=\dfrac{2}{\sqrt{\pi \, L}} \\
    \implies S_{2}&\approx\dfrac{1}{2}\ln(\pi \, L) -\ln(2)
\end{align}
We see that the Rényi entropy has a sub-volume entanglement entropy as opposed to the Volume law entanglement entropy expected for thermal states and matches exactly with the calculation done for the AKLT model~\cite{PhysRevB.98.235156} Although they calculated the Von Neumann entanglement entropy, we believe the scaling should hold true apart from constants. Now, in this way, one can generalize and calculate any $\alpha$ Rényi entropy in principle, but it's not hard to see that the arguments we have used will be exactly the same with the same approximations; the only difference that appears will be in the sum, which will instead become,
\begin{equation}
    e^{(1-\alpha)S_{\alpha}}=Tr((\rho^{(L/2)}(L/2))^{\alpha})=\dfrac{1}{(\Comb{L}{L/2})^{\alpha}}\sum_{r} (\Comb{L/2}{r})^{\alpha}
\end{equation}
where we are calculating $\alpha$ Rényi entropy for half a chain with a scar in the middle of the spectrum. From here, knowing the reduced density matrix, we can calculate almost anything we want, at least in principle.

\subsection{Extension to the bonded bimagnon case}
The bonded bimagon scars as introduced in the Spin 1 XY model by Schecter et al.~\cite{PhysRevLett.123.147201} for the Periodic Boundary conditions(PBC), are defined as,
\begin{equation}
  |\tilde{S}_n\rangle \;=\; \mathcal N_n\, \sum_{i_{1} \neq i_{2} \neq ...\neq i_{n}} S^{+}_{i_{1}}S^{+}_{i_{1}+1}S^{+}_{i_{2}}S^{+}_{i_{2}+1}...S^{+}_{i_{n}}S^{+}_{i_{n}+1} | \Omega\rangle  
  \label{eq:bimagnonscar-tower}
\end{equation}
where $|\Omega\rangle=|-1,-1,\dots,-1 \rangle$ be the fully polarized vacuum. It was shown that the bonded bimagnons can be understood as emergent SU(2) algebra~\cite{PhysRevB.101.174308} of objects that live on the bonds in particular the operator $Q^{\dagger} = \sum_{i} (-1)^{i}S^{+}_{i}S^{+}_{i+1}$, and hence the states can now again be written in the notation introduced above, where again the extra phases are removed by a local unitary acting only on a bipartition of the lattice, i.e., only on the odd sites. The local unitary transformation that does this job is $U=\prod_{j\in odd} e^{i\pi \sigma^{z}_{j}}$, from here we use the notation above, 
\begin{align}
|-1\,\cdots\,0\,\underbrace{\hspace{0.1pt}}_{j_{1}}\,0\,\cdots\,0\,
\underbrace{\hspace{0.1pt}}_{j_{k}}\,1\,
\underbrace{\hspace{0.1pt}}_{j_{k+1}}\,0\cdots -1\rangle
\equiv | j_{1}\cdots j_{k} j_{k+1}\cdots j_{n}\rangle
\end{align}

where the index $j_{k}$ indicates that the bonded bigmanon is present at the $k^{th}$ bond or between sites $k$ and $k+1$. In this notation, up to the local unitary transformation, the scar states are simply permutation symmetric states, i.e.
\begin{align}
    |\tilde{S}_{n}\rangle = \mathcal{N}(n)\sum_{\{j\}}|j_{1}\cdots j_{n}\rangle
\end{align}
 From here, the exact analysis goes through for the calculation of the reduced density matrices and the Rényi entanglement entropies. Also, this implies that the half-chain Bipartite entanglement entropy of these bonded bimagnon states scales as $S \sim ln(L)$
 
\section{Analytical calculation of QFI}\label{app:QFI for operators}
We will be only focusing on operators, which are extensive local operators with momentum $\pi$, i.e., $\sum_{i} e^{i\pi} \hat{o}_{i}$, those with a site support of either 1 or 2 sites, but this analysis works for any generic operator with a larger site support or also for intensive local operators.\newline
For operators with support on 1 site, we show the super extensive scaling by 2 ways, first using the properties of the SGA and second using the explicit permutation symmetry of scars and the explicit expression of reduced density matrices, which we have derived in the previous section. The advantage of using the reduced density matrix method is that can find the QFI scaling for any operator since the expressions for the reduced density matrix are available. Before that we just mention the spin 1 matrices that we are going to use, 
\begin{align}\label{eq:spin 1 matrices}
    \underbrace{\dfrac{1}{\sqrt{2}} \begin{pmatrix}
        0 & 1 & 0 \\
        1 & 0 & 1 \\
        0 & 1 & 0
    \end{pmatrix}}_{\Large{S^{x}}}, \quad
    \underbrace{ \dfrac{1}{\sqrt{2}}\begin{pmatrix}
        0 & -i & 0 \\
        i & 0 & -i \\
        0 & i & 0
    \end{pmatrix}}_{\Large{S^{y}}}, \quad
        \underbrace{\begin{pmatrix}
        1 & 0 & 0 \\
        0 & 0 & 0 \\
        0 & 0 & -1
    \end{pmatrix}}_{\Large{S^{z}}}, \quad
    \underbrace{\sqrt{2} \begin{pmatrix}
        0 & 1 & 0 \\
        0 & 0 & 1 \\
        0 & 0 & 0
    \end{pmatrix}}_{\Large{S^{+}}}, \quad
    \underbrace{\sqrt{2}   \begin{pmatrix}
        0 & 0 & 0 \\
        1 & 0 & 0 \\
        0 & 1 & 0
    \end{pmatrix}}_{\Large{S^{-}}}
\end{align}

\subsection{QFI scaling from SGA}
Recall that we will be working in the rotated basis were we have eliminated the phase factors by a local unitary transformation. So $Q^{\dagger}=\sum_{i} (S^{+}_{i})^{2}$, also
recall that the SGA is defined as,
\begin{align*}
    ([H, Q^{\dagger}]-2hQ^{\dagger})W=0
\end{align*}
Here W is the scar subspace. Firstly, within the scar subspace we have a $SU(2)$ algebra by defining $J^{+}=Q^{\dagger}/2$, $J^{-}=Q/2$ and $J^{z}=H/2h$, we see that the following relations hold,
\begin{align}\label{eq:SU2 algebra}
    \nonumber [J^{z}, J^{\pm}]=\pm J^{\pm}  \\
    [J^{+}, J^{-}]=2 J^{z}
\end{align}
 the first expression is simply the SGA in the scar subspace, whereas the second expression comes from the fact the scar states are annihilated by the hopping terms $T_{i,i+1} = S^{+}_{i}S^{-}_{i+1} + h.c.$ and by acting $T_{i,i+1}$ on $|\Omega\rangle$ we see that it is zero. The only terms that act within the scar subspace is effectively $H=h\sum_{i} (-1)^{i}S^{z}_{i}$ (the extra phases of $(-1)^{i}$ arise from the local unitary transformation we did), cause the anisotropy term is simply a constant within the scar subspace, and it is from here we can complete the $SU(2)$ algebra in~\eqref{eq:SU2 algebra}. We would like to highlight here that for the case for bonded bimagnons we also have an emergent $SU(2)$ Spectrum Generating Algebra(SGA) as illustrated first in~\cite{PhysRevB.101.174308}.\newline 
 Here we just write some known results of the $SU(2)$ algebra,
 \begin{align}
     J^{\pm}|m\rangle&=\sqrt{j(j+1)-m(m\pm1)} \, |m\pm1\rangle \\
     J^{z}|m\rangle&=m|m\rangle
 \end{align}
Here $|m\rangle$ is just the scar state that has m $1$'s i.e. $|S_{m}\rangle$.
Now, we would like to define a basis of $3\times3$ hermitian matrices that would serve as an operator basis for one site operators.
\begin{align} \label{app, eq: basis for fundamental operators}
    \underbrace{\begin{pmatrix}
        0 & 0 & 1 \\
        0 & 0 & 0 \\
        1 & 0 & 0
    \end{pmatrix}}_{\Large{\Bar{\sigma}^{x}}},
        \underbrace{\begin{pmatrix}
        0 & 0 & -i \\
        0 & 0 & 0 \\
        i & 0 & 0
    \end{pmatrix}}_{\Bar{\sigma}^{y}},
        \underbrace{\begin{pmatrix}
        1 & 0 & 0 \\
        0 & 0 & 0 \\
        0 & 0 & -1
    \end{pmatrix}}_{\Bar{\sigma}^{z}},
        \underbrace{\begin{pmatrix}
        1 & 0 & 0 \\
        0 & 0 & 0 \\
        0 & 0 & 1
    \end{pmatrix}}_{\Bar{\mathds{1}}}
\end{align}
\begin{align*}
\underbrace{\begin{pmatrix}
        0 & 1 & 0 \\
        1 & 0 & 0 \\
        0 & 0 & 0
    \end{pmatrix}}_{\gamma^{x}}
    ,
    \underbrace{\begin{pmatrix}
        0 & -i & 0 \\
        i & 0 & 0 \\
        0 & 0 & 0
    \end{pmatrix}}_{\bar{\gamma}^{x}}
    ,
    \underbrace{\begin{pmatrix}
        0 & 0 & 0 \\
        0 & 0 & 1 \\
        0 & 1 & 0
    \end{pmatrix}}_{\gamma^{y}}
    ,
    \underbrace{\begin{pmatrix}
        0 & 0 & 0 \\
        0 & 0 & -i \\
        0 & i & 0
    \end{pmatrix}}_{\bar{\gamma}^{y}}
        ,
        \underbrace{\begin{pmatrix}
        0 & 0 & 0 \\
        0 & 1 & 0 \\
        0 & 0 & 0
    \end{pmatrix}}_{\gamma^{z}}
\end{align*}
One can check that the set of matrices provided are linearly independent and also act as basis of the operator basis for a $S=1$ spin. The nice thing about this choice of basis is that the defined matrices $\Bar{\sigma}^{x},\Bar{\sigma}^{y},\Bar{\sigma}^{z},\Bar{\mathds{1}}$ form a basis for the operators that doesn't introduce a $0$ state locally.
\newline
So now, we consider an arbitrary on site operator of $S=1$ spin on chain, and we can express that arbitrary operator in this basis and then compute the averages with respect to the scar states, but note that for the 5 matrices the averages on scar states would go to zero since they introduce a $0$ in the state, So consider, the following decomposition,
\begin{align*}
    \hat{o}=n_{x} \bar{\sigma}^{x}+n_{y} \bar{\sigma}^{y}+n_{z} \bar{\sigma}^{z}+n_{0} \bar{\mathds{1}}+p_{x} \gamma^{x}+p_{\bar{x}} \bar{\gamma}^{x}+p_{y} \gamma^{y}+p_{\bar{y}} \bar{\gamma}^{y}+p_{z} \gamma^{z}
\end{align*}
Here, the coefficients can be, in principle, obtained from the inner product between the operators, but that is not necessary for the formalism, $\hat{o}$ is some operator on a single site, and we have just abused the notation by not writing the site index just for convenience. Note that the scar states have only $-1$, $1$ on each site, so this automatically gives me that the averages over the bottom 5, i.e., $\gamma$ matrices, is 0. From the local operators we create a translationally invariant operator $\hat{O} = \sum_{i} \hat{o}_{i}$, and the decomposition of the average of this operator in terms of the collective translationally invariant operators is,
\begin{align}
    \nonumber \langle O \rangle&=\langle\sum_{i} \hat{o}_{i}\rangle=n_{x}\langle\sum_{i}\bar{\sigma}^{x}_{i}\rangle +n_{y}\langle\sum_{i}\bar{\sigma}^{y}_{i}\rangle+n_{z}\langle\sum_{i}\bar{\sigma}^{z}_{i}\rangle+n_{0}\langle\sum_{i}\bar{\mathds{1}}_{i}\rangle   \\
    \implies \langle O \rangle&=2n_{x} \langle J^{x} \rangle+ 2n_{y} \langle J^{y}\rangle +2n_{z} \langle J^{z}\rangle +n_{0}
\end{align}
Note that $(S^{+}_{k})^{2}=\bar{\sigma}^{x}_{k}+i\bar{\sigma}^{y}_{k}$ and also $J^{+}=1/2\sum_{i} (S^{+}_{i})^{2}=1/2\sum_{i} \bar{\sigma}^{+}_{i}$ this is not a coincidence but rather of because we define the basis matrices in such a way that the dynamics between the states look like dynamics between states having 2 states per site, because it effectively has 2 states per site which are $-1$, $1$. We can equivalently write the above expression as,
\begin{equation}\label{eq: expt value for 1 site collective op}
    \langle O \rangle=n_{-}\langle J^{+}\rangle +n_{+}\langle J^{-}\rangle +2n_{z}\langle J^{z}\rangle +n_{0}=2n_{z}\langle J^{z}\rangle +n_{0}=n_{z}(2m-L)+n_{0}
\end{equation}
where $n^{+}=n^{x}+in^{y}$ and $n^{-}=n^{x}-in^{y}$. Now we move on to compute the expectation value of $O^{2}$, and note that it has a relatively complex expression. We try to simplify this expression by considering that $O$ is a $\pi $ translationally invariant operator so,
\begin{align}
    \nonumber O=&n_{x} \sum_{i} \bar{\sigma}^{x}_{i}+n_{y} \sum_{i} \bar{\sigma}^{y}_{i}+n_{z} \sum_{i} \bar{\sigma}^{z}_{i}+n_{0} \sum_{i} \bar{\mathds{1}}_{i}+p_{x} \sum_{i} {\gamma}^{x}_{i}+p'_{x} \sum_{i} \bar{\gamma}^{x}_{i}+p_{y} \sum_{i} {\gamma}^{y}_{i}+p'_{y} \sum_{i} \bar{\gamma}^{y}_{i}+p_{z} \sum_{i} {\gamma}^{z}_{i}  \\
    O=&n_{-} J^{+}+n_{+} J^{-} +2n_{z} J^{z} +n_{0} \mathds{1} +p_{x} \sum_{i} {\gamma}^{x}_{i}+p'_{x} \sum_{i} \bar{\gamma}^{x}_{i}+p_{y} \sum_{i} {\gamma}^{y}_{i}+p'_{y} \sum_{i} \bar{\gamma}^{y}_{i}+p_{z} \sum_{i} {\gamma}^{z}_{i}
\end{align}
From the above decomposition we see that $O^{2}$ contains 3 types of terms $1^{st}$ the type of terms formed by the generators of SGA i.e. $J^{\alpha}, \, \alpha=+,-,z$, $2^{nd}$ type of term is formed by the terms involving one of the generators of SGA and then one term from the $\gamma$ matrices, and note that the expectation values of such terms would be zero, as only 1 $\gamma$ matrix would create a zero at any site and since scar states don't have any zero local state, we would get the inner product to be 0. The $3^{rd}$ type of terms arise from the product of $\gamma$ matrices, which in principle may have a non-zero expectation value. Let's first consider the case when the coefficients of the $\gamma$ matrices i.e. $p_{x}=p'_{x}=p_{y}=p'_{y}=p_{z}=0$ are zero, then
\begin{align*}
    \langle O \rangle=n_{z}(2m-L) +n_{0}
\end{align*}
\begin{align*}
    \langle O^{2} \rangle=n_{{+}}n_{{-}}\langle J^{+}J^{-}+J^{-}J^{+} \rangle +4n_{z}^{2}\langle (J^{Z})^{2}\rangle+4n_{z}n_{0}\langle J^{Z}\rangle+n_{0}^{2}
\end{align*}
where in considering the expectation value of $O^{2}$ with respect to the scar state $|S_{n}\rangle$ we can just drop the terms which have only one raising or lowering terms as they change the scar state from $|S_{n}\rangle$ to $|S_{n\pm1}\rangle$, such terms arise by the product of $J^{\pm}$ and $\mathds{1}$ and this state is orthogonal to $|S_{n}\rangle$, so the inner product is zero. We can find the value of the expectation value of $J^{+}J^{-}+J^{-}J^{+}$ by using the fact that the generators of the SGA have a $SU(2)$ algebra so, 
\begin{align*}
    J^{-}J^{+}+J^{+}J^{-}&=2(J^{2}-(J^{z})^{2}) \\
    \implies \langle J^{-}J^{+}+J^{+}J^{-}\rangle&=2\langle(J^{2}-(J^{z})^{2})\rangle  \\
     &=2\left(\dfrac{L}{2}\left(\dfrac{L}{2}+1\right)-\left(m-\dfrac{L}{2}\right)^{2}\right)  \\
     &=2m(L-m)+L
\end{align*}
Using this, we get
\begin{align*}
    \langle O^{2} \rangle=&n_{{+}}n_{{-}}\left(2m(L-m)+L\right) +4n_{z}^{2}\left(m-\dfrac{L}{2}\right)^{2}+4n_{z}n_{0}\left(m-\dfrac{L}{2}\right)+n_{0}^{2} \\
    \langle O^{2} \rangle=&n_{{+}}n_{{-}}\left(2m(L-m)+L\right)+\left(n_{z}\left(2m-L\right)+n_{0}\right)^{2} 
\end{align*}
\begin{align}\label{eq:Variance without gamma}
    (\Delta O)^{2}=n_{+}n_{-}\left(2m(L-m)+L\right)
\end{align}
for states in middle of the spectrum we see that $m\sim L/2$ which means QFI scales as $N^{2}$. From here, one can see that any operator that preserves the scar subspace will have an extensive QFI density. The $SU(2)$ scar subspace is very important in getting the extensive QFI density. If we don't have this structure, there will be no scars and also no extensive QFI density.
As an example, we calculate the QFI for the perturbation in the main text, which is $V=\sum_{i} (-1)^{i}(S^{x}_{i})^{2}$ which becomes $\tilde{V}=\sum_{i} (S^{x}_{i})^{2}$ after the local unitary transformation, we break the operator $S^{x}_{i}$ as $S^{x}_{i}=(S^{+}_{i}+S^{-}_{i})/2$ and do the decomposition of the perturbation in the SGA operators. 
\begin{align*}
    \tilde{V} &= \dfrac{1}{4}\sum_{i} (S^{+}_{i}+S^{-}_{i})^{2} = \dfrac{1}{4} \sum_{i} \left((S^{+}_{i})^{2}+(S^{-}_{i})^{2} + S^{+}_{i}S^{-}_{i} +S^{-}_{i}S^{+}_{i} \right) \\
    \tilde{V} &= \dfrac{1}{2} J^{+} + \dfrac{1}{2} J^{-} +\dfrac{1}{4}\sum_{i}\left( S^{+}_{i}S^{-}_{i} +S^{-}_{i}S^{+}_{i}\right) \\
    \tilde{V} &= \dfrac{1}{2} J^{+} + \dfrac{1}{2} J^{-} +\dfrac{1}{2}\sum_{i}\left(S^{2}_{i}-(S^{z}_{i})^{2}\right) \\
    \tilde{V} &= \dfrac{1}{2}J^{+} + \dfrac{1}{2} J^{-} +\dfrac{1}{2}\mathds{1} + \gamma^{z}
\end{align*}
Here, $\gamma^{z}$ is a shorthand notation for $\sum_{i} \gamma_{i}^{z}$, and $\mathds{1}$ is shorthand for $\sum_{i} \Bar{\mathds{1}}_{i}$, Now when we do the averaging with respect to a scar state $|S_{m}\rangle$ with $J_{z}|S_{m}\rangle = m|S_{m}\rangle$, note that $\langle \gamma^{z}\rangle = 0$ since $\gamma^{z}$ gives zero when local states $|1\rangle, |-1\rangle$ act on it ( A mathematical explanation of this is explained in the next section where we calculate the expectation values using the reduced density matrices)  and also $(\gamma^{z})^{2} = \gamma^{z}$, hence the $\gamma^{z}$ term wont contribute to the variance of V, and following the same derivation as above, we can see that, 
\begin{align}
    4(\Delta V)^{2} = 2m(L-m)+L
\end{align}
where the variance is with respect to $|S_{m}\rangle$ state, for the scar state in the middle of the spectrum i.e., $m=L/2$ for even system size, and $m=(L-1)/2$ for odd system size, gives,
\[
f_V =
\begin{cases}
\dfrac{L}{2} + 1, & L \ \text{is even}, \\[6pt]
\dfrac{L}{2} + 1 - \dfrac{1}{2L}, & L \ \text{is odd}.
\end{cases}
\]
These analytical expressions match exactly with the QFI we get numerically for the above perturbation.

\begin{figure}
    \centering
    \includegraphics[width = 0.6\linewidth]{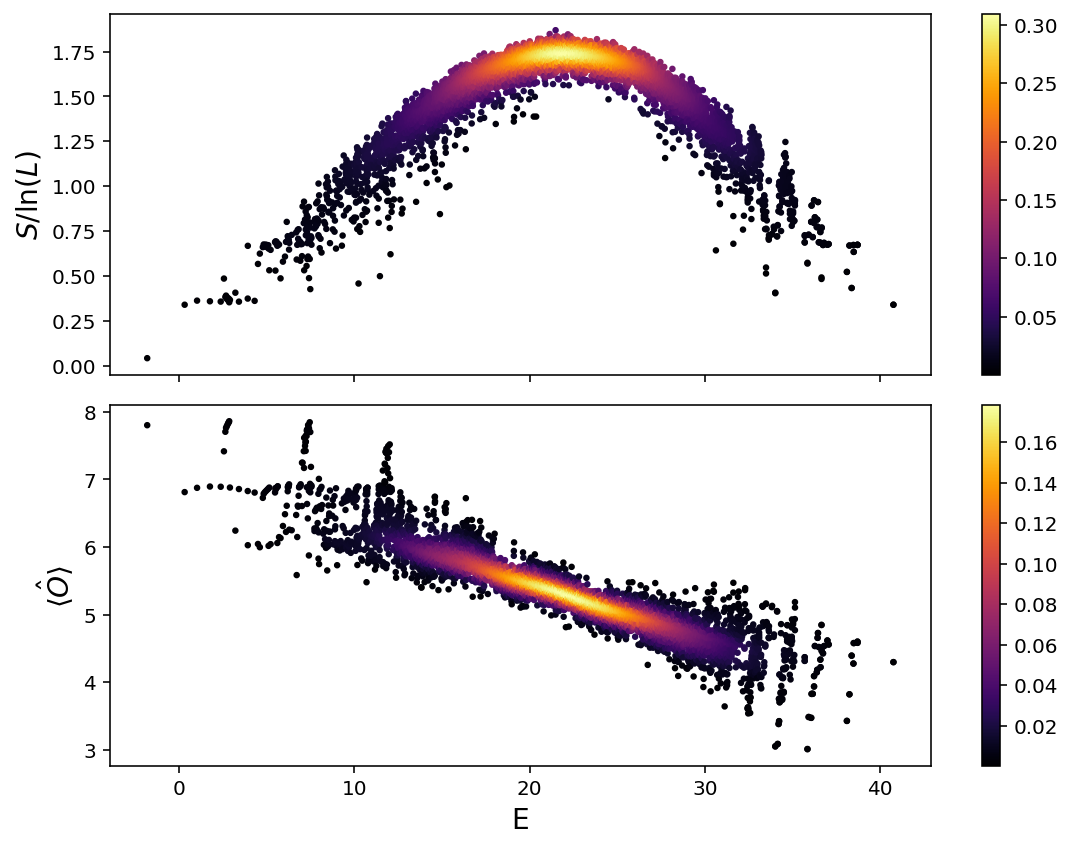}
    \caption{The Entanglement entropy and diagonal ETH for the spin 1XY model + the perturbation $V_{2}=\sum_{i} (S^{x}_{i})^{2}$, for system size $L=8$}
    \label{fig:sum X^2 all same bond}
\end{figure}

Important thing to note here is that the perturbations that preserve the permutation symmetry and also dont introduce the $|0\rangle$ state locally will only have overlap with the SGA operators and hence have a QFI density scaling which is extensive. If we break the permutation symmetry by considering a perturbation $V_{2}=\sum_{i} (S^{x}_{i})^{2}$ which transforms to $\tilde{V_{2}}=\sum_{i} (-1)^{i}(S^{x}_{i})^{2}$ under the local unitary transformation, and this perturbation has no permutation symmetry because of the phases, and we can easily see that it has no over lap with the SGA, and hence should break the scar structure, to supplement this, we numerically add the perturbation and diagonalize and see there are indeed no scars as shown in Fig.~\eqref{fig:sum X^2 all same bond}.

We hereby, claim that any perturbation constructed out of the local site matrices, $\bar{\sigma}_{x}, \bar{\sigma}_{y}, \bar{\sigma}_{z}, \bar{\mathds{1}}, \gamma_{z}$ with a $\pi$ momentum, i.e. $\hat{o} = a_{x}\bar{\sigma}_{x} + a_{y}\bar{\sigma}_{x} +a_{z}\bar{\sigma}_{z}+ a_{0} \bar{\mathds{1}}+a_{\gamma} \gamma_{z} $. Note that this also includes set of perturbations that are non-hermitian, hence implying the presence of scars even in certain non-hermitian systems.

\subsection{Using the expression for the reduced density matrix}
The other way to calculate the QFI is to rely on the reduced density matrix expressions, and from here one can calculate the QFI for any type of operator, including a non-translationally invariant operator, which cannot be calculated by the above SGA technique. This method of evaluating the QFI for operators is much more general and useful than the previous way using SGA.
\newline
We would again like to point out that we are working in the rotated basis as mentioned above. Note that for a pure state,
\begin{align*}
    \rho&=|\psi\rangle\langle\psi|=\sum c^{i'_{1} i'_{2} ... i'_{N}}_{i_{1} i_{2} ... i_{N}} |i_{1} i_{2} ... i_{N} \rangle \langle i'_{1} i'_{2} ... i'_{N}|  \\
    \implies Tr(\rho \hat{o_{j}})&=Tr \left(\sum c^{i'_{1} i'_{2} ... i'_{N}}_{i_{1} i_{2} ... i_{N}} \hat{o}_{j}|i_{1} i_{2} ... i_{N} \rangle \langle i'_{1} i'_{2} ... i'_{N}|\right) \\
    \implies Tr(\rho \hat{o_{j}})&=\sum c^{i'_{1} i'_{2} ... i'_{N}}_{i_{1} i_{2} ... i_{N}} Tr\left( \hat{o}_{j}|i_{1} i_{2} ... i_{N} \rangle \langle i'_{1} i'_{2} ... i'_{N}| \right) = \sum c^{i'_{1} i'_{2} ... i'_{N}}_{i_{1} i_{2} ... i_{N}} Tr\left( \langle i'_{j}| \hat{o}_{j}|i_{j}\rangle|i_{1} i_{2} ... i_{N} \rangle \langle i'_{1} i'_{2} ... i'_{N}| \right)  \\
    \implies Tr(\rho \hat{o_{j}})&=\sum c^{i'_{1} i'_{2} ... i'_{N}}_{i_{1} i_{2} ... i_{N}} Tr_{j}\left(  \hat{o}_{j}|i_{j}\rangle\langle i'_{j}|\right) \left(Tr_{\text{sites except j}}|i_{1} i_{2} ... i_{N} \rangle \langle i'_{1} i'_{2} ... i'_{N}| \right) \\
    &= \sum_{i_{j}, i'_{j}}Tr_{j}\left(  \hat{o}_{j}|i_{j}\rangle\langle i'_{j}|\right)
    \underbrace{\left(Tr_{\text{sites except j}}\sum c^{i'_{1} i'_{2} ... i'_{N}}_{i_{1} i_{2} ... i_{N}}|i_{1} i_{2} ... i_{N} \rangle \langle i'_{1} i'_{2} ... i'_{N}| \right)}_{\rho^{(1)}_{j}}
\end{align*}
We see that, $Tr(\rho \hat{o}_{j})=Tr(\rho^{(1)} \hat{o}_{j})$ and now for collective operators, we simply sum up the individual expectation values. We now calculate the expectation values of operators in scarred states $|S_{m}\rangle$ using this. We again consider the decomposition of $\hat{o}_{i}$ and see that,
\begin{align*}
    Tr(\rho \, \hat{o}_{j})=2n_{+}Tr(\rho^{(1)}(m)\bar{\sigma}^{+}_{j})+2n_{-}Tr(\rho^{(1)}(m)\bar{\sigma}^{-}_{j})+2n_{z}Tr(\rho^{(1)}(m)\bar{\sigma}^{z}_{j})+n_{0}Tr(\rho^{(1)} (m)\mathds{1}_{j})\\
    +p_{x}Tr(\rho^{(1)}(m)\gamma^{x}_{j})+p'_{x}Tr(\rho^{(1)}(m)\bar{\gamma}^{x}_{j})+p_{y}Tr(\rho^{(1)}(m)\gamma^{y}_{j})+p'_{y}Tr(\rho^{(1)}(m)\gamma^{y}_{j})+p_{z}Tr(\rho^{(1)}(m)\gamma^{z}_{j})
\end{align*}
Now we have the expression for $\rho^{(1)}(m)$ which is the 1-body reduced density matrix for the scar state $|S_{m}\rangle$, performing those traces using~\eqref{eq: 1 body reduced density matrix} we get,
\begin{align}
    \nonumber \langle \hat{o}_{j}\rangle&=\dfrac{n_{z}}{L}(2m-L)+\dfrac{n_{0}}{L}  \\
    \implies \langle \sum_{j} \hat{o}_{j}\rangle&=n_{z}(2m-L)+n_{0}
\end{align}
which matches exactly with the calculation that we did in the first approach, i.e.~\eqref{eq: expt value for 1 site collective op}. Also note that the term $Tr(\rho^{(1)}\gamma^{z}_{j}) =0$, which is what we gave a explanation for in the previous section, but here it is apparent from the reduced density matrix structure as to why the expectation value vanishes. Next we calculate $\langle O^{2} \rangle$,
\begin{multline*}
    O^{2}= n_{-}^{2} (J^{+})^{2}+n_{+}n_{-} (J^{+}J^{-}+J^{-}J^{+})+n_{-}n_{0} J^{+}+ 2n_{-}n_{z} (J^{+}J^{z}+J^{z}J^{+})+4n_{z}^{2}(J^{z})^{2} +n_{0}^{2}+4n_{z}n_{0}J^{z} \\
    +... (\text{Terms comprised of $\gamma$ matrices)}
\end{multline*}
We have not written the whole expansion of $O^{2}$ we just wrote a few terms which represent the type of terms that one will encounter, but now different types of terms will vanish when we take expectation value of $O^{2}$ like $(J^{+})^{2}$ so we see that, 
\begin{align*}
    \langle O^{2} \rangle=&n_{+}n_{-}\langle J^{+}J^{-}+J^{-}J^{+} \rangle+4n_{z}^{2}\langle(J^{z})^{2}\rangle+4n_{z}n_{0}\langle J^{z}\rangle+n_{0}^{2}+... \text{Terms due to $\gamma$ matrices}
\end{align*}
From here one can evaluate each of the expectation values by simply expanding the collective operators and then computing the expectation w.r.t. the 2-body reduced density matrix, to illustrate this, consider,
\begin{align*}
    4J^{+}J^{-}=&\sum_{i,j, i\neq j}\bar{\sigma}^{+}_{i}\bar{\sigma}^{-}_{j}+\sum_{i} \bar{\sigma}^{+}_{i}\bar{\sigma}^{-}_{i} \\
    4\langle J^{+}J^{-} \rangle=& \sum_{i,j, i\neq j}Tr(\rho^{(2)}(m)\bar{\sigma}^{+}_{i}\bar{\sigma}^{-}_{j})+\sum_{i} Tr(\rho^{(1)}(m)\bar{\sigma}^{+}_{i}\bar{\sigma}^{-}_{i}) \\
    4\langle J^{+}J^{-} \rangle=&\sum_{i}\dfrac{4m}{L}+\sum_{i,j, i\neq j} \dfrac{4m(L-m)}{L(L-1)} =4\left(m+m(L-m)\right) \\
    \implies \langle J^{+}J^{-}\rangle=& m(L-m)+m
\end{align*}
Now we do the same the same thing with the operator $J^{-}J^{+}$ and we get,
\begin{align*}
    4J^{-}J^{+}=&\sum_{i,j, i\neq j}\bar{\sigma}^{-}_{i}\bar{\sigma}^{+}_{j}+\sum_{i} \bar{\sigma}^{-}_{i}\bar{\sigma}^{+}_{i} \\
    4\langle J^{-}J^{+} \rangle=& \sum_{i,j, i\neq j}Tr(\rho^{(2)}(m)\bar{\sigma}^{-}_{i}\bar{\sigma}^{+}_{j})+\sum_{i} Tr(\rho^{(1)}(m)\bar{\sigma}^{-}_{i}\bar{\sigma}^{+}_{i}) \\
    4\langle J^{-}J^{+} \rangle=&\sum_{i}\dfrac{4(L-m)}{L}+\sum_{i,j, i\neq j} \dfrac{4m(L-m)}{L(L-1)} =4\left(L-m+m(L-m)\right) \\
    \implies \langle J^{-}J^{+}\rangle=& m(L-m)+L-m
\end{align*}
Using the above expressions, we get
\begin{align*}
    \langle J^{+}J^{-}+J^{-}J^{+}\rangle=2m(L-m)+L
\end{align*}
which again matches with the exact expression from the earlier SGA method. Rest of the arguments go the same way like the only terms contributing would be the terms that bring me back to the same scar state, this expression matches exactly with the previous method. Also, it can be used to show the super-extensive nature of the Quantum Fisher information.\newline
Again, as an example here, we take the 2 types of perturbations mentioned in the text when discussing about the 3 different scaling regimes i.e. $V_{3} = \sum_{i} S^{x}_{i}$ and $V_{4} = S^{x}_{4}$.
Starting with $V_{4}$, and using the expression for the reduced density matrix eq.~\eqref{eq: 1 body reduced density matrix}, and using the expression of $S^{x}$ operator eq.\eqref{eq:spin 1 matrices}
\begin{align*}
    \langle S^{x}_{4}\rangle &= Tr(\rho^{(1)} S^{x}_{4}) = \dfrac{1}{\sqrt{2}}Tr\left(
    \begin{pmatrix}
    1/2 & 0 & 0 \\
    0 & 0 & 0 \\
    0 & 0 & 1/2
    \end{pmatrix}
    \begin{pmatrix}
    0 & 1 & 0 \\
    1 & 0 & 1 \\
    0 & 1 & 0
    \end{pmatrix}
    \right)   \\
    \implies \langle S^{x}_{4}\rangle &= 0
\end{align*}
Now we calculate $\langle (S^{x}_{4})^{2}\rangle$, 
\begin{align}\label{eq:sx^2 expectation}
    \nonumber \langle (S^{x}_{4})^{2}\rangle &= Tr(\rho^{(1)} S^{x}_{4}) = \dfrac{1}{2}Tr\left(
    \begin{pmatrix}
    1/2 & 0 & 0 \\
    0 & 0 & 0 \\
    0 & 0 & 1/2
    \end{pmatrix}
    \begin{pmatrix}
    1 & 0 & 1 \\
    0 & 2 & 0 \\
    1 & 0 & 1
    \end{pmatrix}
    \right)   \\
    \implies \langle (S^{x}_{4})^{2}\rangle &=  \dfrac{1}{2}
\end{align} 
This implies that the variance for the perturbation is, 
\begin{align}
    (\Delta V_{4})^{2} = \langle (S^{x}_{4})^{2}\rangle - (\langle S^{x}_{4}\rangle)^{2} = \dfrac{1}{2}
\end{align}
This means that QFI, i.e.,$F_{V_{4}}=4(\Delta V_{4})^{2}=2$, or the QFI density is $f_{V_{4}}=2/L$, which is what we see in the figure in the main text too.
Next, for the perturbation $V_{3}=\sum_{i} S^{x}_{i}$, we will apply the local unitary transformation to remove the phases of alternating $-1$ in the scar states, in that bases $V_{3}$ is $\tilde{V}_{3} = \sum_{i} (-1)^{i}S^{x}_{i}$, from the above analysis it is clear that, $\langle \tilde{V}_{3}\rangle = 0$, we now want $\langle (\tilde{V}_{3})^{2}\rangle$, to get this, we expand as,
\begin{align}
    \langle \tilde{V}_{3}^{2} \rangle &= \sum_{i, j} \langle S^{x}_{i}S^{x}_{j} \rangle = \sum_{i,j, i\neq j} \langle S^{x}_{i}S^{x}_{j}\rangle + \sum_{i} \langle (S^{x}_{i})^{2}\rangle   \\
    \implies \langle V_{3}^{2} \rangle &= \sum_{i, j, i\neq j} Tr\left(\rho^{(2)}S^{x}_{i}S^{x}_{j}\right) + \sum_{i} Tr \left( \rho^{(1)} (S^{x}_{i})^{2}\right)
\end{align}
Now we will use the fact that the scar states are permutation symmetric which means that the 2 body reduced density matrix is the same for any 2 sites, and then using eq.\eqref{eq: 2body reduced density matrix} and eq.\eqref{eq:spin 1 matrices} we compute the trace, and see that $\langle S^{x}_{i}S^{x}_{j}\rangle = 0$, and from the above example using eq.~\eqref{eq:sx^2 expectation}, we can see that, 
\begin{align}
    4(\Delta V_{3})^{2} &= 4\left(\langle V_{3}^{2}\rangle-(\langle V_{3} \rangle)^{2}\right) \\
    F_{V_{3}} &= 2L
\end{align}
This is exactly what we see for the perturbation $V_{3}$ in the main text: the QFI density is constant and is 2, i.e., $f_{V_{3}} = 2$. Note that here we only focused on operators that have overlap with only the SGA generators; there is still a possibility that the decomposition may involve SGA generators as well as operators outside SGA, and may still give an extensive QFI

\section{Non-integrability of the long-range Spin 1 XY model}
\begin{figure}[h]
  \centering
  \includegraphics[width=0.6\linewidth]{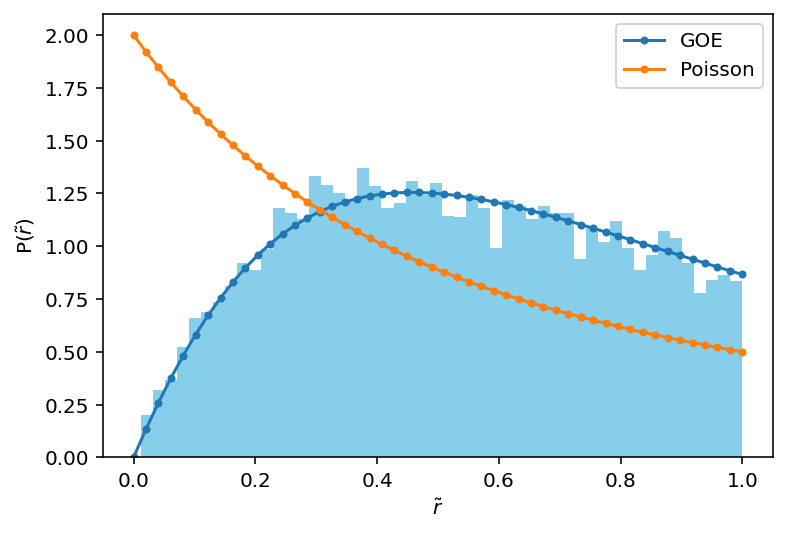}
  \caption{Level-ratio statistics for the long-range Spin-1 XY model for parameters $J_1=1$, $J_3=0.1$, $J_5=0.2$, $ J_7=0.5$, $J_9= 0.4$, $J_{11}=0.6$, $J_{13}=0.9$, $ D=0.1$, $h=1 $ and system size $L=14$, magnetization $m=-8$, state parity $I=-1$ under chain inversion.}
  \label{fig:level-ratio_long range}
\end{figure}

The long-range Spin 1XY model, as mentioned in the main text, is non-integrable. This can be seen from the level-ratio statistics data in Fig.~\eqref{fig:level-ratio_long range}. Within a \(U(1)\) magnetization sector, we find Wigner–Dyson (GOE-like) level statistics, as diagnosed by the adjacent-gap ratio \(r_n=\min(\delta_n,\delta_{n+1})/\max(\delta_n,\delta_{n+1})\) with \(\delta_n=E_{n+1}-E_n\)~\cite{PhysRevB.75.155111, PhysRevLett.110.084101}. The GOE benchmark is \(\langle r\rangle\simeq 0.5307\). Our data (Fig.~\ref{fig:level-ratio_long range}) yield \(\langle r\rangle\approx 0.528\), consistent with GOE predictions in the symmetry-resolved sector. 

\end{document}